\documentclass[floats,twocolumn,pra]{revtex4}%
\usepackage{amssymb}
\usepackage{graphicx}
\usepackage{amsmath}
\usepackage{amsfonts}
\usepackage{accents}
\newtheorem{theorem}{Theorem}
\newtheorem{lemma}[theorem]{Lemma}
\newtheorem{proposition}[theorem]{Proposition}
\newtheorem{corollary}[theorem]{Corollary}
\newtheorem{definition}[theorem]{Definition}
\newtheorem{remark}[theorem]{Remark}

\begin{document}
\title{Capacities of repeater-assisted quantum communications}
\author{Stefano Pirandola}
\affiliation{Computer Science and York Centre for Quantum Technologies, University of York,
York YO10 5GH, United Kingdom}

\begin{abstract}
We consider quantum and private communications assisted by repeaters, from the
basic scenario of a single repeater chain to the general case of an
arbitrarily-complex quantum network, where systems may be routed through
single or multiple paths. In this context, we investigate the ultimate rates
at which two end-parties may transmit quantum information, distribute
entanglement, or generate secret keys. These end-to-end capacities are defined
by optimizing over the most general adaptive protocols that are allowed by
quantum mechanics. Combining techniques from quantum information and classical
network theory, we derive single-letter upper bounds for the end-to-end
capacities in repeater chains and quantum networks connected by arbitrary
quantum channels, establishing exact formulas under basic decoherence models,
including bosonic lossy channels, quantum-limited amplifiers, dephasing and
erasure channels. For the converse part, we adopt a teleportation-inspired
simulation of a quantum network which leads to upper bounds in terms of the
relative entropy of entanglement. For the lower bounds we combine
point-to-point quantum protocols with classical network algorithms. Depending
on the type of routing (single or multiple), optimal strategies corresponds to
finding the widest path or the maximum flow in the quantum network. Our theory
can also be extended to simultaneous quantum communication between multiple
senders and receivers.

\end{abstract}
\maketitle

\section{Introduction}

The field of quantum communications is gradually evolving into network
implementations. In this scenario, a well developed area is certainly quantum
key distribution (QKD)~\cite{BB84,Ekert,GisinRMP,Scaranim}, which has been the
first quantum task to be extended to network
architectures~\cite{net1,net3,net4,To1,To2,To4}, with high-rate prototypes
recently devised for metropolitan
applications~\cite{MDI1,CVMDIQKD,Correspondence,metro1,metro2}. Quantum
teleportation~\cite{teleBENNETT,Samtele,Samtele2} is another fundamental
protocol with remarkable advances~\cite{telereview}. It is considered to be
the best option to transfer quantum information among the nodes of a future
quantum Internet~\cite{Kimble,Rod}, most likely based on the integration of
different substrates and technologies~\cite{HybridINTERNET}.

The construction of a quantum network not only aims at connecting and
delivering quantum services to multiple users, but also addresses a precise
physical problem: Extending the range of the quantum communications. In fact,
because quantum signals are fragile to loss and noise, the performances of any
direct point-to-point protocol is limited in both distance and rate. Closing
an investigation started in 2009~\cite{RevCohINFO,ReverseCAP,TGW},
Ref.~\cite{QKDpaper} has finally established the maximum rates at which two
remote parties can transmit quantum information, distill entanglement or
generate secret keys over a lossy communication line with transmissivity
$\eta$. These are all equal to~\cite{QKDpaper}
\begin{equation}
\mathcal{C}(\eta)=-\log_{2}(1-\eta)\simeq1.44\eta\text{ bits/use for }%
\eta\simeq0\text{.} \label{PLOBbound}%
\end{equation}

Here $\mathcal{C}(\eta)$ is a repeaterless bound that cannot be violated by
any point-to-point protocol, even if the parties exploit the most general
local operations (LOs) assisted by unlimited and two-way classical
communication (CC), briefly called \textquotedblleft adaptive
LOCCs\textquotedblright~\cite{QKDpaper,GenBenchmarks}. In quantum information
theory, $\mathcal{C}(\eta)$ is what is called a \textquotedblleft two-way
assisted capacity\textquotedblright. Depending on the task, it may represent a
two-way quantum capacity $Q_{2}$ (transmission of qubits), a two-way
entanglement distribution capacity $D_{2}$ (distribution of entanglement bits,
i.e., ebits), or a secret key capacity $K$ (generation of secret bits).

From a methodological point of view, Ref.~\cite{QKDpaper} has rigorously
devised the technique of teleportation stretching (also known as `reduction by
teleportation') by extending and generalizing the precursory (but restrictive)
simulation tools of Ref.~\cite{B2main}. For the first time,
Ref.~\cite{QKDpaper} showed how to apply this technique to completely simplify
adaptive protocols of private communication at any dimension and how to
combine this reduction with the properties of the relative entropy of
entanglement (REE)~\cite{RMPrelent,VedFORMm,Pleniom}. This insight led to
establish a single-letter upper bound for the two-way capacities of an
arbitrary quantum channel, discovering many of these capacities after the
first investigations started about 20 years ago~\cite{ErasureChannel}. The
crucial insight of Ref.~\cite{QKDpaper} has been exploited in a number of
other works, e.g., see the various applications in
Refs.~\cite{boradLAU,FiniteSTRET,nonPauli,Gu,Metro}.

To overcome the limitations imposed by Eq.~(\ref{PLOBbound}), one needs to
design a multi-hop quantum network which exploits the assistance of quantum
repeaters~\cite{Briegel,Rep2,Rep3,Rep4,Rep5,Rep6,Rep7,Rep8,Rep9,Rep10,Rep12,Rep13,Rep13bis,Rep14,Rep15,Rep16}%
. The advantage of introducing a quantum repeater can be explained with a
simple example. Suppose that Alice and Bob are connected by an optical fiber
with transmissivity $\eta$, such that the two-way capacity $\mathcal{C}(\eta)$
is zero. Split the fiber in two identical portions and introduce a middle
relay. Because each segment is a lossy channel with transmissivity $\sqrt
{\eta}$, both Alice-relay and relay-Bob can reach the capacity $\mathcal{C}%
(\sqrt{\eta})>\mathcal{C}(\eta)$. Combining the outputs, e.g., composing keys
or swapping entanglement, $\mathcal{C}(\sqrt{\eta})>0$ becomes an achievable
rate for the entire repeater-assisted communication between Alice and Bob.

Once understood that quantum repeaters may increase rates (and therefore
distances), it is fundamental to ask the next crucial question: \textit{What
are the ultimate rates that are achievable with their assistance}? In general,
we may consider increasing forms of repeater-assisted quantum communications,
from the basic case of a repeater chain to an arbitrarily-complex quantum
network, where systems may be routed through single or multiple paths. In
these scenarios, it is an open problem to determine the optimal rates that are
achievable by two end-users, i.e., to establish their end-to-end capacities
for transmitting qubits, distributing ebits and generating secret keys,
assuming the most general network protocols.

In this work, we address this fundamental question by combining methods from
quantum information theory~\cite{N&C02,RMP,BraRMP,HolevoBOOK} and classical
network theory~\cite{Slepian,Schrijver,Gamal,Cover&Thomas,netflow}. First of
all, we derive upper bounds for the end-to-end capacities in repeater chains
and, more generally, arbitrary quantum networks, with single- or multiple-path
routings of systems. These bounds are derived assuming completely arbitrary
quantum channels connecting the repeaters (nodes of the quantum network). The
key methodology relies in a suitable generalization of teleportation
stretching~\cite{QKDpaper}, where an entire quantum network (or a repeater
chain) can be simulated by replacing quantum channels with corresponding
resource states. Combining this representation with suitable entanglement cuts
of the network (or the chain), we derive single-letter upper bounds in terms
of the REE~\cite{RMPrelent,VedFORMm,Pleniom}. Such bounds are further
simplified if the quantum channels are teleportation-covariant~\cite{QKDpaper}%
, i.e., suitably commutes with random unitaries induced by teleportation.
Furthermore, these bounds can also be extended to simultaneous communication
between multiple end-users.

Most importantly, for two end-users we show that the upper bounds are
achievable under fundamental noise models for discrete variable (DV) and
continuous variable (CV) systems, including bosonic loss which is the most
relevant in quantum optical communications. More precisely, the achievability
is proven in networks (or chains) which are connected by bosonic lossy
channels, quantum-limited amplifiers, dephasing or erasure channels in
arbitrary finite dimension. In these cases, we therefore establish exact
formulas for the various end-to-end quantum and private capacities. Depending
on the type of routing in the quantum network (single- or multi-path), optimal
strategies can be found by solving the widest path~\cite{Pollack,MITp} or the
maximum flow
problem~\cite{Harris,Ford,ShannonFLOW,netflow,Karp,Dinic,Alon,Ahuja,Cheriyan,King,Orlin}
suitably extended from the classical to the quantum setting.

The manuscript has the following structure. In Sec.~\ref{mainRES} we summarize
our main results for two end-users in a repeater chain or a quantum network.
Subsequent Secs.~\ref{GeneralSECTION}-\ref{secMULTI} are technical and provide
full background and detailed proofs. We start in Sec.~\ref{GeneralSECTION}
with a review of preliminary methods for simplifying quantum and private
communications~\cite{QKDpaper}. Then, in Sec.~\ref{SECrepeaters} we discuss
the detailed methods and results for chains of quantum repeaters. In following
Secs.~\ref{SECnetworks} and \ref{SecSTRETCHINGNET}, we provide exact
definitions and tools for quantum networks. Bounds and formulas for their
capacities are given in Secs.~\ref{secSINGLE} and~\ref{secMULTI}, under
different types of routing. Then, in Sec.~\ref{SECmultipleNETs}\ we extend the
theory beyond the results of Sec.~\ref{mainRES}, considering networks with
multiple end-users. Sec.~\ref{SECconclusions} is for conclusions.

\section{Main results for two end-users\label{mainRES}}

\subsection{Ultimate limits of repeater chains}

\begin{figure}[ptbh]
\vspace{-2.5cm}
\par
\begin{center}
\includegraphics[width=0.48\textwidth] {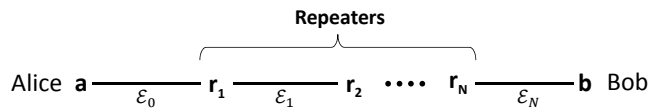} \vspace{-3.2cm}
\end{center}
\caption{Linear chain of $N$ quantum repeaters $\mathbf{r}_{1},\ldots
,\mathbf{r}_{N}$ between the two end-users, Alice $\mathbf{a}:=\mathbf{r}_{0}$
and Bob $\mathbf{b}:=\mathbf{r}_{N+1}$. The chain is connected by $N+1$
quantum channels $\{\mathcal{E}_{i}\}$.}%
\label{intropicmain}%
\end{figure}

Consider Alice $\mathbf{a}$ and Bob $\mathbf{b}$ at the two ends of a linear
chain of $N$ quantum repeaters, labeled by $\mathbf{r}_{1},\ldots
,\mathbf{r}_{N}$. Each point has a local register of quantum systems which may
be augmented with incoming systems or depleted by outgoing ones. As depicted
in Fig.~\ref{intropicmain}, the chain is connected by $N+1$ quantum channels
$\{\mathcal{E}_{0},\ldots,\mathcal{E}_{i},\ldots,\mathcal{E}_{N}\}$ through
which systems are sequentially transmitted. In a single end-to-end
transmission or use of the chain, all the channels are used exactly once.
Assume that the end-points aim to share target bits, which may be ebits or
private bits~\cite{KD1}. The most general quantum distribution protocol
$\mathcal{P}_{\text{\textrm{chain}}}$ involves transmissions which are
interleaved by adaptive LOCCs among all parties, i.e., LOs assisted by two-way
CCs among end-points and repeaters.

After $n$ adaptive uses of the chain, the end-points share an output state
$\rho_{\mathbf{ab}}^{n}$ with $nR_{n}$ target bits. By optimizing the
asymptotic rate $\lim_{n}R_{n}$ over all protocols $\mathcal{P}%
_{\text{\textrm{chain}}}$, we define the generic two-way\ capacity of the
chain $\mathcal{C}(\{\mathcal{E}_{i}\})$. If the target are ebits, the
repeater-assisted capacity $\mathcal{C}$ is an entanglement-distribution
capacity $D_{2}$. The latter coincides with a quantum capacity $Q_{2}$,
because distributing an ebit is equivalent to transmitting a qubit under
two-way CCs. If the target are private bits, then $\mathcal{C}$ is a
secret-key capacity $K\geq D_{2}$ (because ebits are specific types of private
bits).

Exact definitions and relevant methods are explained in
Sec.~\ref{SECrepeaters}. In order to state our upper bound for $\mathcal{C}%
(\{\mathcal{E}_{i}\})$ we need to introduce the notion of channel simulation.
Recall that any quantum channel $\mathcal{E}$ can be simulated by an LOCC
$\mathcal{T}$ applied to the input state $\rho$ and some bipartite resource
state $\sigma$, so that $\mathcal{E}(\rho)=\mathcal{T}(\rho\otimes\sigma
)$~\cite{QKDpaper}. The pair $S:=(\mathcal{T},\sigma)$ represents a possible
\textquotedblleft LOCC\ simulation\textquotedblright\ of the channel. In
particular, for channels that suitably commute with the random unitaries of
teleportation, called \textquotedblleft
teleportation-covariant\textquotedblright\ channels~\cite{QKDpaper}, one finds
that $\mathcal{T}$ is teleportation and $\sigma$ is their Choi matrix
$\sigma_{\mathcal{E}}:=\mathcal{I}\otimes\mathcal{E}(\Phi)$, with $\Phi$ being
a maximally-entangled state. The latter is also known as \textquotedblleft
teleportation simulation\textquotedblright. For bosonic channels, the Choi
matrix is energy-unbounded, so that simulations and functionals need to be
formulated in an asymptotic fashion. Channel simulation is at the core of
teleportation stretching~\cite{QKDpaper} which transforms any adaptive
protocol over a channel $\mathcal{E}$ into a block protocol over copies of its
resource state $\sigma$.

The other notion to introduce is that of entanglement cut between Alice and
Bob. In the setting of a linear chain, a cut \textquotedblleft$i$%
\textquotedblright\ disconnects channel $\mathcal{E}_{i}$ between repeater
$\mathbf{r}_{i}$ and $\mathbf{r}_{i+1}$. Such channel can be replaced by a
simulation $S_{i}$ with some resource state $\sigma_{i}$. Then, we may write
\begin{equation}
\mathcal{C}(\{\mathcal{E}_{i}\})\leq\min_{i}E_{\mathrm{R}}(\sigma_{i}),
\label{chain1}%
\end{equation}
where $E_{\mathrm{R}}(\cdot)$ is the relative entropy of entanglement
(REE)~\cite{RMPrelent} and the minimization is over all possible cuts in the
chain. The tightest bound in Eq.~(\ref{chain1}) is obtained by minimizing over
all possible channel simulations, i.e., by enforcing $\min_{i}\rightarrow
\min_{i}\min_{S_{i}}$. For a chain of teleportation-covariant channels, we may
use their teleportation simulation over Choi matrices and write
\begin{equation}
\mathcal{C}(\{\mathcal{E}_{i}\})\leq\min_{i}E_{\mathrm{R}}(\sigma
_{\mathcal{E}_{i}}). \label{teleKKK}%
\end{equation}

Note that the class of teleportation-covariant channels is very wide,
including Pauli channels (at any dimension)~\cite{N&C02} and bosonic Gaussian
channels~\cite{RMP}. Within such a class, there are channels $\mathcal{E}$
whose generic two-way capacity $\mathcal{C}=Q_{2}$, $D_{2}$ or $K$ satisfies
\begin{equation}
\mathcal{C}(\mathcal{E})=E_{\mathrm{R}}(\sigma_{\mathcal{E}})=D_{1}%
(\sigma_{\mathcal{E}}), \label{one-wayCC}%
\end{equation}
where the latter is the one-way (forward or backward) distillable entanglement
of the Choi matrix. These \textquotedblleft distillable
channels\textquotedblright\ include bosonic lossy channels, quantum-limited
amplifiers, dephasing and erasure channels~\cite{QKDpaper}. For a chain of
distillable channels, we may exactly establish the capacity as%
\begin{equation}
\mathcal{C}(\{\mathcal{E}_{i}\})=\min_{i}\mathcal{C}(\mathcal{E}_{i})=\min
_{i}E_{\mathrm{R}}(\sigma_{\mathcal{E}_{i}}). \label{DistillableCAPP}%
\end{equation}
In fact the upper bound ($\leq$) follows from Eqs.~(\ref{teleKKK})
and~(\ref{DistillableCAPP}). The lower bound ($\geq$) relies on the fact that
an achievable rate for end-to-end entanglement distillation consists in: (i)
each pair of neighbor repeaters, $\mathbf{r}_{i}$ and $\mathbf{r}_{i+1}$,
exchanging $D_{1}(\sigma_{\mathcal{E}_{i}})$ ebits over $\mathcal{E}_{i}$; and
(ii) performing entanglement swapping on the distilled ebits. In this way, at
least $\min_{i}D_{1}(\sigma_{\mathcal{E}_{i}})$ ebits are shared between Alice
and Bob. Thanks to Eq.~(\ref{one-wayCC}), the capacity of distillable chains
is achievable by just using one-way CCs.

\begin{figure}[ptbh]
\begin{center}
\vspace{+0.1cm} \includegraphics[width=0.38\textwidth] {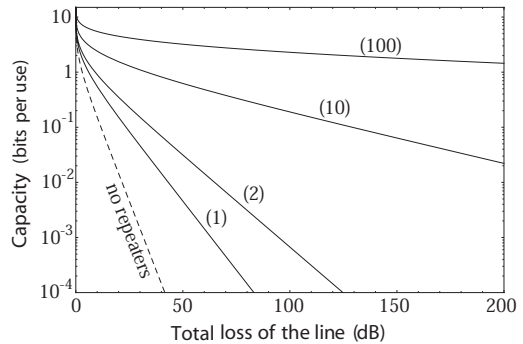}
\vspace{-0.5cm}
\end{center}
\caption{Capacity (target bits per chain use) versus total loss of the line
(decibels, dB) for $N=1,2,10$ and $100$ equidistant repeaters. Compare the
repeater-assisted capacities (solid curves) with the point-to-point bound
(dashed curve).}%
\label{figg}%
\end{figure}

Let us specify Eq.~(\ref{DistillableCAPP}) to basic examples. For a chain of
quantum repeaters connected by lossy channels with transmissivities
$\{\eta_{i}\}$, we find the capacity%
\begin{equation}
\mathcal{C}_{\text{loss}}=-\log_{2}(1-\eta_{\text{min}}),~~\eta_{\text{min}%
}:=\min_{i}\eta_{i}~. \label{EqLOSSY2}%
\end{equation}
Thus, the minimum transmissivity within the lossy chain establishes the
ultimate rate for repeater-assisted quantum/private communications between the
end-users. For instance, consider an optical fiber with transmissivity $\eta$
and insert $N$ repeaters so that the fiber is split into $N+1$ lossy channels.
The optimal configuration corresponds to equidistant repeaters, so that
$\eta_{\text{min}}=\sqrt[N+1]{\eta}$ and the maximum capacity of the lossy
chain is%
\begin{equation}
\mathcal{C}_{\text{loss}}(\eta,N)=-\log_{2}\left(  1-\sqrt[N+1]{\eta}\right)
~. \label{optLOSScap0}%
\end{equation}
This capacity is plotted in Fig.~\ref{figg} and compared with the
point-to-point bound $\mathcal{C}(\eta)=\mathcal{C}_{\text{loss}}(\eta,0)$.
Note that if we want to guarantee a performance of $1$ target bit per use of
the chain, we need at least $\eta_{\text{min}}=1/2$, which corresponds to
$3$dB of maximum loss in each link. This \textquotedblleft$3$dB
rule\textquotedblright\ implies that $1$ bit rate communication can only occur
in a repeater chain whose maximum point-to-point distance is 15km in standard
optical fiber (loss rate of $0.2$dB/km).

In a chain whose repeaters are connected by quantum-limited amplifiers with
gains $\{g_{i}\}$, we find
\begin{equation}
\mathcal{C}_{\text{amp}}=-\log_{2}\left(  1-g_{\max}^{-1}\right)  ,~~g_{\max
}:=\max_{i}g_{i}~.
\end{equation}
For a spin chain where the state transfer between the $i$th spin and the next
one is modeled by a dephasing channel $\mathcal{E}_{i}$ with probability
$p_{i}\leq1/2$, we find%
\begin{equation}
\mathcal{C}_{\text{deph}}=1-H_{2}(p_{\max}),~~p_{\max}:=\max_{i}p_{i}~,
\end{equation}
where $H_{2}$ is the binary Shannon entropy. Finally, for spin chains
connected by erasure channels $\{\mathcal{E}_{i}\}$ with probabilities
$\{p_{i}\}$ we determine $\mathcal{C}_{\text{erase}}=1-p_{\max}$.

\subsection{Quantum networks with single-path routing}

\begin{figure}[ptbh]
\vspace{-1.7cm}
\par
\begin{center}
\includegraphics[width=0.49\textwidth] {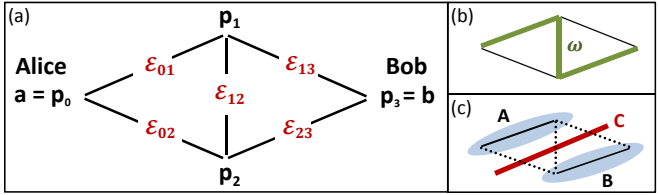} \vspace{-2.4cm}
\end{center}
\caption{Diamond quantum network $\mathcal{N}^{\Diamond}$.~\textbf{(a)}~This
is an elementary quantum network of four points $P=\{\mathbf{p}_{0}%
,\mathbf{p}_{1},\mathbf{p}_{2},\mathbf{p}_{3}\}$, with end-points
$\mathbf{p}_{0}=\mathbf{a}$ (Alice) and $\mathbf{p}_{3}=\mathbf{b}$ (Bob). Two
points $\mathbf{p}_{i}$ and $\mathbf{p}_{j}$ are connected by an edge
$(\mathbf{p}_{i},\mathbf{p}_{j})$ if there is an associated quantum channel
$\mathcal{E}_{ij}$. This channel has a corresponding resource state
$\sigma_{ij}$ in a simulation $S(\mathcal{N})$ of the network. There are four
(simple) routes: $1:\mathbf{a}-\mathbf{p}_{1}-\mathbf{b}$, $2:\mathbf{a}%
-\mathbf{p}_{2}-\mathbf{b}$, $3:\mathbf{a}-\mathbf{p}_{2}-\mathbf{p}%
_{1}-\mathbf{b}$, and $4:\mathbf{a}-\mathbf{p}_{1}-\mathbf{p}_{2}-\mathbf{b}$.
As an example, route $4$ involves the transmission through the sequence of
quantum channels $\{\mathcal{E}_{k}^{4}\}$ which is defined by $\mathcal{E}%
_{0}^{4}:=\mathcal{E}_{01}$, $\mathcal{E}_{1}^{4}:=\mathcal{E}_{12}\ $and
$\mathcal{E}_{2}^{4}:=\mathcal{E}_{23}$. \textbf{(b)}~We explicitly show route
$\omega=4$. In a sequential protocol, each use of the network corresponds to
using a single route $\omega$ between the two end-points, with some
probability $p_{\omega}$. \textbf{(c)}~We show an entanglement cut $C$ of the
network separating the ensembles of points $\mathbf{A}$ and $\mathbf{B}$.
These are connected by the cut-set $\tilde{C}$ composed by the dotted edges.}%
\label{DiamondLL}%
\end{figure}

A quantum communication network can be represented by an undirected finite
graph~\cite{Slepian} $\mathcal{N}=(P,E)$, where $P$ is the set of points of
the network and $E$ is the set of all edges. Each point $\mathbf{p}$\ has a
local register of quantum systems. Two points $\mathbf{p}_{i}$\textbf{
}and\textbf{ }$\mathbf{p}_{j}$ are connected by an edge $(\mathbf{p}%
_{i},\mathbf{p}_{j})\in E$ if there is a quantum channel $\mathcal{E}%
_{ij}:=\mathcal{E}_{\mathbf{p}_{i}\mathbf{p}_{j}}$ between them. A route is an
undirected path $\mathbf{a}-\mathbf{p}_{i}-\cdots-\mathbf{p}_{j}-\mathbf{b}$
between the two end-points, Alice $\mathbf{a}$ and Bob $\mathbf{b}$. These are
connected by an ensemble of possible routes $\Omega=\{1,\ldots,\omega
,\ldots\}$, with the generic route $\omega$ involving the transmission through
a sequence of quantum channels $\{\mathcal{E}_{0}^{\omega},\ldots
,\mathcal{E}_{k}^{\omega}\ldots\}$. Finally, an entanglement cut $C$ is a
bipartition $(\mathbf{A},\mathbf{B})$ of the points $P$ such that
$\mathbf{a}\in\mathbf{A}$ and $\mathbf{b}\in\mathbf{B}$. Any such cut $C$
identifies a super Alice $\mathbf{A}$ and a super Bob $\mathbf{B}$, which are
connected by the cut-set $\tilde{C}=\{(\mathbf{x},\mathbf{y})\in
E:\mathbf{x}\in\mathbf{A},\mathbf{y}\in\mathbf{B}\}$. See Fig.~\ref{DiamondLL}%
. 

\begin{figure}[ptbh]
\vspace{-0.3cm}
\par
\begin{center}
\includegraphics[width=0.49\textwidth] {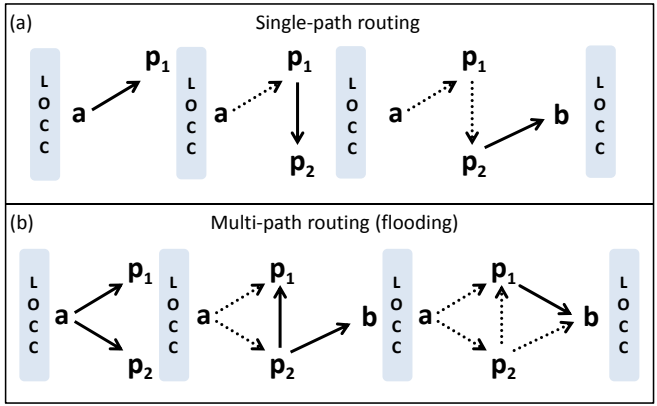} \vspace{-1.1cm}
\end{center}
\caption{Network protocols of quantum/private communication. (a)~In a
sequential protocol, systems are routed from Alice to Bob through a single
path which is probabilistically chosen by the points. Here this is
$\mathbf{a}-\mathbf{p}_{1}-\mathbf{p}_{2}-\mathbf{b}$. Each transmission
occurs between two adaptive LOCCs, where all points of the network perform LOs
assisted by two-way CC. (b)~In a flooding protocol, systems are simultaneously
routed from Alice to Bob through a sequence of multicasts in such a way that
each edge of the network is used exactly once in each end-to-end transmission.
Here we show a possible sequence $\mathbf{a}\rightarrow\{\mathbf{p}%
_{1},\mathbf{p}_{2}\}$, $\mathbf{p}_{2}\rightarrow\{\mathbf{p}_{1}%
,\mathbf{b}\}$, $\mathbf{p}_{1}\rightarrow\{\mathbf{b}\}$. Each multicast
occurs between two adaptive LOCCs.}%
\label{NETproto}%
\end{figure}

In a sequential protocol, the whole network is initialized by a preliminary
network LOCC, where all the points communicate with each other via unlimited
two-way CCs and perform adaptive LOs on their local quantum systems. With some
probability, Alice exchanges a quantum system with repeater $\mathbf{p}_{i}$,
followed by a second network LOCC; then repeater $\mathbf{p}_{i}$ exchanges a
system with repeater $\mathbf{p}_{j}$, followed by a third network LOCC and so
on, until Bob is reached through some route (see Fig.~\ref{NETproto}). For
large $n$ uses of the network, there is a probability distribution associated
with the ensemble $\Omega$, with the generic route $\omega$ being used
$np_{\omega}$ times. Alice and Bob's output state $\rho_{\mathbf{ab}}^{n}$
will asymptotically approximate a target state with $nR_{n}$ bits. By
optimizing over the sequential protocols and taking the limit of large $n$, we
define the sequential or single-path capacity of the network $\mathcal{C}%
(\mathcal{N})$.

Exact definitions and relevant methods are explained in
Secs.~\ref{SECnetworks} and \ref{SecSTRETCHINGNET}. In order to state our
upper bound we need to introduce the\textit{ flow} of REE through a cut. Given
an entanglement cut $C$ of the network, consider its cut-set $\tilde{C}$. For
each edge $(\mathbf{x},\mathbf{y})$ in $\tilde{C}$, we have a channel
$\mathcal{E}_{\mathbf{xy}}$ and a corresponding resource state $\sigma
_{\mathbf{xy}}$ associated with a simulation. Then we define the single-edge
flow of REE across cut $C$ as%
\begin{equation}
E_{\mathrm{R}}(C):=\max_{(\mathbf{x},\mathbf{y})\in\tilde{C}}E_{\mathrm{R}%
}(\sigma_{\mathbf{xy}}). \label{ubseqNN0}%
\end{equation}
The minimization of this quantity over all entanglement cuts provides our
upper bound for the single-path capacity of the network, i.e.,
\begin{equation}
\mathcal{C}(\mathcal{N})\leq\min_{C}E_{\mathrm{R}}(C), \label{ubseqNN}%
\end{equation}
which is the network generalization of Eq.~(\ref{chain1}).

For a network of teleportation-covariant channels, the resource state
$\sigma_{\mathbf{xy}}$ in Eq.~(\ref{ubseqNN0}) is the Choi matrix
$\sigma_{\mathcal{E}_{\mathbf{xy}}}$ of the channel $\mathcal{E}_{\mathbf{xy}%
}$. In particular, for a network of distillable channels, we may also set
\begin{equation}
\mathcal{C}(\mathcal{E}_{\mathbf{xy}})=E_{\mathrm{R}}(\sigma_{\mathcal{E}%
_{\mathbf{xy}}})=D_{1}(\sigma_{\mathcal{E}_{\mathbf{xy}}}), \label{disNETtool}%
\end{equation}
for any edge $(\mathbf{x},\mathbf{y})$. Therefore, we may write the bound of
Eq.~(\ref{ubseqNN}) with $E_{\mathrm{R}}(C)=\mathcal{C}(C)$ where
\begin{equation}
\mathcal{C}(C):=\max_{(\mathbf{x},\mathbf{y})\in\tilde{C}}\mathcal{C}%
(\mathcal{E}_{\mathbf{xy}})
\end{equation}
is the single-edge capacity of a cut. To show the achievability of the upper
bound, we first prove that $\min_{C}\mathcal{C}(C)=\max_{\omega}%
\mathcal{C}(\omega)$, where $\mathcal{C}(\omega):=\min_{i}\mathcal{C}%
(\mathcal{E}_{i}^{\omega})$ is the capacity of route $\omega$. Then, we
observe that $\mathcal{C}(\omega)$ is achievable by the end-users. In fact,
any two consecutive points on route $\omega$ may first communicate at the rate
$\mathcal{C}(\mathcal{E}_{i}^{\omega})$; then, the point-to-point outputs can
be distributed to the end-users via entanglement swapping or key composition
at the minimum rate $\min_{i}\mathcal{C}(\mathcal{E}_{i}^{\omega})$.

Thus, for a distillable network, we exactly establish the single-path capacity
as%
\begin{equation}
\mathcal{C}(\mathcal{N})=\min_{C}\mathcal{C}(C)=\max_{\omega}\mathcal{C}%
(\omega)~. \label{distillableCCCC}%
\end{equation}
Finding the optimal route $\omega_{\ast}$ corresponds to solving the widest
path problem~\cite{Pollack} where the weights of the edges $(\mathbf{x}%
,\mathbf{y})$ are the two-way capacities $\mathcal{C}(\mathcal{E}%
_{\mathbf{xy}})$. Route $\omega_{\ast}$ can be found via modified Dijkstra's
shortest path algorithm~\cite{MITp}, which works in time $O(\left\vert
E\right\vert \log_{2}\left\vert P\right\vert )$, where $\left\vert
E\right\vert $ is the number of edges and $\left\vert P\right\vert $ is the
number of points. Over route $\omega_{\ast}$, a capacity-achieving protocol is
non adaptive and based on one-way CCs, with point-to-point sessions of one-way
entanglement distillation followed by entanglement swapping.

An important example is an optical lossy network $\mathcal{N}_{\text{loss}}$
where any route $\omega$ is composed of lossy channels with transmissivities
$\{\eta_{i}^{\omega}\}$. Denote by $\eta_{\omega}:=\min_{i}\eta_{i}^{\omega}$
the end-to-end transmissivity of route $\omega$. The single-path capacity is
given by the route with maximum transmissivity%
\begin{equation}
\mathcal{C}(\mathcal{N}_{\text{loss}})=-\log_{2}(1-\tilde{\eta}),~~\tilde
{\eta}:=\max_{\omega\in\Omega}\eta_{\omega}. \label{lossSINGLE}%
\end{equation}
In particular, this is the ultimate rate at which the two end-points may
generate secret bits per sequential use of the lossy network. See
Sec.~\ref{secSINGLE} for full details and results on single-path routing.

\subsection{Quantum networks with multipath routing}

In a network we may consider a more powerful routing strategy, where systems
are transmitted through a sequence of multicasts. For instance, as shown in
the example of Fig.~\ref{NETproto}, Alice may simultaneously sends systems to
repeaters $\mathbf{p}_{1}$ and $\mathbf{p}_{2}$, which is denoted by
$\mathbf{a}\rightarrow\{\mathbf{p}_{1},\mathbf{p}_{2}\}$. Then, repeater
$\mathbf{p}_{1}$ may communicate with repeater $\mathbf{p}_{2}$ and Bob
$\mathbf{b}$, i.e., $\mathbf{p}_{1}\rightarrow\{\mathbf{p}_{2},\mathbf{p}%
_{3}\}$. Finally, repeater $\mathbf{p}_{2}$ may communicate with Bob, i.e.,
$\mathbf{p}_{2}\rightarrow\mathbf{b}$. Note that each edge of the network is
used exactly once during the end-to-end transmission, a strategy known as
\textquotedblleft flooding\textquotedblright\ in computer
networks~\cite{flooding}. This is achieved by non-overlapping multicasts,
where the receiving repeaters choose unused edges for the next transmissions.

Thus, in a flooding protocol, the network is initialized by a preliminary
network LOCC. Then, Alice $\mathbf{a}$ broadcasts quantum systems to all her
neighbor repeaters $\mathbf{a}\rightarrow\{\mathbf{p}_{k}\}$. This is followed
by another network LOCC. Then, each receiving repeater multicasts systems to
neighbor repeaters through unused edges, and so on. Each multicast is
interleaved by network LOCCs and may distribute multi-partite entanglement.
Eventually, Bob is reached as an end-point in the first use of the network. In
the limit of many uses $n$ and optimizing over the protocols, we define the
multipath capacity of the network $\mathcal{C}^{\text{m}}(\mathcal{N})$.

Exact definitions and relevant methods are explained in
Secs.~\ref{SECnetworks} and \ref{SecSTRETCHINGNET}. As before, given an
entanglement cut $C$, consider its cut-set $\tilde{C}$. For each edge
$(\mathbf{x},\mathbf{y})$ in $\tilde{C}$, there is a channel $\mathcal{E}%
_{\mathbf{xy}}$ with a corresponding resource state $\sigma_{\mathbf{xy}}$. We
define the multi-edge flow of REE through $C$ as%
\begin{equation}
E_{\mathrm{R}}^{\text{m}}(C):=%
{\textstyle\sum\limits_{(\mathbf{x},\mathbf{y})\in\tilde{C}}}
E_{\mathrm{R}}(\sigma_{\mathbf{xy}}).
\end{equation}
The minimization of this quantity over all entanglement cuts provides our
upper bound for the multipath capacity of the network, i.e.,
\begin{equation}
\mathcal{C}^{\text{m}}(\mathcal{N})\leq\min_{C}E_{\mathrm{R}}^{\text{m}}(C),
\label{mpBOUND}%
\end{equation}
which is the multipath generalization of Eq.~(\ref{ubseqNN}). In a
teleportation-covariant network we may set $\sigma_{\mathbf{xy}}%
=\sigma_{\mathcal{E}_{\mathbf{xy}}}$. For a distillable network, we may also
use Eq.~(\ref{disNETtool}) and write Eq.~(\ref{mpBOUND}) with $E_{\mathrm{R}%
}^{\text{m}}(C)=\mathcal{C}^{\text{m}}(C)$, where
\begin{equation}
\mathcal{C}^{\text{m}}(C):=%
{\textstyle\sum\limits_{(\mathbf{x},\mathbf{y})\in\tilde{C}}}
\mathcal{C}(\mathcal{E}_{\mathbf{xy}})
\end{equation}
is the multi-edge capacity of a cut.

To show the achievability of the bound for a distillable network, we need to
determine the optimal flow of qubits from Alice to Bob. First of all, from the
knowledge of the capacities $\mathcal{C}(\mathcal{E}_{\mathbf{xy}})$, the
parties solve a classical problem of maximum flow~\cite{netflow} which
provides an optimal orientation for the network and rates $R_{\mathbf{xy}}%
\leq\mathcal{C}(\mathcal{E}_{\mathbf{xy}})$. Then, any pair of neighbor
points, $\mathbf{x}$ and $\mathbf{y}$, distill $nR_{\mathbf{xy}}$ ebits via
one-way CCs. Such ebits are used to teleport $nR_{\mathbf{xy}}$ qubits from
$\mathbf{x}$ to $\mathbf{y}$ according to the optimal orientation. In this
way, $nR$ of qubits are teleported from Alice to Bob, generating a flow of
quantum information through the network. Using the max-flow min-cut
theorem~\cite{Harris,Ford,ShannonFLOW,netflow,Karp,Dinic,Alon,Ahuja,Cheriyan,King,Orlin}%
, the maximum flow is $n\mathcal{C}^{\text{m}}(C_{\text{min}})$ where
$C_{\text{min}}$ is the minimum cut, i.e., $\mathcal{C}^{\text{m}%
}(C_{\text{min}})=\min_{C}\mathcal{C}^{\text{m}}(C)$. Thus, that for a
distillable $\mathcal{N}$, we find%
\begin{equation}
\mathcal{C}^{\text{m}}(\mathcal{N})=\min_{C}\mathcal{C}^{\text{m}}(C),
\end{equation}
which is the multipath version of Eq.~(\ref{distillableCCCC}). This is
achievable using one-way CCs and the optimal routing is given by Orlin's
algorithm~\cite{Orlin} in $O(|P|\times|E|)$ time.

As an example, consider again a lossy optical network $\mathcal{N}%
_{\text{loss}}$ whose generic edge $(\mathbf{x},\mathbf{y})$ has
transmissivity $\eta_{\mathbf{xy}}$. Given a cut $C$, consider its loss
$l(C):=%
{\textstyle\prod\nolimits_{(\mathbf{x},\mathbf{y})\in\tilde{C}}}
(1-\eta_{\mathbf{xy}})$ and define the total loss of the network as the
maximization $l(\mathcal{N}_{\text{loss}}):=\max_{C}l(C)$. We find that the
multipath capacity is just given by%
\begin{equation}
\mathcal{C}^{\text{m}}(\mathcal{N}_{\text{loss}})=-\log_{2}l(\mathcal{N}%
_{\text{loss}}).
\end{equation}
It is interesting to make a direct comparison between the performance of
single- and multi-path strategies. For this purpose, consider a diamond
network $\mathcal{N}_{\text{loss}}^{\Diamond}$ whose links are lossy channels
with the same transmissivity $\eta$. In this case, we easily see that the
multipath capacity doubles the single-path capacity of the network, i.e.,%
\begin{equation}
\mathcal{C}^{\text{m}}(\mathcal{N}_{\text{loss}}^{\Diamond})=2\mathcal{C}%
(\mathcal{N}_{\text{loss}}^{\Diamond})=-2\log_{2}(1-\eta).
\end{equation}
As expected the parallel use of the quantum network is more powerful than the
sequential use. See Sec.~\ref{secMULTI} for details and other results on
multi-path routing.

\newpage

\section{Preliminaries\label{GeneralSECTION}}

In this preliminary section, we review techniques and results developed for
point-to-point quantum and private communications~\cite{QKDpaper}. These
notions will be later generalized and combined with other tools from classical
network theory when we discuss repeater chains in Sec.~\ref{SECrepeaters} and
then quantum networks from Sec.~\ref{SECnetworks}. The expert reader may skip
this section and directly go to Sec.~\ref{SECrepeaters}.

\subsection{General definitions}

\begin{figure}[pth]
\vspace{-2.3cm}
\par
\begin{center}
\includegraphics[width=0.50\textwidth]{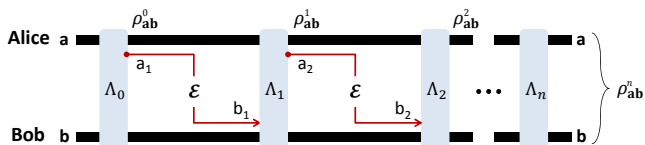} \vspace{-2.9cm}
\end{center}
\caption{Point-to-point adaptive protocol. Each transmission $a_{i}\rightarrow
b_{i}$ through the quantum channel $\mathcal{E}$ is interleaved by two
adaptive LOCCs, $\Lambda_{i-1}$ and $\Lambda_{i}$, applied to Alice's and
Bob's local registers $\mathbf{a}$ and $\mathbf{b}$. After $n$ transmissions,
Alice and Bob share an output state $\rho_{\mathbf{ab}}^{n}$ close to some
target state $\phi^{n}$.}%
\label{longPIC}%
\end{figure}

Let us start by defining an adaptive point-to-point protocol $\mathcal{P}$
through a quantum channel $\mathcal{E}$. Assume that Alice has register
$\mathbf{a}$ and Bob has register $\mathbf{b}$. These registers are
(countable) sets of quantum systems which are prepared in some state
$\rho_{\mathbf{ab}}^{0}$ by an adaptive LOCC $\Lambda_{0}$ applied to some
fundamental separable state $\rho_{\mathbf{a}}^{0}\otimes\rho_{\mathbf{b}}%
^{0}$. Then, for the first transmission, Alice picks a system $a_{1}%
\in\mathbf{a}$ and sends it through channel $\mathcal{E}$; at the output, Bob
receives a system $b_{1}$ which is included in his register $b_{1}%
\mathbf{b}\rightarrow\mathbf{b}$. Another adaptive LOCC $\Lambda_{1}$ is
applied to the registers. Then, there is the second transmission
$\mathbf{a}\ni a_{2}\rightarrow b_{2}$ through $\mathcal{E}$, followed by
another LOCC $\Lambda_{2}$ and so on (see Fig.~\ref{longPIC}). After $n$ uses,
Alice and Bob share an output state $\rho_{\mathbf{ab}}^{n}$ which is
epsilon-close to some target state $\phi^{n}$ with $nR_{n}$ bits. This means
that, for any $\varepsilon>0$, one has $\left\Vert \rho_{\mathbf{ab}}^{n}%
-\phi^{n}\right\Vert \leq\varepsilon$ in trace norm. This is also called an
($n,R_{n},\varepsilon$)-protocol, but we omit this technical notation for
simplicity. Operationally, the protocol $\mathcal{P}$ is completely
characterized by the sequence of adaptive LOCCs $\mathcal{L}=\{\Lambda
_{0},\Lambda_{1}\ldots\}$. 

The (generic) two-way assisted capacity of the quantum channel is defined by
taking the limit of the asymptotic rate $\lim_{n}R_{n}$ and maximizing over
all adaptive protocols $\mathcal{P}$, i.e.,%
\begin{equation}
\mathcal{C}(\mathcal{E}):=\sup_{\mathcal{P}}\lim_{n}R_{n}.
\end{equation}
The specification of the target state $\phi^{n}$\ identifies a corresponding
type of two-way capacity. If $\phi^{n}$ is a maximally-entangled state, then
we have the two-way entanglement-distribution capacity $D_{2}(\mathcal{E})$.
The latter is in turn equal to the two-way quantum capacity $Q_{2}%
(\mathcal{E})$, because transmitting qubits is equivalent to distributing
ebits under two-way CCs. If $\phi^{n}$ is a private state~\cite{KD1}, then we
have the secret key capacity $K(\mathcal{E})$ and we have $K(\mathcal{E})\geq
D_{2}(\mathcal{E})$, because a maximally-entangled state is a particular type
of private state. Also note that $K(\mathcal{E})=P_{2}(\mathcal{E})$, where
$P_{2}$ is the two-way private capacity, i.e., the maximum rate at which Alice
may \textit{deterministically} transmit secret bits~\cite{Devetak}. Thus, we
may write the chain of (in)equalities
\begin{equation}
D_{2}(\mathcal{E})=Q_{2}(\mathcal{E})\leq K(\mathcal{E})=P_{2}(\mathcal{E}).
\label{hierarchy}%
\end{equation}

\subsection{Weak converse bound}

The two-way capacity $\mathcal{C}(\mathcal{E})$ [i.e., any of the capacities
in Eq.~(\ref{hierarchy})] can be bounded by a general expression in terms of
the REE~\cite{RMPrelent}. First of all, recall that the REE of a quantum state
$\sigma$ is given by
\begin{equation}
E_{\mathrm{R}}(\sigma)=\inf_{\gamma\in\text{\textrm{SEP}}}S(\sigma||\gamma),
\label{REEbona}%
\end{equation}
where $\gamma$ is a separable state and $S$ is the quantum relative entropy,
defined by~\cite{RMPrelent}
\begin{equation}
S(\sigma||\gamma):=\mathrm{Tr}\left[  \sigma(\log_{2}\sigma-\log_{2}%
\gamma)\right]  .
\end{equation}

The notion of REE can be extended to an asymptotic state $\sigma:=\lim_{\mu
}\sigma^{\mu}$, which is defined as a limit of a sequence of states
$\sigma^{\mu}$ (e.g., this is the case for energy unbounded states of
CV\ systems). In this case, we may modify Eq.~(\ref{REEbona}) into the
following expression~\cite{QKDpaper}
\begin{equation}
E_{\mathrm{R}}(\sigma):=\inf_{\gamma^{\mu}}\underset{\mu\rightarrow+\infty
}{\lim\inf}S(\sigma^{\mu}||\gamma^{\mu}), \label{REE_weaker}%
\end{equation}
where $\gamma^{\mu}$ is sequence of separable states that converges in
trace-norm, i.e., such that $||\gamma^{\mu}-\gamma||\overset{\mu}{\rightarrow
}0$ for some separable $\gamma$, and the inferior limit comes from the lower
semi-continuity of the quantum relative entropy (valid at any dimension,
including for CV\ systems~\cite{HolevoBOOK}).

With these notions in hand, we may write~\cite{QKDpaper,Notaa1}.

\begin{theorem}
[Weak converse]\label{generalWEAK}For any quantum channel $\mathcal{E}$ (at
any dimension, finite or infinite), we may write%
\begin{equation}
\mathcal{C}(\mathcal{E})\leq E_{\mathrm{R}}^{\bigstar}(\mathcal{E}%
):=\sup_{\mathcal{P}}\underset{n}{\lim}\frac{E_{R}(\rho_{\mathbf{ab}}^{n})}%
{n}~, \label{mainweak}%
\end{equation}
where the bound $E_{\mathrm{R}}^{\bigstar}(\mathcal{E})$\ is defined computing
the REE of the output state $\rho_{\mathbf{ab}}^{n}$, taking the limit for
many channels uses, and optimizing over all the adaptive protocols
$\mathcal{P}$.
\end{theorem}

\noindent To simplify $E_{\mathrm{R}}^{\bigstar}(\mathcal{E})$ into a
single-letter quantity, Ref.~\cite{QKDpaper} devised a general technique,
dubbed \textquotedblleft teleportation stretching\textquotedblright. A
preliminary step consists in using a suitable simulation of the quantum
channel, which may be replaced by a corresponding resource state. Then, this
simulation argument can be exploited to reduce the adaptive protocol into a
much simpler block-type protocol, where the output is decomposed into a tensor
product of resource states up to a trace-preserving LOCC.

\subsection{LOCC simulation of quantum channels}

\begin{figure}[pth]
\vspace{0.1cm}
\par
\begin{center}
\includegraphics[width=0.28\textwidth]{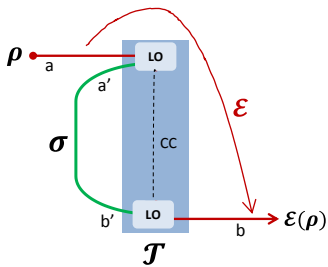} \vspace{-0.6cm}
\end{center}
\caption{LOCC simulation of an arbitrary quantum channel $\mathcal{E}$ by
means of an LOCC $\mathcal{T}$ applied to the input state $\rho$ and a
resource state $\sigma$, according to Eq.~(\ref{sigma}). For asymptotic
simulation, we have the approximate channel $\mathcal{E}^{\mu}$ which is
simulated by $(\mathcal{T}^{\mu},\sigma^{\mu})$. We then take the point-wise
limit for infinite $\mu$, which defines the asymptotic channel $\mathcal{E}$
as in Eq.~(\ref{asymptotic}).}%
\label{simulationPIC}%
\end{figure}

Given an arbitrary quantum channel $\mathcal{E}$, we may consider a
corresponding simulation $S(\mathcal{E})=(\mathcal{T},\sigma)$ based on some
LOCC $\mathcal{T}$ and resource state $\sigma$. This simulation is such that,
for any input state $\rho$, the output of the channel can be expressed
as~\cite{QKDpaper}
\begin{equation}
\mathcal{E}(\rho)=\mathcal{T}(\rho\otimes\sigma). \label{sigma}%
\end{equation}
See also Fig.~\ref{simulationPIC}. A channel $\mathcal{E}$ which is simulable
as in Eq.~(\ref{sigma}) is also called \textquotedblleft$\sigma$%
-stretchable\textquotedblright. Note that there are different simulations for
the same channel. One is trivial because it just corresponds to choosing
$\sigma$ as a maximally-entangled state and $\mathcal{T}$ as teleportation
followed by $\mathcal{E}$ completely pushed in Bob's LO. Therefore, it is
implicitly understood that one has to carry out an optimization over these
simulations, which also depends on the specific functional under study.

Furthermore, the simulation can also be asymptotic, i.e., we may consider
sequences of LOCCs $\mathcal{T}^{\mu}$ and resource states $\sigma^{\mu}$ such
that~\cite{QKDpaper}%
\begin{equation}
\mathcal{E}(\rho)=\lim_{\mu}\mathcal{E}^{\mu}(\rho),~~\mathcal{E}^{\mu}%
(\rho):=\mathcal{T}^{\mu}(\rho\otimes\sigma^{\mu}). \label{asymptotic}%
\end{equation}
In other words a quantum channel $\mathcal{E}$ may be defined as a point-wise
limit of a sequence of approximating channels $\mathcal{E}^{\mu}$ which are
simulable as in Eq.~(\ref{asymptotic}). We therefore call $(\mathcal{T}%
,\sigma):=\lim_{\mu}(\mathcal{T}^{\mu},\sigma^{\mu})$ the asymptotic
simulation of $\mathcal{E}$. This generalization is important for bosonic
channels and the amplitude damping channel. See Ref.~\cite{Notaa2} for a
discussion on the literature of channel simulation.

\subsection{Teleportation covariance and simulability}

There exist an important class of quantum channels, called teleportation
covariant, for which the LOCC simulation takes a convenient form.

\begin{definition}
[tele-covariance]A quantum channel $\mathcal{E}$ is called teleportation
covariant if, for any teleportation unitary $U$ (Pauli operators in DVs,
phase-space displacements in CVs~\cite{telereview}),\ we may write%
\begin{equation}
\mathcal{E}(U\rho U^{\dagger})=V\mathcal{E}(\rho)V^{\dagger}~,
\label{stretchability}%
\end{equation}
for another (generally-different) unitary $V$.
\end{definition}

\noindent Note that this is a wide family which includes Pauli channels (e.g.,
depolarizing or dephasing), erasure channels and bosonic Gaussian channels.

Thanks to the property in Eq.~(\ref{stretchability}), the random corrections
of the teleportation protocol can be pushed at the output of these channels.
For this reason, they may be simulated by teleportation, as first shown for DV
systems~\cite{Leung}, and then extended to any dimension~\cite{QKDpaper}.

\begin{lemma}
[Tele-covariance and simulability]A teleportation-covariant channel
$\mathcal{E}$ can be simulated as
\begin{equation}
\mathcal{E}(\rho)=\mathcal{T}_{\text{tele}}(\rho\otimes\sigma_{\mathcal{E}}),
\label{kkkll}%
\end{equation}
where $\mathcal{T}_{\text{tele}}$\ is teleportation (Bell detection and
conditional unitaries) and $\sigma_{\mathcal{E}}$ is the Choi matrix of the
channel, defined as $\sigma_{\mathcal{E}}:=\mathcal{I}\otimes\mathcal{E}%
(\Phi)$, with $\Phi$ being a maximally entangled state. For single-mode
bosonic channels, we may write the asymptotic simulation
\begin{equation}
\mathcal{E}(\rho)=\lim_{\mu}\mathcal{T}_{\text{tele}}^{\mu}(\rho\otimes
\sigma_{\mathcal{E}}^{\mu}), \label{asymptotics}%
\end{equation}
where $\mathcal{T}_{\text{tele}}^{\mu}$ is a sequence of teleportation-LOCCs
(based on finite-energy versions of the ideal CV\ Bell detection) and
$\sigma_{\mathcal{E}}^{\mu}$ is a sequence of Choi-approximating states of the
form $\sigma_{\mathcal{E}}^{\mu}:=\mathcal{I}\otimes\mathcal{E}(\Phi^{\mu})$,
where $\Phi^{\mu}$ is a two-mode squeezed vacuum (TMSV) state~\cite{RMP} with
$\bar{n}=\mu-1/2$ mean thermal photons in each mode.
\end{lemma}

When a quantum channel can be simulated as in Eq.~(\ref{kkkll}) or
(\ref{asymptotics}) is also known as \textquotedblleft
Choi-stretchable\textquotedblright\ via teleportation or \textquotedblleft
teleportation simulable\textquotedblright. Therefore, the content of the
previous lemma can be simply stated by saying that a teleportation-covariant
channel is teleportation simulable, at any dimension~\cite{QKDpaper}.

\subsection{Teleportation stretching of an adaptive protocol\label{BosonSTTT}}

\begin{figure*}[pth]
\vspace{-4.5cm}
\par
\begin{center}
\includegraphics[width=0.98\textwidth]{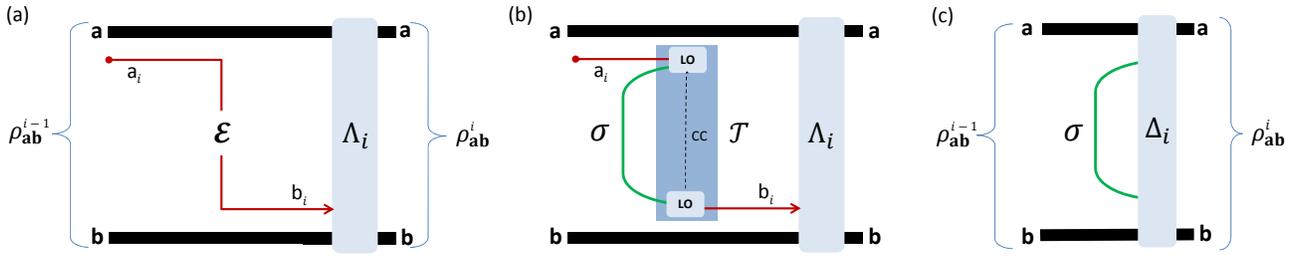} \vspace{-4.8cm}
\end{center}
\caption{Teleportation stretching of the $i$th transmission of an adaptive
protocol. (a)~We depict the original transmission through the channel
$\mathcal{E}$ which transforms the register state $\rho_{\mathbf{ab}}%
^{i-1}:=\rho_{\mathbf{a}a_{i}\mathbf{b}}$ into the output $\rho_{\mathbf{ab}%
}^{i}$. (b)~We simulate the channel by means of an LOCC $\mathcal{T}$ and a
resource state $\sigma$, as in previous Fig.~\ref{simulationPIC}. (c)~We
collapse $\mathcal{T}$ and the adaptive LOCC $\Lambda_{i}$ into a single LOCC
$\Delta_{i}$ applied to the tensor product $\rho_{\mathbf{ab}}^{i-1}%
\otimes\sigma$, as in Eq.~(\ref{toite}).}%
\label{pppPIC}%
\end{figure*}

By exploiting the LOCC simulation $S(\mathcal{E})=(\mathcal{T},\sigma)$ of a
quantum channel $\mathcal{E}$, we may completely simplify an adaptive
protocol. In fact, the output state $\rho_{\mathbf{ab}}^{n}$ can be decomposed
into a tensor-product of resources states $\sigma^{\otimes n}$ up to a
trace-preserving LOCC $\bar{\Lambda}$. In other words, we may
write~\cite{QKDpaper}
\begin{equation}
\rho_{\mathbf{ab}}^{n}=\bar{\Lambda}\left(  \sigma^{\otimes n}\right)  .
\label{StretchingMAIN}%
\end{equation}
For non-asymptotic simulations the proof goes as follows. As shown in
Fig.~\ref{pppPIC}, for the generic $i$th transmission, we replace the original
quantum channel $\mathcal{E}$ with a simulation $S(\mathcal{E})=(\mathcal{T}%
,\sigma)$. Then, we collapse the LOCC $\mathcal{T}$ into the adaptive LOCC
$\Lambda_{i}$ to form the composite LOCC $\Delta_{i}$. As a result, the
pre-transmission state $\rho_{\mathbf{ab}}^{i-1}:=\rho_{\mathbf{a}%
a_{i}\mathbf{b}}$ is transformed into the following post-transmission state
\begin{equation}
\rho_{\mathbf{ab}}^{i}=\Delta_{i}\left(  \rho_{\mathbf{ab}}^{i-1}\otimes
\sigma\right)  . \label{toite}%
\end{equation}
The next step is to iterate Eq.~(\ref{toite}). One finds%
\begin{equation}
\rho_{\mathbf{ab}}^{n}=(\Delta_{n}\circ\cdots\circ\Delta_{1})(\rho
_{\mathbf{ab}}^{0}\otimes\sigma^{\otimes n}).
\end{equation}
Because $\rho_{\mathbf{ab}}^{0}$ is separable, its preparation may be included
in the LOCCs and we get Eq.~(\ref{StretchingMAIN}) for a complicated but
single trace-preserving LOCC $\bar{\Lambda}$.

For a bosonic channel with asymptotic simulation as in Eq.~(\ref{asymptotic}),
the procedure is more involved. One first considers an imperfect channel
simulation $\mathcal{E}^{\mu}(\rho):=\mathcal{T}^{\mu}(\rho\otimes\sigma^{\mu
})$ in each transmission. By adopting this simulation, we realize an imperfect
stretching of the protocol, with output state $\rho_{\mathbf{ab}}^{\mu
,n}:=\bar{\Lambda}_{\mu}\left(  \sigma^{\mu\otimes n}\right)  $\ for a
trace-preserving LOCC $\bar{\Lambda}_{\mu}$. This is done similarly to the
steps in Fig.~\ref{pppPIC}, but considering $\mathcal{E}^{\mu}$ in the place
of the original channel $\mathcal{E}$. A crucial point is now the estimation
of the error in the channel simulation, which must be suitably controlled and
propagated to the output state.

Assume that, during the $n$ transmissions of the protocol, the total mean
number of photons in the registers is bounded by some large but finite value
$\bar{E}$. We may therefore define the set of energy-constrained states
\begin{equation}
\mathcal{D}_{\bar{E}}:=\{\rho_{\mathbf{ab}}~|~\mathrm{Tr}(\hat{H}%
\rho_{\mathbf{ab}})\leq\bar{E}\},
\end{equation}
where $\hat{H}$ is the multi-mode number operator. For the $i$th transmission
$a_{i}\mathbf{\rightarrow}b_{i}$, the simulation error\ may be quantified in
terms of the energy-bounded diamond norm~\cite{QKDpaper}
\begin{align}
\varepsilon_{\bar{E}}  &  =\left\Vert \mathcal{E}-\mathcal{E}^{\mu}\right\Vert
_{\diamond\bar{E}}:=\label{bDIAMONDnorm}\\
&  \sup_{\rho_{a_{i}\mathbf{ab}}\in\mathcal{D}_{\bar{E}}}\left\Vert
\mathcal{E}\otimes\mathcal{I}_{\mathbf{ab}}(\rho_{a_{i}\mathbf{ab}%
})-\mathcal{E}^{\mu}\otimes\mathcal{I}_{\mathbf{ab}}(\rho_{a_{i}\mathbf{ab}%
})\right\Vert .\nonumber
\end{align}
Because $\mathcal{D}_{\bar{E}}$ is compact and channel $\mathcal{E}$ is
defined by the point-wise limit $\mathcal{E}(\rho)=\lim_{\mu}\mathcal{E}^{\mu
}(\rho)$, we may write the following uniform limit
\begin{equation}
\varepsilon_{\bar{E}}\overset{\mu}{\rightarrow}0\mathrm{~~}\text{for~any }%
\bar{E}\text{.}%
\end{equation}

This error has to be propagated to the output state, so that we can suitably
bound the trace distance between the actual output $\rho_{\mathbf{ab}}^{n}$
and the simulated output $\rho_{\mathbf{ab}}^{n,\mu}$. By using basic
properties of the trace distance (triangle inequality and monotonicity under
maps), Ref.~\cite{QKDpaper} showed that the simulation error in the output
state satisfies%
\begin{equation}
\left\Vert \rho_{\mathbf{ab}}^{n}-\rho_{\mathbf{ab}}^{n,\mu}\right\Vert \leq
n\left\Vert \mathcal{E}-\mathcal{E}^{\mu}\right\Vert _{\diamond\bar{E}}~.
\end{equation}
Therefore, for any $\bar{E}$, we may write the trace-norm limit
\begin{equation}
\left\Vert \rho_{\mathbf{ab}}^{n}-\bar{\Lambda}_{\mu}\left(  \sigma
^{\mu\otimes n}\right)  \right\Vert \overset{\mu}{\rightarrow}0,
\label{stretch2}%
\end{equation}
i.e., the asymptotic stretching $\rho_{\mathbf{ab}}^{n}=\lim_{\mu}\bar
{\Lambda}_{\mu}(\sigma^{\mu\otimes n})$. This is true for any energy bound
$\bar{E}$ which can be implicitly relaxed at the very end of the
calculations~\cite{QKDpaper}.

\begin{remark}
Note that teleportation stretching simplifies an \textit{arbitrary} adaptive
protocol over an \textit{arbitrary} channel at \textit{any} dimension, finite
or infinite.\ In particular, it works by maintaining the original
communication task. This means that\ an adaptive protocol of quantum
communication (QC), entanglement distribution (ED) or key generation (KG), is
reduced to a corresponding block protocol with exactly the same original task
(QC, ED, or KG), but with the output state being decomposed in the form of
Eq.~(\ref{StretchingMAIN}) or Eq.~(\ref{stretch2}). In the literature, there
were some precursory but restricted arguments, as those in
Refs.~\cite{B2main,Niset}. These were limited to the transformation of a
protocol of QC into a protocol of ED, over specific classes of channels (e.g.,
Pauli channels in Ref.~\cite{B2main}). Furthermore, no control of the
simulation error was considered in previous literature~\cite{Niset}, while
this is crucial for the rigorous simulation of bosonic channels.
\end{remark}

\subsection{Single-letter upper bound for two-way capacities}

The most crucial insight of Ref.~\cite{QKDpaper} has been the combination of
the previous two ingredients, i.e., channel's REE and teleportation
stretching, which is the key observation leading to a single-letter upper
bound for all the two-way capacities of a quantum channel. In fact, let us
compute the REE\ of the output state decomposed as in
Eq.~(\ref{StretchingMAIN}). We derive%
\begin{equation}
E_{\mathrm{R}}(\rho_{\mathbf{ab}}^{n})\overset{(1)}{\leq}E_{\mathrm{R}}%
(\sigma^{\otimes n})\overset{(2)}{\leq}nE_{\mathrm{R}}(\sigma)~, \label{toREP}%
\end{equation}
using (1) the monotonicity of the REE under trace-preserving LOCCs and (2) its
subadditive over tensor products. By replacing Eq.~(\ref{toREP}) in
Eq.~(\ref{mainweak}), we then find the single-letter upper
bound~\cite{QKDpaper}
\begin{equation}
\mathcal{C}(\mathcal{E})\leq E_{\mathrm{R}}(\sigma)~. \label{UB1}%
\end{equation}
In particular, if the channel $\mathcal{E}$ is teleportation-covariant, it is
Choi-stretchable, and we may write~\cite{QKDpaper}
\begin{equation}
\mathcal{C}(\mathcal{E})\leq E_{\mathrm{R}}(\sigma_{\mathcal{E}}). \label{UB2}%
\end{equation}
These results are suitable extended to asymptotic simulations. In particular,
using the weaker definition in Eq.~(\ref{REE_weaker}), Ref.~\cite{QKDpaper}
showed that Eqs.~(\ref{UB1}) and~(\ref{UB2}) are valid for bosonic channels
with asymptotic simulations.

\subsection{Bounds for teleportation-covariant channels\label{teleBOUNDS}}

The upper bound of Eq.~(\ref{UB2}) is valid for any teleportation-covariant
channel, in particular for Pauli channels and bosonic Gaussian channels. In
particular, consider a qubit Pauli channel $\mathcal{E}_{\text{Pauli}}$ with
probability distribution $\{p_{k}\}$, i.e.,
\begin{equation}
\mathcal{E}_{\text{Pauli}}(\rho)=p_{0}\rho+p_{1}X\rho X+p_{2}Y\rho
Y+p_{3}Z\rho Z,
\end{equation}
where $X$, $Y$, and $Z$ are Pauli operators~\cite{N&C02}. Let us call $H_{2}$
the binary Shannon entropy and $p_{\max}:=\max\{p_{k}\}$. Then, we may
write~\cite{QKDpaper}%
\begin{equation}
\mathcal{C}(\mathcal{E}_{\text{Pauli}})\leq\left\{
\begin{array}
[c]{c}%
1-H_{2}(p_{\max}),~~\mathrm{if~~}p_{\max}\geq1/2,\\
\\
0,\text{~~}\mathrm{if~~}p_{\max}<1/2,\text{~~~~~~~~~~~~~~~}%
\end{array}
\right.
\end{equation}
which can be easily generalized to arbitrary finite dimension
(qudits)~\cite{QKDpaper}.

Consider now phase-insensitive Gaussian channels. The most important is the
thermal-loss channel $\mathcal{E}_{\eta,\bar{n}}$ which transforms input
quadratures $\mathbf{\hat{x}}=(\hat{q},\hat{p})^{T}$ as $\mathbf{\hat{x}%
}\rightarrow\sqrt{\eta}\mathbf{\hat{x}}+\sqrt{1-\eta}\mathbf{\hat{x}}_{E}$,
where $\eta\in(0,1)$ is the transmissivity and $E$ is the thermal environment
with $\bar{n}$ mean photons. For this channel, we may derive~\cite{QKDpaper}
\begin{equation}
\mathcal{C}(\mathcal{E}_{\eta,\bar{n}})\leq\left\{
\begin{array}
[c]{c}%
-\log_{2}\left[  (1-\eta)\eta^{\bar{n}}\right]  -h(\bar{n}),~~\mathrm{if~}%
\bar{n}<\frac{\eta}{1-\eta},\\
\\
0,\text{~~}\mathrm{if~~}\bar{n}\geq\frac{\eta}{1-\eta}%
,\text{~~~~~~~~~~~~~~~~~~~~~~~~~~~}%
\end{array}
\right.
\end{equation}
where we have set
\begin{equation}
h(x):=(x+1)\log_{2}(x+1)-x\log_{2}x. \label{hEntropyMAIN}%
\end{equation}

For a noisy quantum amplifier $\mathcal{E}_{g,\bar{n}}$ we have the
transformation $\mathbf{\hat{x}}\rightarrow\sqrt{g}\mathbf{\hat{x}}+\sqrt
{g-1}\mathbf{\hat{x}}_{E}$, where $g>1$ is the gain and $E$ is the thermal
environment with $\bar{n}$ mean photons. In this case, we may
compute~\cite{QKDpaper}%
\begin{equation}
\mathcal{C}(\mathcal{E}_{\eta,\bar{n}})\leq\left\{
\begin{array}
[c]{c}%
\log_{2}\left(  \dfrac{g^{\bar{n}+1}}{g-1}\right)  -h(\bar{n}),~~\mathrm{if~~}%
\bar{n}<(g-1)^{-1},\\
\\
0,\text{~~}\mathrm{if~~}\bar{n}\geq(g-1)^{-1}.\text{~~~~~~~~~~~~~~~~~~~~~~~~~}%
\end{array}
\right.
\end{equation}
Finally, for an additive-noise Gaussian channel $\mathcal{E}_{\xi}$, we have
$\mathbf{\hat{x}}\rightarrow\mathbf{\hat{x}}+(z,z)^{T}$ where $z$ is a
classical Gaussian variable with zero mean and variance $\xi\geq0$. In this
case, we have the bound~\cite{QKDpaper}%
\begin{equation}
\mathcal{C}(\mathcal{E}_{\xi})\leq\left\{
\begin{array}
[c]{c}%
\frac{\xi-1}{\ln2}-\log_{2}\xi,~~\mathrm{if~~}\xi<1,\\
\\
0,\text{~~}\mathrm{if~~}\xi\geq1\text{~~~~~~~~~~~~~~}%
\end{array}
\right.
\end{equation}

\subsection{Two-way capacities for distillable channels\label{teleDISTILLABLE}%
}

Within the class of teleportation-covariant channels, there is a sub-class for
which the upper bound $E_{\mathrm{R}}(\sigma_{\mathcal{E}})$ in Eq.~(\ref{UB2}%
) coincides with an achievable rate for one-way entanglement distillation.
These \textquotedblleft distillable channels\textquotedblright~\cite{QKDpaper}
are those for which we may write%
\begin{equation}
E_{\mathrm{R}}(\sigma_{\mathcal{E}})=D_{1}(\sigma_{\mathcal{E}}),
\label{coinCCC}%
\end{equation}
where $D_{1}(\sigma_{\mathcal{E}})$ is the distillable entanglement of the
Choi matrix $\sigma_{\mathcal{E}}$ via one-way CCs, forward or backward (this
quantity is also suitably extended to asymptotic Choi matrices in the case of
bosonic channels~\cite{QKDpaper}).

The equality in Eq.~(\ref{coinCCC}) is a remarkable coincidence for three reasons:

\begin{enumerate}
\item Since $D_{1}(\sigma_{\mathcal{E}})$ is a lower bound to $D_{2}%
(\mathcal{E})$, all the two-way capacities of these channels coincide
($D_{2}=Q_{2}=K=P_{2}$) and are fully established as%
\begin{equation}
\mathcal{C}(\mathcal{E})=E_{\mathrm{R}}(\sigma_{\mathcal{E}})=D_{1}%
(\sigma_{\mathcal{E}}). \label{coincidence}%
\end{equation}

\item The two-way capacities are achieved by means of rounds of one-way CCs,
so that adaptiveness is not needed and CCs are limited.

\item Because of the hashing inequality, we have
\begin{equation}
D_{1}(\sigma_{\mathcal{E}})\geq\max\{I_{\mathrm{C}}(\sigma_{\mathcal{E}%
}),I_{\mathrm{RC}}(\sigma_{\mathcal{E}})\},
\end{equation}
where $I_{\mathrm{C}}$ and $I_{\mathrm{RC}}$ are the
coherent~\cite{Schu96,Lloyd97} and reverse
coherent~\cite{RevCohINFO,ReverseCAP} information of the Choi matrix. Such
quantities (and their asymptotic versions) are easily computable and may be
used to show the coincidence in Eq.~(\ref{coincidence}).
\end{enumerate}

\noindent These elements were combined in Ref.~\cite{QKDpaper} to determine
strikingly simple formulas for the two-way capacities of the most fundamental
quantum channels, such as the lossy channel, the quantum-limited amplifier,
the dephasing and erasure channels (all distillable channels).

In particular, for a bosonic lossy channel $\mathcal{E}_{\eta}$\ with
transmissivity $\eta$ (and zero thermal noise $\bar{n}=0$), one
has~\cite{QKDpaper}
\begin{equation}
\mathcal{C}(\eta)=-\log_{2}(1-\eta)~. \label{formCloss}%
\end{equation}
The secret-key capacity\ of the lossy channel $K(\eta)$ determines the maximum
rate achievable by any QKD protocol. At high loss $\eta\simeq0$, one has the
optimal rate-loss scaling of $K\simeq1.44\eta$ secret bits per channel use.
Because Eq.~(\ref{formCloss}) establishes the upper limit of any
point-to-point quantum optical communication, it also establishes a
\textquotedblleft repetearless bound\textquotedblright, i.e., a benchmark for
quantum repeaters, also known as Pirandola-Laurenza-Ottaviani-Banchi (PLOB)
bound. Note that the PLOB bound can be extended to a multiband lossy channel,
for which we write $\mathcal{C}=-\sum_{i}\log_{2}(1-\eta_{i})$, where
$\eta_{i}$ are the transmissivities of the various bands or frequency
components. For instance, for a multimode telecom fibre with constant
transmissivity $\eta$ and bandwidth $W$, we have%
\begin{equation}
\mathcal{C}=-W\log_{2}(1-\eta).
\end{equation}

Now consider the other distillable channels. For a quantum-limited amplifier
$\mathcal{E}_{g}$ with gain $g>1$ (and zero thermal noise $\bar{n}=0$), one
finds~\cite{QKDpaper}%
\begin{equation}
\mathcal{C}(g)=-\log_{2}\left(  1-g^{-1}\right)  ~. \label{Campli}%
\end{equation}
For a qubit dephasing channel $\mathcal{E}_{p}^{\text{deph}}$ with dephasing
probability $p$, one finds~\cite{QKDpaper}%
\begin{equation}
\mathcal{C}(p)=1-H_{2}(p)~, \label{dep2}%
\end{equation}
where $H_{2}$ is the binary Shannon entropy. This can be extended to arbitrary
dimension $d$, so that~\cite{QKDpaper}
\begin{equation}
\mathcal{C}(p,d)=\log_{2}d-H(\{P_{i}\})~, \label{depDgen}%
\end{equation}
where $H$ is the Shannon entropy and $P_{i}$ is the probability of $i$ phase
flips. Finally, for the qudit erasure channel $\mathcal{E}_{p,d}%
^{\text{erase}}$ with erasure probability $p$, one finds~\cite{QKDpaper}
\begin{equation}
\mathcal{C}(p)=(1-p)\log_{2}d~. \label{erase2}%
\end{equation}
For this channel, only $Q_{2}$ was previously known~\cite{ErasureChannel},
while \cite{QKDpaper,GEW} co-established $K$.

\section{Chains of quantum repeaters\label{SECrepeaters}}

Let us go beyond point-to-point quantum communications. The first non-trivial
extension is a linear chain of quantum repeaters between the two remote
parties, which is the simplest example of a multi-hop quantum network. Thus,
consider Alice and Bob to be end-points of a chain of $N+2$ points with $N$
repeaters in the middle. For $i=0,\ldots,N$ we assume that point $i$ is
connected with point $i+1$ by a quantum channel $\mathcal{E}_{i}$ which can be
forward or backward, for a total of $N+1$ channels $\{\mathcal{E}_{0}%
,\ldots\mathcal{E}_{i},\ldots\mathcal{E}_{N}\}$. Each point has a local
register which is a countable ensemble of quantum systems, denoted by
$\mathbf{r}_{i}$ for the $i$-th point. In particular, we set $\mathbf{a}%
=\mathbf{r}_{0}$ for Alice and $\mathbf{b}=\mathbf{r}_{N+1}$ for Bob.
Registers are updated. For instance, if Alice sends a system $a$, then we
update $\mathbf{a}\rightarrow\mathbf{a}a$; if Bob receives a system $b$, then
we update $b\mathbf{b}\rightarrow\mathbf{b}$.

The most general distribution protocol over the chain is based on adaptive LOs
and unlimited two-way CC involving all the points in the chain. In other
words, each point broadcasts classical information and receives classical
feedback from all the other points, which is used to perform conditional LOs
on the local registers. In the following we always assume these
\textquotedblleft network\textquotedblright\ adaptive LOCCs, unless we specify
otherwise. The first step is the preparation of the registers by an LOCC
$\Lambda_{0}$ whose application to some fundamental state provides an initial
separable state $\sigma_{\mathbf{ar}_{1}\cdots\mathbf{r}_{N}\mathbf{b}}$.
Then, Alice and the first repeater exchange a quantum system through channel
$\mathcal{E}_{0}$ (via forward or backward transmission). This is followed by
an LOCC $\Lambda_{1}$ on the updated registers $\mathbf{ar}_{1}\mathbf{r}%
_{2}\ldots\mathbf{r}_{N}\mathbf{b}$. Next, the first and the second repeaters
exchange another quantum system through channel $\mathcal{E}_{1}$ followed by
another LOCC $\Lambda_{2}$, and so on. Finally, Bob exchanges a system with
the $N$th repeater through channel $\mathcal{E}_{N}$ and the final LOCC
$\Lambda_{N+1}$ provides the output state $\rho_{\mathbf{ar}_{1}%
\cdots\mathbf{r}_{N}\mathbf{b}}$.

\begin{figure}[pth]
\vspace{-0.9cm}
\par
\begin{center}
\includegraphics[width=0.5\textwidth]{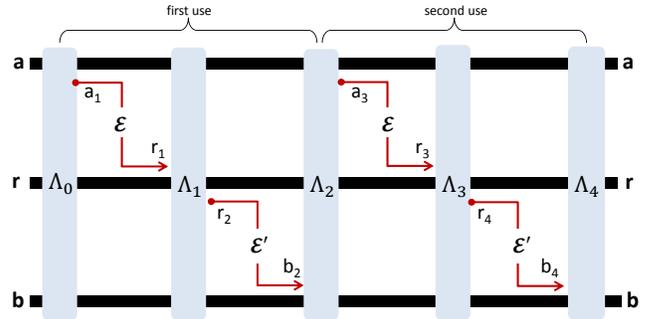} \vspace{-1.5cm}
\end{center}
\caption{Chain with a single repeater $\mathbf{r}$ and connected by two
forward channels $\mathcal{E}$ and $\mathcal{E}^{\prime}$. Each transmission
$k$ through one of the two channels occurs between two adaptive LOCC
$\Lambda_{k-1}$ and $\Lambda_{k}$. In particular, here we show two uses of the
chain, with total output state $\rho_{\mathbf{arb}}^{2}$. Note that, if the
parties want to distribute ebits or private bits, they may also use a
different order of transmissions in each use. For instance, in the first use,
the first transmission could be between the repeater $\mathbf{r}$ and Bob
$\mathbf{b}$, followed by that between Alice $\mathbf{a}$ and the repeater
$\mathbf{r}$. The order of the transmissions and the sequence of LOCCs defines
the adaptive protocol $\mathcal{P}$ over the chain.}%
\label{repeaterSCHEME}%
\end{figure}

This procedure completes the first use of the chain. In the second use, the
initial state is the (non-separable) output state of the first round
$\sigma_{\mathbf{ar}_{1}\cdots\mathbf{r}_{N}\mathbf{b}}^{2}=\rho
_{\mathbf{ar}_{1}\cdots\mathbf{r}_{N}\mathbf{b}}^{1}$. The protocol goes as
before with each pair of points $i$ and $i+1$ exchanging one system between
two LOCCs. The second use ends with the output state $\rho_{\mathbf{ar}%
_{1}\cdots\mathbf{r}_{N}\mathbf{b}}^{2}$ which is the input for the third use
and so on. After $n$ uses, the points share an output state $\rho
_{\mathbf{ar}_{1}\cdots\mathbf{r}_{N}\mathbf{b}}^{n}$. By tracing out the
repeaters, we get Alice and Bob's final state $\rho_{\mathbf{ab}}^{n}$, which
depends on the sequence of LOCCs $\mathcal{L}=\{\Lambda_{0},\cdots
,\Lambda_{n(N+1)}\}$. In general, in each use of the chain, the order of the
transmissions can also be permuted. Both the order of these transmissions and
the sequence of LOCCs $\mathcal{L}$ defines the adaptive protocol
$\mathcal{P}_{\text{\textrm{chain}}}$ generating the output $\rho
_{\mathbf{ab}}^{n}$. See Fig.~\ref{repeaterSCHEME} for an example.

We say that an adaptive protocol $\mathcal{P}_{\text{\textrm{chain}}}$ has
rate $R_{n}$ if $\left\Vert \rho_{\mathbf{ab}}^{n}-\phi_{n}\right\Vert
\leq\varepsilon$, where $\phi_{n}$ is a target state with $nR_{n}$ bits. By
taking the limit of $n\rightarrow+\infty$\ and optimizing over $\mathcal{P}%
_{\text{\textrm{chain}}}$, we define the generic two-way capacity of the
chain, i.e.,
\begin{equation}
\mathcal{C}(\{\mathcal{E}_{i}\}):=\sup_{\mathcal{P}_{\text{\textrm{chain}}}%
}\lim_{n}R_{n}~.
\end{equation}
This capacity has different nature depending on the task of the distribution
protocol. For QKD, the target state is a private state~\cite{KD1} with secret
key rate $R_{n}^{\text{key}}$ (bits per chain use). In this case
$\mathcal{C}(\{\mathcal{E}_{i}\})$ is the secret key capacity of the chain
$K(\{\mathcal{E}_{i}\})$. Under two-way CCs, this is also equal to the maximum
rate at which Alice can deterministically send a secret message to Bob through
the chain, i.e., its two-way private capacity $P_{2}(\{\mathcal{E}_{i}\})$.
For entanglement distribution (ED), the target state is a maximally-entangled
state with rate $R_{n}^{\text{ED}}\leq R_{n}^{\text{key}}$ (ebits per chain
use). In this case, $\mathcal{C}(\{\mathcal{E}_{i}\})$ is an
entanglement-distribution capacity $D_{2}(\{\mathcal{E}_{i}\})\leq
K(\{\mathcal{E}_{i}\})$. Under two-way CCs, $D_{2}$ is equal to the maximum
rate at which Alice can reliably send a qubits to Bob through the chain, i.e.,
its two-way quantum capacity $Q_{2}(\{\mathcal{E}_{i}\})$.

We can build an upper bound for all the previous capacities, i.e., for the
generic $\mathcal{C}(\{\mathcal{E}_{i}\})$. In fact, the general weak converse
of Eq.~(\ref{mainweak}) can be directly extended to the present scenario
proviso that we replace the supremum over point-to-point adaptive protocols
$\mathcal{P}$ with a supremum over the adaptive protocols $\mathcal{P}%
_{\text{\textrm{chain}}}$ over the chain. The other elements in the formula of
Eq.~(\ref{mainweak}) are the same because it is derived under the assumption
that the output state $\rho_{\mathbf{ab}}^{n}$ is epsilon-close to a target
private state with $nR_{n}^{\text{key}}$ bits, no matter how the output state
has been generated. Thus, we may write the REE weak converse bound
\begin{equation}
\mathcal{C}(\{\mathcal{E}_{i}\})\leq E_{\mathrm{R}}^{\star}(\{\mathcal{E}%
_{i}\}):=\sup_{\mathcal{P}_{\text{\textrm{chain}}}}\underset{n}{\lim}%
~n^{-1}E_{\mathrm{R}}(\rho_{\mathbf{ab}}^{n}). \label{hard}%
\end{equation}

In order to reduce the latter bound to a single-letter quantity we simulate
the chain, by replacing each channel $\mathcal{E}_{i}$ with a simulation
$S_{i}=(\mathcal{T}_{i},\sigma_{i})$ for some LOCC $\mathcal{T}_{i}$ and
resource state $\sigma_{i}$. The next step is to use teleportation stretching
to re-organize the adaptive protocol into a block version, where the output
state is expressed in terms of a tensor product of resource states. A direct
application of this procedure will allow us to write
\begin{equation}
\rho_{\mathbf{ab}}^{n}=\bar{\Lambda}_{\mathbf{ab}}\left(  \otimes_{i=0}%
^{N}~\sigma_{i}^{\otimes n}\right)  ~, \label{UBnot}%
\end{equation}
for a trace-preserving LOCC $\bar{\Lambda}_{\mathbf{ab}}$ (this reduction is
proven afterwards). By using Eq.~(\ref{UBnot}), we may then write
$E_{\mathrm{R}}(\rho_{\mathbf{ab}}^{n})\leq n\Pi_{i=0}^{N}E_{\mathrm{R}%
}(\sigma_{i})$, leading to the upper bound%
\begin{equation}
E_{\mathrm{R}}^{\star}(\{\mathcal{E}_{i}\})\leq\Pi_{i=0}^{N}E_{\mathrm{R}%
}(\sigma_{i})~.
\end{equation}

Unfortunately, this bound is too large. To improve it, we need to perform cuts
of the chain, such that Alice and Bob end up to be disconnected. In a linear
chain, the situation is particularly simple, because any cut disconnects the
two end-points. The refined procedure consists of cutting channel
$\mathcal{E}_{i}$, stretching the protocol with respect to that channel and
finally minimizing over all cuts. Let us start with the formal definition of
cut of a chain.

\begin{definition}
[Cut of a chain]\label{defCUT}Consider a chain of $N$ repeaters $\{\mathbf{r}%
_{1},\ldots,\mathbf{r}_{N}\}$ connecting Alice $\mathbf{a}=\mathbf{r}_{0}%
$\ and Bob $\mathbf{b}=\mathbf{r}_{N+1}$\ by means of $N+1$ quantum channels
$\{\mathcal{E}_{0},\ldots,\mathcal{E}_{i},\ldots,\mathcal{E}_{N}\}$ as in
Fig.~\ref{intropicmain}. An entanglement cut \textquotedblleft$i$%
\textquotedblright\ disconnects channel $\mathcal{E}_{i}$ and induces a
bipartition $(\mathbf{A},\mathbf{B})$, where the set of points $\mathbf{A}%
=\{\mathbf{r}_{0},\ldots,\mathbf{r}_{i}\}$ is \textquotedblleft
super-Alice\textquotedblright\ and $\mathbf{B}=\{\mathbf{r}_{i+1}%
,\ldots,\mathbf{r}_{N}\}$ is \textquotedblleft super-Bob\textquotedblright.
\end{definition}

By performing entanglement cuts in the chain, we may state the following
result which correctly extends teleportation stretching to chains of quantum repeaters

\begin{lemma}
[Chain stretching]\label{TheoCHAIN}Consider a chain of $N$ repeaters as in
Definition~\ref{defCUT}. Given an arbitrary entanglement cut $i$, consider the
disconnected channel $\mathcal{E}_{i}$\ and its simulation via some resource
state $\sigma_{i}$. For any such cut $i=0,\ldots,N$ the output of the most
general adaptive protocol $\mathcal{P}_{\text{\textrm{chain}}}$ over $n$ uses
of the chain can be decomposed as
\begin{equation}
\rho_{\mathbf{ab}}^{n}=\bar{\Lambda}_{i}\left(  \sigma_{i}^{\otimes n}\right)
, \label{REPde}%
\end{equation}
where $\bar{\Lambda}_{i}$ is a trace-preserving LOCC. In particular, for a
chain of teleportation-covariant channels, we may write Eq.~(\ref{REPde})
using the Choi-matrices $\sigma_{\mathcal{E}_{i}}$ (with asymptotic
formulations for bosonic channels).
\end{lemma}

\begin{figure}[ptbh]
\vspace{-0.5cm}
\par
\begin{center}
\vspace{0.05cm} \vspace{+0.5cm}
\includegraphics[width=0.65\textwidth] {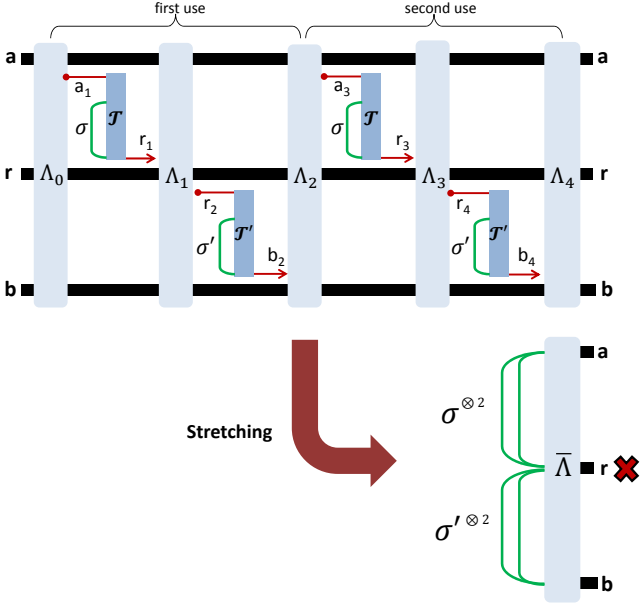}
\end{center}
\par
\vspace{-0.4cm}\caption{Teleportation stretching of a repeater. See text.}%
\label{RePIC}%
\end{figure}

\textbf{Proof.}~For simplicity let us start with the simple case of a
$3$-point chain ($N=1$), where Alice $\mathbf{a}$ and Bob $\mathbf{b}$ are
connected with a middle repeater $\mathbf{r}$ by means of two channels
$\mathcal{E}$ and $\mathcal{E}^{\prime}$ as in Fig.~\ref{repeaterSCHEME} (the
direction of the channels may be different as well as the order in which they
are used). Assume two adaptive uses of the chain ($n=2$) starting from a
fundamental state $\rho_{\mathbf{a}}^{0}\otimes\rho_{\mathbf{r}}^{0}%
\otimes\rho_{\mathbf{b}}^{0}$. As depicted in Fig.~\ref{RePIC}, we replace
each channel with a corresponding simulation: $\mathcal{E}\rightarrow
(\mathcal{T},\sigma)$ and $\mathcal{E}^{\prime}\rightarrow(\mathcal{T}%
^{\prime},\sigma^{\prime})$. Then, the resource states are stretched back in
time before the LOCCs which are all collapsed into a single LOCC $\bar
{\Lambda}$ (trace-preserving after averaging over all measurements). After two
uses of the repeater we have the output state $\rho_{\mathbf{arb}}^{2}%
=\bar{\Lambda}\left(  \sigma^{\otimes2}\otimes\sigma^{\prime\otimes2}\right)
$. By tracing the repeater $\mathbf{r}$, we derive $\rho_{\mathbf{ab}}%
^{2}=\bar{\Lambda}_{\mathbf{ab}}\left(  \sigma^{\otimes2}\otimes\sigma
^{\prime\otimes2}\right)  $ up to re-defining the LOCC. By extending the
procedure to an arbitrary number of repeaters $N$ and uses $n$, we get%
\begin{equation}
\rho_{\mathbf{ar}_{1}\ldots\mathbf{r}_{N}\mathbf{b}}^{n}=\bar{\Lambda}\left(
\otimes_{i=0}^{N}~\sigma_{i}^{\otimes n}\right)  , \label{decoPROOF}%
\end{equation}
and tracing out all the repeaters, we derive Eq.~(\ref{UBnot}).

Therefore, thanks to teleportation stretching, the quantum transmissions
between each pair of near-neighbor points have been replaced with
tensor-products of resource states, followed by a single but complicated
trace-preserving LOCC. In this reduction, the resource states are responsible
for distributing entanglement between the points of the chain. In order to get
tight upper bounds we need to perform entanglement cuts.\begin{figure}[ptbh]
\vspace{-1.2cm}
\par
\begin{center}
\includegraphics[width=0.52\textwidth] {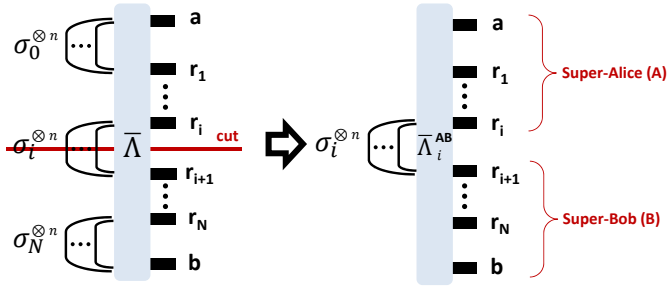}
\end{center}
\par
\vspace{-1.7cm}\caption{Reduction of the stretched scenario. See text.}%
\label{Reduction}%
\end{figure}

Let us perform a cut \textquotedblleft$i$\textquotedblright\ of the chain, so
that channel $\mathcal{E}_{i}$ is disconnected between $\mathbf{r}_{i}$ and
$\mathbf{r}_{i+1}$. This cut can be done directly on the stretched chain as in
Fig.~\ref{Reduction}. This cut defines super-Alice $\mathbf{A}$ and super-Bob
$\mathbf{B}$. Now, let us include all the resource states $\sigma_{k}^{\otimes
n}$ with $k<i$ in the LOs of super-Alice, and all the resource states with
$k>i+1$ in the LOs of super-Bob. This operation has two outcomes: (i) it
defines a novel trace-preserving LOCC $\bar{\Lambda}_{i}$ which is local with
respect to the super-parties; and (ii) it leaves with a reduced number of
resource states $\sigma_{i}^{\otimes n}$, i.e., only those associated with the
cut. For the super-parties, we may write $\rho_{\mathbf{AB}}^{n}=\bar{\Lambda
}_{\mathbf{AB}}^{i}(\sigma_{i}^{\otimes n})$. By tracing out all the middle
repeaters $\mathbf{r}_{1}\mathbf{r}_{2}\ldots\mathbf{r}_{N}$, the resulting
LOCC $\bar{\Lambda}_{i}$ remains local with respect to $\mathbf{a}$ and
$\mathbf{b}$, and we get the end-to-end output $\rho_{\mathbf{ab}}^{n}$ as in
Eq.~(\ref{REPde}), for any cut $i$.

The extension of the proof to bosonic channels exploits asymptotic simulations
as explained in Sec.~\ref{BosonSTTT}. For each channel $\mathcal{E}_{i}$ in
the chain we may consider its approximation $\mathcal{E}_{i}^{\mu}$ with
simulation $(\mathcal{T}_{i}^{\mu},\sigma_{i}^{\mu})$. This leads to the
output state $\rho_{\mathbf{ab}}^{n,\mu}=\bar{\Lambda}_{i}^{\mu}(\sigma
_{i}^{\mu\otimes n})$ for a trace-preserving LOCC $\bar{\Lambda}_{i}^{\mu}$.
Since $\mathcal{E}_{i}$ is the point-wise limit of $\mathcal{E}_{i}^{\mu}$ for
large $\mu$, if we consider the energy-bounded diamond norm $\varepsilon
_{\bar{E}}^{i}:=\left\Vert \mathcal{E}_{i}-\mathcal{E}_{i}^{\mu}\right\Vert
_{\diamond\bar{E}}$, we have $\varepsilon_{\bar{E}}^{i}\rightarrow0$ for any
energy $\bar{E}$ and cut $i$. By directly extending a \textquotedblleft
peeling\textquotedblright\ argument given in Ref.~\cite[Methods, Eq.
(103)]{QKDpaper}, we easily show that the trace-distance between the actual
output $\rho_{\mathbf{ab}}^{n}$ and the simulated one $\rho_{\mathbf{ab}%
}^{n,\mu}$ is controlled as follows%
\begin{equation}
\left\Vert \rho_{\mathbf{ab}}^{n}-\rho_{\mathbf{ab}}^{n,\mu}\right\Vert \leq n%
{\textstyle\sum\limits_{i=0}^{N}}
\left\Vert \mathcal{E}_{i}-\mathcal{E}_{i}^{\mu}\right\Vert _{\diamond\bar{E}%
}.
\end{equation}
Clearly, this distance goes to zero in $\mu$, for any number of uses $n$,
number of repeaters $N$ and energy $\bar{E}$. In other words, given an
arbitrary cut $i$ we have%
\begin{equation}
\left\Vert \rho_{\mathbf{ab}}^{n}-\bar{\Lambda}_{i}^{\mu}(\sigma_{i}%
^{\mu\otimes n})\right\Vert \overset{\mu}{\rightarrow}0, \label{stretcvv}%
\end{equation}
or, more compactly,\
\begin{equation}
\rho_{\mathbf{ab}}^{n}=\lim_{\mu}\bar{\Lambda}_{i}^{\mu}(\sigma_{i}%
^{\mu\otimes n}), \label{stretcvvv}%
\end{equation}
for any number of uses $n$, repeaters $N$, and energy $\bar{E}$.$~\blacksquare
$

By using the previous lemma, we can now prove the following result which
establishes a single-letter REE\ upper bound for the generic two-way capacity
$\mathcal{C}(\{\mathcal{E}_{i}\})$ of a chain of quantum repeaters. This is a
bound for the maximal rates for entanglement distribution ($D_{2}$), quantum
communication ($Q_{2}$), secret key generation ($K$) and private communication
($P_{2}$) through the repeater chain. The formula simplifies for a
teleportation-covariant chain and even more for a distillable chain, for which
the repeater-assisted capacity is found to be the minimum among the two-way
capacities of the individual distillable channels.

\begin{theorem}
[Single-letter REE bound]\label{singleLETTtheorem}Consider a chain of $N$
repeaters as in Definition~\ref{defCUT}. The generic two-way capacity of the
chain must satisfy the following minimization over the entanglement cuts%
\begin{equation}
\mathcal{C}(\{\mathcal{E}_{i}\})\leq\min_{i}E_{\mathrm{R}}(\sigma_{i}),
\label{TheoRE}%
\end{equation}
where $\sigma_{i}$\ is the resource state of an arbitrary LOCC simulation of
$\mathcal{E}_{i}$. For a chain of teleportation-covariant channels (e.g.
Pauli, Gaussian channels), we may write the bound in terms of their Choi
matrices, i.e.,%
\begin{equation}
\mathcal{C}(\{\mathcal{E}_{i}\})\leq\min_{i}E_{\mathrm{R}}(\sigma
_{\mathcal{E}_{i}}), \label{TheoREtelecov}%
\end{equation}
where the REE\ is intended to be asymptotic for bosonic channels. In
particular, for a chain of distillable channels (lossy channels,
quantum-limited amplifiers, dephasing and erasure channels), we establish the
capacity as%
\begin{equation}
\mathcal{C}(\{\mathcal{E}_{i}\})=\min_{i}E_{\mathrm{R}}(\sigma_{\mathcal{E}%
_{i}})=\min_{i}\mathcal{C}(\mathcal{E}_{i}), \label{reppAssCC}%
\end{equation}
where $\mathcal{C}(\mathcal{E}_{i})$ are the individual two-way capacities
associated with each distillable channel $\mathcal{E}_{i}$ in the chain. In
this case, we also have $\mathcal{C}(\{\mathcal{E}_{i}\})=\min_{i}D_{1}%
(\sigma_{\mathcal{E}_{i}})$, so that the capacity may be achieved by using
one-way entanglement distillation followed by entanglement swapping.
\end{theorem}

\textbf{Proof.}~For an arbitrary chain, perform the stretching of the protocol
for any entanglement cut $i$, so that we may write Eq.~(\ref{REPde}). Because
the REE\ is non-decreasing under trace-preserving LOCCs, we get $E_{\mathrm{R}%
}(\rho_{\mathbf{ab}}^{n})\leq E_{\mathrm{R}}(\sigma_{i}^{\otimes n})$. By
replacing the latter inequality in the general weak converse bound of
Eq.~(\ref{hard}), we may drop the supremum over the protocols $\mathcal{P}%
_{\text{\textrm{chain}}}$\ and derive the following bound in terms of the
regularized REE\ of the resource state%
\begin{equation}
\mathcal{C}(\{\mathcal{E}_{i}\})\leq E_{\mathrm{R}}^{\infty}(\sigma_{i}%
):=\lim_{n}~n^{-1}E_{\mathrm{R}}(\sigma_{i}^{\otimes n}). \label{regull}%
\end{equation}
By minimizing over all the entanglement cuts, we get%
\begin{equation}
\mathcal{C}(\{\mathcal{E}_{i}\})\leq\min_{i}E_{\mathrm{R}}^{\infty}(\sigma
_{i})\leq\min_{i}E_{\mathrm{R}}(\sigma_{i}), \label{regull2}%
\end{equation}
where the last inequality is due to the subadditivity of the REE over
tensor-product states.

For teleportation-covariant channels, we may set $\sigma_{i}=\sigma
_{\mathcal{E}_{i}}$, so that Eq.~(\ref{TheoREtelecov}) holds. Then, for
distillable channels, we may also write $\mathcal{C}(\mathcal{E}%
_{i})=E_{\mathrm{R}}(\sigma_{\mathcal{E}_{i}})=D_{1}(\sigma_{\mathcal{E}_{i}%
})$, so that $\mathcal{C}(\{\mathcal{E}_{i}\})\leq\min_{i}D_{1}(\sigma
_{\mathcal{E}_{i}})$. It is clear that $\min_{i}D_{1}(\sigma_{\mathcal{E}_{i}%
})$ is also an achievable lower bound so that it provides the capacity and we
may also write Eq.~(\ref{reppAssCC}). In fact, in the $i$th point-to-point
connection, points $\mathbf{r}_{i}$ and $\mathbf{r}_{i+1}$ may distill
$D_{1}(\sigma_{\mathcal{E}_{i}})$ ebits via one-way CCs. After this is done in
all the connections, sessions of entanglement swapping will transfer at least
$\min_{i}D_{1}(\sigma_{\mathcal{E}_{i}})$ ebits to the end points.

To extend the result to bosonic channels with asymptotic simulations, we adopt
a weaker definition of REE as given in Eq.~(\ref{REE_weaker}). Consider the
asymptotic stretching of the output state $\rho_{\mathbf{ab}}^{n}$\ as in
Eq.~(\ref{stretcvv}) which holds for any number of uses $n$, repeaters $N$,
and energy $\bar{E}$. Then, for any cut $i$, the simplification of the REE
bound $E_{\mathrm{R}}(\rho_{\mathbf{ab}}^{n})$ goes as follows
\begin{align}
E_{\mathrm{R}}(\rho_{\mathbf{ab}}^{n})  &  =\inf_{\gamma\in\mathrm{SEP}}%
S(\rho_{\mathbf{ab}}^{n}||\gamma)\nonumber\\
&  \overset{(1)}{\leq}\inf_{\gamma^{\mu}}S\left[  \lim_{\mu}\bar{\Lambda}%
_{i}^{\mu}(\sigma_{i}^{\mu\otimes n})~||~\lim_{\mu}\gamma^{\mu}\right]
\nonumber\\
&  \overset{(2)}{\leq}\inf_{\gamma^{\mu}}\underset{\mu\rightarrow+\infty}%
{\lim\inf}~S\left[  \bar{\Lambda}_{i}^{\mu}(\sigma_{i}^{\mu\otimes
n})~||~\gamma^{\mu}\right] \nonumber\\
&  \overset{(3)}{\leq}\inf_{\gamma^{\mu}}\underset{\mu\rightarrow+\infty}%
{\lim\inf}~S\left[  \bar{\Lambda}_{i}^{\mu}(\sigma_{i}^{\mu\otimes n}%
)~||~\bar{\Lambda}_{i}^{\mu}(\gamma^{\mu})\right] \nonumber\\
&  \overset{(4)}{\leq}\inf_{\gamma^{\mu}}\underset{\mu\rightarrow+\infty}%
{\lim\inf}~S\left(  \sigma_{i}^{\mu\otimes n}~||~\gamma^{\mu}\right)
\nonumber\\
&  \overset{(5)}{=}E_{\mathrm{R}}(\sigma_{i}^{\otimes n})
\end{align}
where: (1)$~\gamma^{\mu}$ is a generic sequence of separable states converging
in trace norm, i.e., such that there is a separable state $\gamma:=\lim_{\mu
}\gamma^{\mu}$ so that $\Vert\gamma-\gamma^{\mu}\Vert\overset{\mu}%
{\rightarrow}0$; (2)~we use the lower semi-continuity of the relative
entropy~\cite{HolevoBOOK}; (3)~we use that $\bar{\Lambda}_{i}^{\mu}%
(\gamma^{\mu})$ are specific types of converging separable sequences within
the set of all such sequences; (4)~we use the monotonicity of the relative
entropy under trace-preserving LOCCs; and (5)~we use the definition of REE for
asymptotic states.

For any energy $\bar{E}$, we may apply the general weak converse bound of
Eq.~(\ref{hard}), so that we may again write Eq.~(\ref{regull}) in terms of
the regularized REE $E_{\mathrm{R}}^{\infty}(\sigma_{i})$. Since this upper
bound does no longer depend on the protocols $\mathcal{P}%
_{\text{\textrm{chain}}}$, it applies to both energy-constrained and
energy-unconstrained registers (i.e., we may relax the constraint $\bar{E}$).
The proof of the further condition $E_{\mathrm{R}}^{\infty}(\sigma_{i})\leq
E_{\mathrm{R}}(\sigma_{i})$ is based on the subadditivity of the REE over
tensor product states, which holds for asymptotic states too~\cite{QKDpaper}.
Thus, the minimization over the cuts provides again Eq.~(\ref{regull2}). The
remaining steps of the proof for teleportation and distillable channels are
trivially extended to asymptotic simulations. In particular, one can define an
asymptotic notion of one-way distillable entanglement $D_{1}$ for an unbounded
Choi matrix as explained in Ref.~\cite{QKDpaper}.~$\blacksquare$

By using Theorem~\ref{singleLETTtheorem} and the bounds in
Sec.~\ref{teleBOUNDS}, we can easily derive upper bounds for the capacities of
teleportation-covariant chains, which includes chains of Pauli channels (at
any finite dimension) or chains of Gaussian channels, such as thermal-loss
channels, noisy quantum amplifiers or additive-noise Gaussian channels. These
bounds can be then further specified for distillable chains, by combining
Theorem~\ref{singleLETTtheorem} with the results in Sec.~\ref{teleDISTILLABLE}%
. In this case, we exactly establish the repeater assisted capacities deriving
extremely simple formulas that we discuss in the following section.

\subsection{Capacities for distillable chains}

Let us specify our results for various types of distillable chains. Let us
start by considering a lossy chain, where Alice and Bob are connected by $N$
repeaters and each connection $\mathcal{E}_{i}$ is a lossy channel with
transmissivity $\eta_{i}$. By combining Eq.~(\ref{reppAssCC}) of
Theorem~\ref{singleLETTtheorem} with Eq.~(\ref{formCloss}), we find that the
capacity of the lossy chain is given by%
\begin{align}
\mathcal{C}_{\text{loss}}(\{\eta_{i}\})  &  =\min_{i}\mathcal{C}(\eta
_{i})=\min_{i}\left[  -\log_{2}(1-\eta_{i})\right] \nonumber\\
&  =-\log_{2}(1-\eta_{\text{min}}),~~~\eta_{\text{min}}:=\min_{i}\eta_{i}~.
\label{LossyCHAIN}%
\end{align}
Therefore, no matter how many repeaters we use, the minimum transmissivity in
the chain fully determines the ultimate rate of quantum or private
communication between the two end-points. Suppose that we require a minimum
performance of $1$ bit per use of the chain (this could be $1$ secret bit or
$1$ ebit or $1$ qubit). From Eq.~(\ref{LossyCHAIN}), we see that we need to
ensure at least $\eta_{\text{min}}=1/2$, which means at most $3$dB of loss in
each link. This \textquotedblleft$3$dB rule\textquotedblright\ implies that
$1$ bit rate communication can occur in chains whose maximum point-to-point
distance is 15km (assuming fiber connections at the loss rate of 0.2dB/km).

Consider now an amplifying chain, i.e., a chain which is connected by
quantum-limited amplifiers with arbitrary gains $\{g_{i}\}$. Using
Eqs.~(\ref{reppAssCC}) and~(\ref{Campli}), we find that the repeater-assisted
capacity is fully determined by the highest gain $g_{\max}:=\max_{i}g_{i}$, so
that
\begin{equation}
\mathcal{C}_{\text{amp}}(\{g_{i}\})=-\log_{2}\left(  1-g_{\max}^{-1}\right)  .
\end{equation}

In the DV setting, start with a spin chain where the state transfer between
the $i$th spin and the next one is modeled by a dephasing channel with
probability $p_{i}\leq1/2$. Using Eqs.~(\ref{reppAssCC}) and~(\ref{dep2}), we
find the repeater-assisted capacity%
\begin{equation}
\mathcal{C}_{\text{deph}}(\{p_{i}\})=1-H_{2}(p_{\max}),
\end{equation}
where $p_{\max}:=\max_{i}p_{i}$ is the maximum probability of phase flipping
in the chain, and $H_{2}$ is the binary Shannon entropy. When the spins are
connected by erasure channels with probabilities $\{p_{i}\}$, we combine
Eqs.~(\ref{reppAssCC}) and~(\ref{erase2}) for $d=2$, and we derive
\begin{equation}
\mathcal{C}_{\text{erase}}(\{p_{i}\})=1-p_{\max},
\end{equation}
where $p_{\max}$ is the maximum probability of an erasure.

Note that the latter results for the spin chains can be readily extended from
qubits to qudits of arbitrary dimension $d$, by using the two-way capacities
of Eqs.~(\ref{depDgen}) and~(\ref{erase2}). Finally, also note that
Eq.~(\ref{reppAssCC}) of Theorem~\ref{singleLETTtheorem} may be applied to
hybrid distillable chains, where channels are distillable but of different
kind between each pair of repeaters, e.g., we might have erasure channels
alternated with dephasing channels or lossy channels, etc.

\subsection{Quantum repeaters in optical communications\label{OptimalUSE}}

Let us discuss in more detail the use of quantum repeaters in the bosonic
setting. Suppose that we are given a long communication line with
transmissivity $\eta$, such as an optical/telecom fiber. A cut of this line
generates two lossy channels with transmissivities $\eta^{\prime}$ and
$\eta^{\prime\prime}$ such that $\eta=\eta^{\prime}\eta^{\prime\prime}$.
Suppose that we are also given a number $N$ of repeaters that we could
potentially insert along the line. The question is: \textit{What is the
optimal way to cut the line and insert the repeaters?}

From the formula in Eq.~(\ref{LossyCHAIN}), we can immediately see that the
optimal solution is to insert $N$\ equidistant repeaters, so that the
resulting $N+1$ lossy channels have identical transmissivities
\begin{equation}
\eta_{i}=\eta_{\text{min}}=\sqrt[N+1]{\eta}~.
\end{equation}
This leads to the maximum repeater-assisted capacity%
\begin{equation}
\mathcal{C}_{\text{loss}}(\eta,N)=-\log_{2}\left(  1-\sqrt[N+1]{\eta}\right)
~. \label{optLOSScap1}%
\end{equation}
This capacity has been plotted in Fig.~\ref{figg} for increasing number of
repeaters $N$ as a function of the total loss of the line, which is expressed
in decibel (dB) by $\eta_{\text{dB}}:=-10\log_{10}\eta$. In particular, we
compare the repeater-assisted capacity with the point-to-point benchmark,
i.e., the maximum performance achievable in the absence of repeaters (PLOB
bound~\cite{QKDpaper}).

Let us study two opposite regimes that we may call repeater-dominant and
loss-dominant. In the former, we fix the total transmissivity $\eta$ of the
line and use many equidistant repeaters $N\gg1$. We then have%
\begin{equation}
\mathcal{C}_{\text{loss}}(\eta,N\gg1)\simeq\log_{2}N-\log_{2}\ln\frac{1}{\eta
}~,
\end{equation}
which means that the capacity scales logarithmically in the number of
repeaters, independently from the loss. In the second regime (loss-dominant),
we fix the number of repeaters $N$ and we consider high loss $\eta\simeq0$, in
such a way that each link of the chain is very lossy, i.e., we may set
$\sqrt[N+1]{\eta}\simeq0$. We then find%
\begin{equation}
\mathcal{C}_{\text{loss}}(\eta\simeq0,N)\simeq\frac{\sqrt[N+1]{\eta}}{\ln
2}\simeq1.44~\sqrt[N+1]{\eta},
\end{equation}
which is also equal to $\sqrt[N+1]{\eta}$ nats per use. This is the
fundamental rate-loss scaling which affects long-distance repeater-assisted
quantum optical communications.

In the bosonic setting, it is interesting to compare the use of quantum
repeaters with the performance of a multi-band communication, where Alice and
Bob can exploit a communication line which is composed of $M$\ parallel and
independent lossy channels with identical transmissivity $\eta$. For instance,
$M$ can be interpreted as the frequency bandwidth of a multimode optical
fiber. As discussed in Sec.~\ref{teleDISTILLABLE}, the capacity of a multiband
lossy channel is given by~\cite{QKDpaper}
\begin{equation}
\mathcal{C}_{\text{loss}}(\eta,M)=-M\log_{2}(1-\eta). \label{Mband}%
\end{equation}

\begin{figure}[ptbh]
\vspace{0.2cm}
\par
\begin{center}
\includegraphics[width=0.46\textwidth] {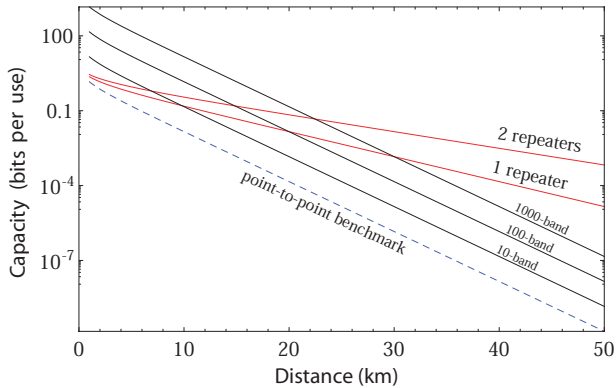} \vspace{2cm}
\vspace{-1.4cm} \vspace{1cm}
\end{center}
\par
\vspace{-2.0cm}\caption{Capacity (bits per use) versus distance (km) assuming
the standard loss rate of $0.2$ dB/km. We compare the use of repeaters
($N=1,2$) with that of a point-to-point multiband communication (for $M=10$,
$100$, and $1000$ bands or parallel channels). Dashed line is the
point-to-point benchmark (single-band, no repeaters). We see how the multiband
strategy increases the capacity in an additive way but it clearly suffers from
a poor long-distance rate-loss scaling with respect to the use of quantum
repeaters. }%
\label{compara}%
\end{figure}

Using Eqs.~(\ref{optLOSScap1}) and~(\ref{Mband}) we may compare the use of $N$
equidistant repeaters with the use of $M$ bands. From Fig.~\ref{compara}, we
clearly see that multiband quantum communication provides an additive effect
on the capacity which is very useful at short-intermediate distances. However,
at long distances, this solution is clearly limited by the same rate-loss
scaling which affects the single-band quantum channel (point-to-point
benchmark) and, therefore, it cannot compete with the long-distance
performance of repeater-assisted quantum communication. 

\subsection{Multiband repeater chains}

In general, the most powerful approach consists of relaying multiband quantum
communication, i.e., combining multiband channels with quantum repeaters. In
this regard, let us first discuss how Theorem~\ref{singleLETTtheorem} can be
easily extended to repeater chains which are connected by multiband quantum
channels. Then, we describe the performances in the bosonic setting.

Consider a multiband channel $\mathcal{E}^{\text{band}}$ which is composed of
$M$ independent channels (or bands) $\mathcal{E}_{k}$, i.e.,
\begin{equation}
\mathcal{E}^{\text{band}}=%
{\textstyle\bigotimes\nolimits_{k=1}^{M}}
\mathcal{E}_{k}~. \label{multibandGEN}%
\end{equation}
Assume that each band $\mathcal{E}_{k}$ can be LOCC-simulated with some
resource state $\sigma_{k}$. From Ref.~\cite{QKDpaper} and the subadditivity
of the REE, we may write the following bound for its two-way capacity
\begin{align}
\mathcal{C}(\mathcal{E}^{\text{band}})  &  \leq E_{\mathrm{R}}\left(
{\textstyle\bigotimes\nolimits_{k=1}^{M}}
\sigma_{k}\right) \nonumber\\
&  \leq%
{\textstyle\sum_{k=1}^{M}}
E_{\mathrm{R}}(\sigma_{k}):=\Psi(\mathcal{E}^{\text{band}}).
\end{align}
A multiband channel $\mathcal{E}^{\text{band}}$ is said\ to be
teleportation-covariant (distillable) if all its components $\mathcal{E}_{k}$
are teleportation-covariant (distillable). In a distillable $\mathcal{E}%
^{\text{band}}$, for each band $\mathcal{E}_{k}$ we may write $\mathcal{C}%
(\mathcal{E}_{k})=D_{1}(\sigma_{\mathcal{E}_{k}})=E_{\mathrm{R}}%
(\sigma_{\mathcal{E}_{k}})$ where $\sigma_{\mathcal{E}_{k}}$ is its Choi
matrix (with suitable asymptotic description in the bosonic case). Then, it is
straightforward to prove that~\cite{QKDpaper}%

\begin{equation}
\mathcal{C}(\mathcal{E}^{\text{band}})=%
{\textstyle\sum_{k=1}^{M}}
\mathcal{C}(\mathcal{E}_{k}).
\end{equation}

Similarly, we can extend Theorem~\ref{singleLETTtheorem}. Consider an adaptive
protocol over a repeater chain connected by multiband channels $\{\mathcal{E}%
_{i}^{\text{band}}\}$. We can define a corresponding two-way capacity for the
multiband chain $\mathcal{C}(\{\mathcal{E}_{i}^{\text{band}}\})$ and derive
the upper bound%
\begin{equation}
\mathcal{C}(\{\mathcal{E}_{i}^{\text{band}}\})\leq\min_{i}\Psi(\mathcal{E}%
_{i}^{\text{band}})~. \label{band1}%
\end{equation}
For a distillable multiband chain, we then have%
\begin{equation}
\mathcal{C}(\{\mathcal{E}_{i}^{\text{band}}\})=\min_{i}\mathcal{C}%
(\mathcal{E}_{i}^{\text{band}}). \label{bandDISTILL11}%
\end{equation}

In the bosonic setting, consider a chain of $N$ quantum repeaters with $N+1$
channels $\{\mathcal{E}_{i}\}$, where $\mathcal{E}_{i}$ is a multiband lossy
channel with $M_{i}$ bands and constant transmissivity $\eta_{i}$ (over the
bands). The two-way capacity of the $i$th link is therefore given by
$\mathcal{C}_{\text{loss}}(\eta_{i},M_{i})$ as specified by Eq.~(\ref{Mband}).
Because multiband lossy channels are distillable, we can apply
Eq.~(\ref{bandDISTILL11}) and derive the following repeater-assisted capacity
of the multiband lossy chain%
\begin{align}
\mathcal{C}_{\text{loss}}(\{\eta_{i},M_{i}\})  &  =\min_{i}\mathcal{C}%
_{\text{loss}}(\eta_{i},M_{i})\nonumber\\
&  =\min_{i}\left[  -M_{i}\log_{2}(1-\eta_{i})\right] \nonumber\\
&  =-\log_{2}\left[  \max_{i}(1-\eta_{i})^{M_{i}}\right] \nonumber\\
&  :=-\log_{2}\theta_{\max}~. \label{preBANDR}%
\end{align}

As before, it is interesting to discuss the symmetric scenario where the $N$
repeaters are equidistant, so that entire communication line is split into
$N+1$ links of the same optical length. Each link \textquotedblleft%
$i$\textquotedblright\ is therefore associated with a multiband lossy channel,
with bandwidth $M_{i}$ and constant transmissivity $\eta_{i}=\sqrt[N+1]{\eta}
$ (equal for all its bands). In this case, we have $\theta_{\max
}=(1-\sqrt[N+1]{\eta})^{\min_{i}M_{i}}$ in previous Eq.~(\ref{preBANDR}). In
other words, the repeater-assisted capacity of the chain becomes%
\[
\mathcal{C}_{\text{loss}}(\eta,N,\{M_{i}\})=-M_{\min}\log_{2}(1-\sqrt[N+1]%
{\eta}),
\]
where $M_{\min}:=\min_{i}M_{i}$ is the minimum bandwidth along the line, as
intuitively expected.

In general, the capacity is determined by an interplay between transmissivity
and bandwidth of each link. This is particularly evident in the regime of high
loss. By setting $\eta_{i}\simeq0$ in Eq.~(\ref{preBANDR}), we in fact derive%
\begin{equation}
\mathcal{C}_{\text{loss}}(\{\eta_{i}\simeq0,M_{i}\})\simeq c~\min_{i}\left(
M_{i}\eta_{i}\right)  ,
\end{equation}
where the constant $c$ is equal to $1.44$ bits or $1$ nat.

\section{Quantum networks\label{SECnetworks}}

We now consider the general case of a quantum network, where two end-users are
connected by an arbitrary ensemble of routes through intermediate points or
repeaters. Assuming the most basic quantum channels for the various
point-to-point connections, we determine the end-to-end capacities for quantum
communication, entanglement distillation and key generation under different
routing strategies. Our analysis combines tools from quantum information
theory (in particular, the generalization of the tools developed in
Ref.~\cite{QKDpaper}, needed for the converse part) and elements from
classical network information theory (necessary for the achievability part).

In this section, we start by introducing the main adaptive protocols based on
sequential (single-path) or parallel (multipath) routing of quantum systems.
We also give the corresponding definitions of network capacities. Then, in
Sec.~\ref{SecSTRETCHINGNET}, we will show how to simulate and
\textquotedblleft stretch\textquotedblright\ quantum networks, so that the
output of an adaptive protocol is completely simplified into a decomposition
of tensor-product states. This tool will be exploited to derive single-letter
REE upper bounds in the subsequent sections. In particular, in
Sec.~\ref{secSINGLE}, we\ will present the results for single-path routing,
while, in Sec.~\ref{secMULTI}, we will present results for multi-path routing.
The upper bounds will be combined with suitable lower bounds, and exact
formulas will be established for quantum networks connected by distillable channels.

\subsection{Notation and general definitions}

Consider a quantum communication network $\mathcal{N}$ whose points are
connected by memoryless quantum channels. The quantum network can be
represented as an undirected finite graph~\cite{Slepian,Acyclic}
$\mathcal{N}=(P,E)$ where $P$ is the finite set of points of the network
(vertices) and $E$ is the set of all connections (edges). Every point $x\in P$
has a local register of quantum systems $\mathbf{x}$ to be used for the
quantum communication. To simplify notation, we identify a point with its
local register $x=\mathbf{x}$. Two points $\mathbf{x},\mathbf{y}\in P$ are
connected by an undirected edge $(\mathbf{x},\mathbf{y})\in E$ if there is a
memoryless quantum channel $\mathcal{E}_{\mathbf{xy}}$ between $\mathbf{x}%
$\ and $\mathbf{y}$, which may be forward $\mathcal{E}_{\mathbf{x\rightarrow
y}}$\ or backward $\mathcal{E}_{\mathbf{y}\rightarrow\mathbf{x}}$.

In general, there may be multiple edges between two points, with each edge
representing an independent quantum channel. For instance, two undirected
edges between $\mathbf{x}$\ and $\mathbf{y}$ represent two channels
$\mathcal{E}_{\mathbf{xy}}\otimes\mathcal{E}_{\mathbf{xy}}^{\prime}$\ and
these may be associated with a double-band quantum communication (in one of
the two directions) or a two-way quantum communication (forward and backward
channels). While we allow for the possibility of multiple edges in the graph
(so that it is more generally a multi-graph) we may also collapse multiple
edges into a single edges to simplify the complexity of the network and
therefore notation.

In the following, we also use the labeled notation $\mathbf{p}_{i}$ for the
generic point of the graphical network, so that two points $\mathbf{p}_{i}$
and $\mathbf{p}_{j}$ are connected by an edge if there is a quantum channel
$\mathcal{E}_{ij}:=\mathcal{E}_{\mathbf{p}_{i}\mathbf{p}_{j}}$. We also adopt
the specific notation $\mathbf{a}$ and $\mathbf{b}$ for the two end-points,
Alice and Bob. An end-to-end route is an undirected path between Alice and
Bob, which is specified by a sequence of edges $\{(\mathbf{a},\mathbf{p}%
_{i}),\cdots,(\mathbf{p}_{j},\mathbf{b})\}$, simply denoted as $\mathbf{a}%
-\mathbf{p}_{i}-\cdots-\mathbf{p}_{j}-\mathbf{b}$. This may be interpreted as
a linear chain of $N$ repeaters between Alice and Bob, connected by a sequence
of $N+1$ channels $\{\mathcal{E}_{k}\}$, i.e.,%
\begin{equation}
\mathbf{a}\overset{\mathcal{E}_{0}}{-}(\mathbf{p}_{i}:=\mathbf{r}_{1}%
)-\cdots\overset{\mathcal{E}_{k}}{-}\cdots-(\mathbf{p}_{j}:=\mathbf{r}%
_{N})\overset{\mathcal{E}_{N}}{-}\mathbf{b},
\end{equation}
where the same repeater may appear at different positions (in particular, this
occurs when the route is not a simple path, so that there are cycles).

In general, the two end-points may transmit quantum systems through an
ensemble of routes $\Omega=\{1,\ldots,\omega,\ldots\}$. Note that this
ensemble is generally large but can always be made finite in a finite network,
by just reducing the routes to be simple paths, void of cycles (without losing
generality). Different routes $\omega$ and $\omega^{\prime}$ may have
collisions, i.e., repeaters and channels in common. Generic route $\omega$
involves the transmission through $N_{\omega}+1$ channels $\{\mathcal{E}%
_{0}^{\omega},\ldots,\mathcal{E}_{k}^{\omega},\ldots,\mathcal{E}_{N_{\omega}%
}^{\omega}\}$. In general, we assume that each quantum transmission through
each channel is alternated with network LOCCs: These are defined as adaptive
LOs performed by all points of the network on their local registers, which are
assisted by unlimited two-way CC involving the entire network.

Finally, we consider two possible fundamental strategies for routing the
quantum systems through the network: Sequential or parallel. In a sequential
or single-path routing, quantum systems are transmitted from Alice to Bob
through a single route for each use of the network. This process is generally
stochastic, so that route $\omega$ is chosen with some probability $p_{\omega
}$. By contrast, in a parallel or multipath routing, systems are
simultaneously transmitted through multiple routes for each use of the
network. This may be seen as a \textquotedblleft broadband
use\textquotedblright\ of the quantum network. We now explain these two
strategies in detail.

\subsection{Sequential (single-path) routing}

The most general network protocol for sequential quantum communication
involves the use of generally-different routes, accessed one after the other.
The network is initialized by means of a first LOCC $\Lambda_{0}$ which
prepares an initial separable state. With probability $\pi_{0}^{1}$, Alice
$\mathbf{a}$ exchanges one system with repeater $\mathbf{p}_{i}$. This is
followed by another LOCC $\Lambda_{1}$. Next, with probability $\pi_{1}^{1}$,
repeater $\mathbf{p}_{i}$ exchanges one system with repeater $\mathbf{p}_{j}$
and so on. Finally, with probability $\pi_{N_{1}}^{1}$, repeater
$\mathbf{p}_{k}$ exchanges one system with Bob $\mathbf{b}$, followed by a
final LOCC $\Lambda_{N_{1}+1}$. Thus, with probability $p_{1}=\Pi_{i}\pi
_{i}^{1}$, the end-points exchange one system which has undergone $N_{1}+1$
transmissions $\{\mathcal{E}_{i}^{1}\}$ along the first route.

The next uses involve generally-different routes. After many uses $n$, the
random process defines a sequential routing table $\mathcal{R}=\{\omega
,p_{\omega}\}$, where route $\omega$ is picked with probability $p_{\omega}$
and involves $N_{\omega}+1$ transmissions $\{\mathcal{E}_{i}^{\omega}\}$.
Thus, we have a total of $N_{\text{tot}}=\Sigma_{\omega}np_{\omega}(N_{\omega
}+1)$ transmissions and a sequence of LOCCs $\mathcal{L}=\{\Lambda_{0}%
,\ldots,\Lambda_{N_{\text{tot}}}\}$, whose output provides Alice and Bob's
final state $\rho_{\mathbf{ab}}^{n}$. Note that we may weaken the previous
description: While maintaining the sequential use of the routes, in each route
we may permute the order of the transmissions (as before for the case of a
linear chain of repeaters).

The sequential network protocol $\mathcal{P}_{\text{\textrm{seq}}}$ is
characterized by $\mathcal{R}$ and $\mathcal{L}$, and its average rate is
$R_{n}$ if $\left\Vert \rho_{\mathbf{ab}}^{n}-\phi_{n}\right\Vert
\leq\varepsilon$, where $\phi_{n}$ is a target state of $nR_{n}$ bits. By
taking the asymptotic rate for large $n$ and optimizing over all the
sequential protocols, we define the sequential or single-path capacity of the
network%
\begin{equation}
\mathcal{C}(\mathcal{N}):=\sup_{\mathcal{P}_{\text{\textrm{seq}}}}\lim
_{n}R_{n}. \label{necCAPdef}%
\end{equation}
The capacity $\mathcal{C}(\mathcal{N})$ provides the maximum number of
(quantum, entanglement, or secret) bits which are distributed per sequential
use of the network or single-path transmission. In particular, by specifying
the target state, we define the corresponding network capacities for quantum
communication, entanglement distillation, key generation and private
communication, which satisfy
\begin{equation}
Q_{2}(\mathcal{N})=D_{2}(\mathcal{N})\leq K(\mathcal{N})=P_{2}(\mathcal{N}).
\end{equation}

\begin{figure}[ptbh]
\vspace{-1.6cm}
\par
\begin{center}
\includegraphics[width=0.5\textwidth] {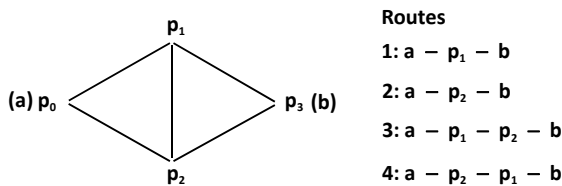}
\end{center}
\par
\vspace{-2.5cm}\caption{Sequential use of a diamond quantum network. Each use
of the network corresponds to routing a quantum system between the two
end-points Alice $\mathbf{a}$ and Bob $\mathbf{b}$. In a diamond network with
four points $\mathbf{p}_{0}=\mathbf{a}$, $\mathbf{p}_{1}$, $\mathbf{p}_{2}$,
and $\mathbf{p}_{3}=\mathbf{b}$, we may identify four basic routes
$\omega=1,2,3,4$ (see list on the right). These are simple paths between Alice
and Bob with the middle points $\mathbf{p}_{1}$ and $\mathbf{p}_{2}$ acting as
quantum repeaters in different succession. For instance, $\mathbf{p}_{1}$ is
the first repeater in route $3$ and the second repeater in route $4$. Note
that we may consider further routes by including loops between $\mathbf{p}%
_{1}$ and $\mathbf{p}_{2}$. These other solutions are non-simple paths that we
may discard without losing generality.}%
\label{NET1}%
\end{figure}

It is important to note that the sequential use is the best practical strategy
when Alice and the other points of the network aim to optimize the use of
their quantum resources.\ In fact, $\mathcal{C}(\mathcal{N})$ can also be
expressed as maximum number of target bits per quantum system routed.
Furthermore, suppose that the end-points have control on the routing, so that
they can adaptively select the best routes based on the CCs received by the
repeaters. Under such hypothesis, they can optimize the protocol on the fly
and adapt the routing table so that it asymptotically converges to the use of
an optimal route $\omega_{\ast}$. See Fig.~\ref{NET1} for an example of
sequential use of a simple network. 

\subsection{Parallel (multipath) routing\label{BroadNET}}

Here we consider a different situation where Alice, Bob and the other points
of the network do not have restrictions or costs associated with the use of
their quantum resources, so that they can optimize the use of the quantum
network without worrying if some of their quantum systems are inefficiently
transmitted or even lost (this may be the practical scenario of many optical
implementations, e.g., based on cheap resources like coherent states). In such
a case, the optimal use of the quantum network is parallel or broadband,
meaning that the quantum systems are simultaneously routed through multiple
paths each time the quantum network is accessed.

In a parallel network protocol, Alice broadcasts quantum systems to all
repeaters she has a connection with. Such a simultaneous transmission to her
\textquotedblleft neighbor\textquotedblright\ repeaters can be denoted by
$\mathbf{a}\rightarrow\{\mathbf{p}_{k}\}$. In turn, each of the receiving
repeaters multicasts quantum systems to another set of neighbor repeaters
$\mathbf{p}_{k}\rightarrow\{\mathbf{p}_{j}\}$ and so on, until Bob
$\mathbf{b}$ is reached as an end-point. This is done in such a way that each
multicast occurs between two network LOCCs, and different multicasts do not
overlap, so that all edges of the network are used exactly once at the end of
each end-to-end transmission. This condition is assured by imposing that
multicasts may only occur though unused connections and is commonly known as
\textquotedblleft flooding\textquotedblright\ strategy~\cite{flooding}.

In general, each multicast must be intended in a weaker sense as a
point-to-multipoint connection where quantum systems may be exchanged through
forward or backward transmissions, depending on the actual physical directions
of the available quantum channels. Independently from the physical directions
of the channels, we may always assign a common sender-receiver direction to
all the edges involved in the process, so that there will be a
\textit{logical} sender-receiver orientation associated with the multicast.
For this reason, the notation $\mathbf{a}\rightarrow\{\mathbf{p}_{k}\}$ must
be generally interpreted as a logical multicast where Alice \textquotedblleft
connects to\textquotedblright\ repeaters $\{\mathbf{p}_{k}\}$. To better
explain this broadband use, let us better formalize the orientations.

\begin{figure}[ptbh]
\vspace{-1.0cm}
\par
\begin{center}
\includegraphics[width=0.5\textwidth] {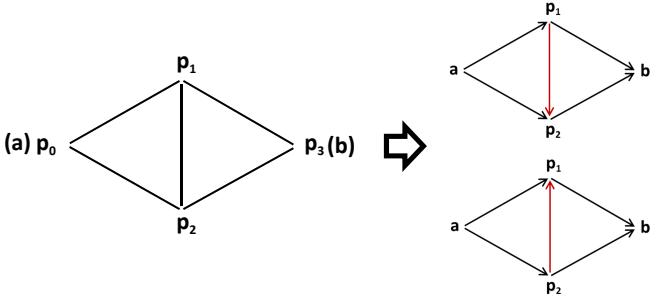}
\end{center}
\par
\vspace{-1.6cm}\caption{Orientations of a diamond quantum network.\ There are
only two possible orientations that transform the original undirected network
(left) into a flow network (right). Within an orientation, there is a
well-defined logical multicast from each point of the network to all its
out-neighborhood (empty for Bob). A multipath routing strategy (flooding) is
defined as a sequence of such multicasts. Therefore, in the upper orientation,
we may identify the basic multipath routing $\mathbf{a}\rightarrow
\{\mathbf{p}_{1},\mathbf{p}_{2}\}$, $\mathbf{p}_{1}\rightarrow\{\mathbf{p}%
_{2},\mathbf{b}\}$, and $\mathbf{p}_{2}\rightarrow\mathbf{b}$. Other routings
are given by permuting these multicasts. For instance, we may have the
different sequence $\mathbf{p}_{1}\rightarrow\{\mathbf{p}_{2},\mathbf{b}\}$,
$\mathbf{p}_{2}\rightarrow\mathbf{b}$ and $\mathbf{a}\rightarrow
\{\mathbf{p}_{1},\mathbf{p}_{2}\}$ for the upper orientation. In the lower
orientation, we have the basic multipath routing $\mathbf{a}\rightarrow
\{\mathbf{p}_{1},\mathbf{p}_{2}\}$, $\mathbf{p}_{2}\rightarrow\{\mathbf{p}%
_{1},\mathbf{b}\}$ and $\mathbf{p}_{1}\rightarrow\mathbf{b}$, plus all the
possible permutations.}%
\label{NET2b}%
\end{figure}

Recall that a directed edge is an ordered pair $(\mathbf{x},\mathbf{y})$,
where the initial vertex $\mathbf{x}$ is called \textquotedblleft
tail\textquotedblright\ and the terminal vertex $\mathbf{y}$ is called
\textquotedblleft head\textquotedblright. Let us transform the undirected
graph of the network $\mathcal{N}=(P,E)$ into a directed graph by randomly
choosing a direction for all the edges, while keeping Alice as tail and Bob as
head. The goal is to represent the quantum network as a flow network where
Alice is the\textit{\ source} and Bob is the\textit{\ sink}~\cite{Dinic,Karp}.
In general, there are many solutions for this random orientation. In fact,
consider the sub-network where Alice and Bob have been disconnected, i.e.,
$\mathcal{N}^{\prime}=(P^{\prime},E^{\prime})$ with $P^{\prime}=P\setminus
\{\mathbf{a,b}\}$. There are $2^{|E^{\prime}|}$ possible directed graphs that
can be generated, where $|E^{\prime}|$ is the number of undirected edges in
$\mathcal{N}^{\prime}$. Thus, we have $2^{|E^{\prime}|}$\ orientations of the
original network $\mathcal{N}$. Each of these orientations defines a flow
network and provides possible strategies for multipath routing. See
Fig.~\ref{NET2b} for a simple example. 

To better formalize the routing strategy, let us exploit the notions of in-
and out-neighborhoods. Given an orientation of $\mathcal{N}$, we have a
corresponding flow network, denoted by $\mathcal{N}_{D}=(P,E_{D})$, where
$E_{D}$ is the set of directed edges. For arbitrary point $\mathbf{p}$, we
define its out-neighborhood as the set of heads going from $\mathbf{p}$%
\begin{equation}
N^{\text{out}}(\mathbf{p})=\{\mathbf{x}\in P:(\mathbf{p},\mathbf{x})\in
E_{D}\},
\end{equation}
and its in-neighborhood as the set of tails going into $\mathbf{p}$%
\begin{equation}
N^{\text{in}}(\mathbf{p})=\{\mathbf{x}\in P:(\mathbf{x},\mathbf{p})\in
E_{D}\}.
\end{equation}
A logical multicast from point $\mathbf{p}$ can be defined as a
point-to-multipoint connection from $\mathbf{p}$ to all its out-neighborhood
$N^{\text{out}}(\mathbf{p})$, i.e., $\mathbf{p}\rightarrow N^{\text{out}%
}(\mathbf{p})$. A multipath routing strategy can therefore be defined as an
ordered sequence of all such multicasts. See Fig.~\ref{NET2b}.

\begin{figure}[ptbh]
\begin{center}
\vspace{-0.5cm} \includegraphics[width=0.45\textwidth] {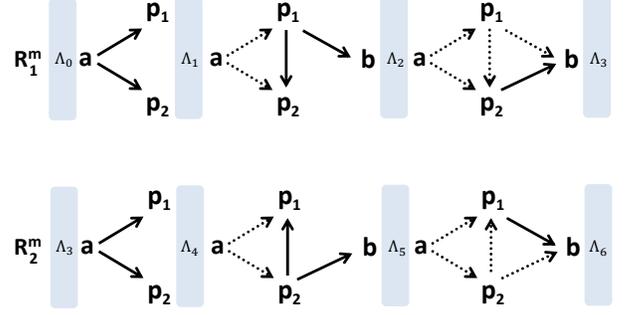}
\end{center}
\par
\vspace{-1.1cm}\caption{Two possible parallel uses of a diamond quantum
network. In the upper multipath routing $R_{1}^{\text{m}}$, after the initial
LOCC $\Lambda_{0}$, there is the first multicast $\mathbf{a}\rightarrow
\{\mathbf{p}_{1},\mathbf{p}_{2}\}$, followed by the LOCC $\Lambda_{1}$. Then,
we have the second multicast $\mathbf{p}_{1}\rightarrow\{\mathbf{b}%
,\mathbf{p}_{2}\}$ followed by $\Lambda_{2}$. Finally, we have $\mathbf{p}%
_{2}\rightarrow\mathbf{b}$ followed by the final LOCC $\Lambda_{3}$. This
completes a single end-to-end transmission. In the lower multipath routing
$R_{2}^{\text{m}}$, the process is similar to $R_{1}^{\text{m}}$ but with
$\mathbf{p}_{1}$ and $\mathbf{p}_{2}$ being inverted.}%
\label{duo}%
\end{figure}

Using these definitions we may easily formalize the multipath network protocol
that we may simply call \textquotedblleft flooding protocol\textquotedblright.
Suppose that we have $|P|=Z+2$ points in the network ($Z$ repeaters plus the
two end-points). The first step of the protocol is the agreement of a
multipath routing strategy $R_{1}^{\text{m}}$ by means of preliminary CCs
among all the points. This is part of an initialization LOCC $\Lambda_{0}$
which prepares an initial separable state for the entire network. Then, Alice
$\mathbf{a}$\ exchanges quantum systems with all her out-neighborhood
$N^{+}(\mathbf{a})$. This multicast is followed by a network LOCC $\Lambda
_{1}$. Next, repeater $\mathbf{p}_{1}\in N^{+}(\mathbf{a})$ exchanges quantum
systems with all its out-neighborhood $N^{+}(\mathbf{p}_{1})$, which is
followed by another LOCC $\Lambda_{2}$ and so on. At some step $Z+1$, Bob
$\mathbf{b}$ will have exchanged quantum systems with all his in-neighborhood
$N^{-}(\mathbf{b})$, after which there is a final LOCC $\Lambda_{Z+1}$. This
completes the first multipath transmission between the end-points by means of
the routing $R_{1}^{\text{m}}$ and the sequence of LOCCs $\{\Lambda_{0}%
,\ldots,\Lambda_{Z+1}\}$. Then, there will be the second use of the network
with a generally different routing strategy $R_{2}^{\text{m}}$, and so on. See
Fig.~\ref{duo}. 

Let us note that the points of the network may generally update their routing
strategy \textquotedblleft on the fly\textquotedblright, i.e., while the
protocol is running; then, the various multicasts may be suitably permuted in
their order. In any case, for large number of uses $n$, we will have a
sequence of multipath routings $\mathcal{R}^{\text{m}}=\{R_{1}^{\text{m}%
},\ldots,R_{n}^{\text{m}}\}$ and network LOCCs $\mathcal{L}=\{\Lambda
_{0},\ldots,\Lambda_{n(Z+1)}\}$ whose output provides Alice and Bob's final
state $\rho_{\mathbf{ab}}^{n}$. The flooding protocol $\mathcal{P}%
_{\mathrm{flood}}$ will be fully described by $\mathcal{R}^{\text{m}}$ and
$\mathcal{L}$. By definition, its average rate is $R_{n}$ if $\left\Vert
\rho_{\mathbf{ab}}^{n}-\phi_{n}\right\Vert \leq\varepsilon$, where $\phi_{n}$
is a target state of $nR_{n}$ bits. The multipath capacity of the network is
defined by optimizing the asymptotic rate over all flooding protocols, i.e.,
\begin{equation}
\mathcal{C}^{\text{m}}(\mathcal{N}):=\sup_{\mathcal{P}_{\mathrm{flood}}}%
\lim_{n}R_{n}. \label{bbnetCAPdef}%
\end{equation}
By specifying the target state, we define corresponding capacities for quantum
communication, entanglement distillation, key generation and private
communication, satisfying
\begin{equation}
Q_{2}^{\text{m}}(\mathcal{N})=D_{2}^{\text{m}}(\mathcal{N})\leq K^{\text{m}%
}(\mathcal{N})=P_{2}^{\text{m}}(\mathcal{N}).
\end{equation}

Before proceeding, some other considerations are in order. Note that the
parallel uses of the network may also be re-arranged in such a way that each
point performs all its multicasts before another point. For instance, in the
example of Fig.~\ref{duo}, we may consider Alice performing all her $n$
multicasts $\mathbf{a}\rightarrow\{\mathbf{p}_{1},\mathbf{p}_{2}\}$ as a first
step. Suppose that routes $R_{1}^{\text{m}}$ and $R_{2}^{\text{m}}$\ are
chosen with probability $p$ and $1-p$. Then, after Alice has finished,\ point
$\mathbf{p}_{1}$ performs its $np$ multicasts and $\mathbf{p}_{2}$\ performs
its $n(1-p)$ multicasts, and so on. We may always re-arrange the protocol and
adapt the LOCC sequence $\mathcal{L}$ to include this variant.

Then, there is a simplified formulation to keep in mind. In fact, a special
case is when the various multicasts within the same routing strategy are not
alternated with network LOCCs but they are all performed simultaneously, with
only the initial and final LOCCs to be applied. For instance, for the routing
$R_{1}^{\text{m}}$ of Fig.~\ref{duo}, this means to set $\Lambda_{1}%
=\Lambda_{2}=I$ and assume that the multicasts $\mathbf{a}\rightarrow
\{\mathbf{p}_{1},\mathbf{p}_{2}\}$, $\mathbf{p}_{1}\rightarrow\{\mathbf{b}%
,\mathbf{p}_{2}\}$ and $\mathbf{p}_{2}\rightarrow\mathbf{b}$ occur
simultaneously, after the initialization $\Lambda_{0}$ and before $\Lambda
_{3}$. In general, any variant of the protocol may be considered as long as
each quantum channel (edge) is used exactly $n$ times at the end of the
communication, i.e., after $n$ uses of the quantum network.

In the following section, we show how to simulate a quantum network and then
exploit teleportation stretching to reduce adaptive protocols (based on
single- or multi-path routings) into much simpler block versions. By combining
this technique with entanglement cuts of the quantum network, we will derive
very useful decompositions for Alice and Bob's output state. These
decompositions will be later exploited in Secs.~\ref{secSINGLE}
and~\ref{secMULTI} to derive single-letter upper bounds for the network
capacities $\mathcal{C}(\mathcal{N})$ and $\mathcal{C}^{\text{m}}%
(\mathcal{N})$. Corresponding lower bounds will also be derived by combining
point-to-point quantum protocols with classical routing strategies, with exact
results for distillable networks.

\section{Simulation and stretching of a quantum
network\label{SecSTRETCHINGNET}}

\subsection{General approach}

Consider a quantum network $\mathcal{N}$\ which is connected by arbitrary
quantum channels. Given two points $\mathbf{x}$\ and $\mathbf{y}$ connected by
channel $\mathcal{E}_{\mathbf{xy}}$, we consider its simulation
$S_{\mathbf{xy}}=(\mathcal{T}_{\mathbf{xy}},\sigma_{\mathbf{xy}})$ for some
LOCC $\mathcal{T}_{\mathbf{xy}}$ and resource state $\sigma_{\mathbf{xy}}$.
Repeating this for all connected points $(\mathbf{x},\mathbf{y})\in E$, we
define an LOCC simulation of the entire network $S(\mathcal{N}%
)=\{S_{\mathbf{xy}}\}_{(\mathbf{x},\mathbf{y})\in E}$ and a corresponding
resource representation of the network $\sigma(\mathcal{N})=\{\sigma
_{\mathbf{xy}}\}_{(\mathbf{x},\mathbf{y})\in E}$. For a network of
teleportation-covariant channels, its simulation $S(\mathcal{N})$ is based on
teleportation over Choi matrices, so that we may consider $\sigma
(\mathcal{N})=\{\sigma_{\mathcal{E}_{\mathbf{xy}}}\}_{(\mathbf{x}%
,\mathbf{y})\in E}$, i.e., we have a \textquotedblleft
Choi-representation\textquotedblright\ of the network. Note that the
simulation may be asymptotic for a network of bosonic channels, following the
same treatment previously explained for a linear chain of repeaters .

\begin{figure}[ptbh]
\vspace{-1.6cm}
\par
\begin{center}
\includegraphics[width=0.52\textwidth] {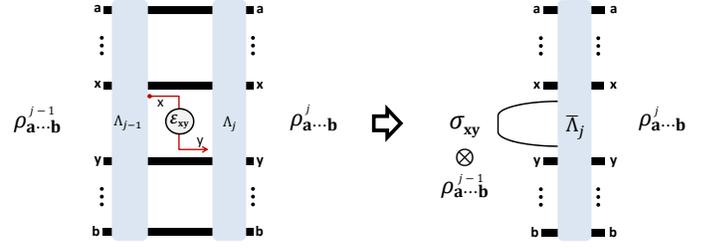}
\end{center}
\par
\vspace{-1.9cm}\caption{Stretching of a network. Consider the $j$th
transmission between points $\mathbf{x}$ and $\mathbf{y}$, so that the network
state $\rho_{\mathbf{a\ldots b}}^{j-1}$ is transformed into $\rho
_{\mathbf{a\ldots b}}^{j}$. By introducing the simulation $(\mathcal{T}%
_{\mathbf{xy}},\sigma_{\mathbf{xy}})$ of channel $\mathcal{E}_{\mathbf{xy}}$,
we may stretch the resource state $\sigma_{\mathbf{xy}}$ out of the LOCCs and
collapse $\Lambda_{j-1}$, $\mathcal{T}_{\mathbf{xy}}$ and $\Lambda_{j}$ into a
single LOCC $\bar{\Lambda}_{j}$ applied to $\rho_{\mathbf{a\ldots b}}%
^{j-1}\otimes\sigma_{\mathbf{xy}}$, as in Eq.~(\ref{Net_IT}).}%
\label{NET3}%
\end{figure}

By adopting a network simulation $S(\mathcal{N})$, we may apply simplify
adaptive protocols via teleportation stretching, by extending the procedure
employed for a linear chain of quantum repeaters, with the important
difference that we now have many possible chains (the network routes) and
these may also have collisions, i.e., repeaters and channels in common. The
stretching of a quantum network is performed iteratively, i.e., transmission
after transmission. Suppose that the $j$th transmission in the network occurs
between points $\mathbf{x}$ and $\mathbf{y}$ via channel $\mathcal{E}%
_{\mathbf{xy}}$ with associated resource state $\sigma_{\mathbf{xy}}$. Call
$\rho_{\mathbf{a\ldots b}}^{j}$ the global state of the network after this
transmission. Then, we may write%
\begin{equation}
\rho_{\mathbf{a\ldots b}}^{j}=\bar{\Lambda}_{j}\left(  \rho_{\mathbf{a\ldots
b}}^{j-1}\otimes\sigma_{\mathbf{xy}}\right)  , \label{Net_IT}%
\end{equation}
where $\bar{\Lambda}_{j}$\ is a trace-preserving LOCC (see also
Fig.~\ref{NET3}). 

By iterating Eq.~(\ref{Net_IT}) and considering that the initial state of
network $\rho_{\mathbf{a\ldots b}}^{0}$ is separable, we may then write the
network output state after $n$ transmissions as%
\begin{equation}
\rho_{\mathbf{a\ldots b}}^{n}=\bar{\Lambda}\left[  \underset{(\mathbf{x}%
,\mathbf{y})\in E}{%
{\textstyle\bigotimes}
}~\sigma_{\mathbf{xy}}^{\otimes n_{\mathbf{xy}}}\right]  ,
\label{NetREDUCTION}%
\end{equation}
where $n_{\mathbf{xy}}$ is the number of uses of channel $\mathcal{E}%
_{\mathbf{xy}}$ or, equivalently, edge $(\mathbf{x},\mathbf{y})$. Then, by
tracing out all the points but Alice and Bob, we get their final shared state
\begin{equation}
\rho_{\mathbf{ab}}^{n}=\bar{\Lambda}_{\mathbf{ab}}\left[  \underset
{(\mathbf{x},\mathbf{y})\in E}{%
{\textstyle\bigotimes}
}~\sigma_{\mathbf{xy}}^{\otimes n_{\mathbf{xy}}}\right]  , \label{NET2RED}%
\end{equation}
for another trace-preserving LOCC $\bar{\Lambda}_{\mathbf{ab}}$.

\begin{figure}[ptbh]
\vspace{-1.3cm}
\par
\begin{center}
\includegraphics[width=0.5\textwidth] {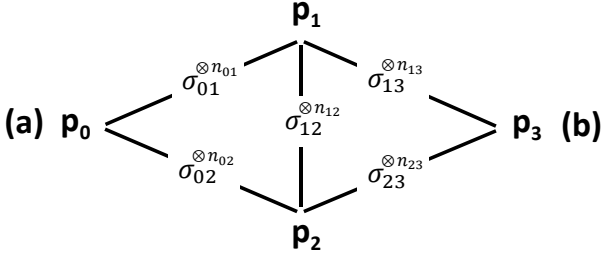} \vspace{-2.2cm}
\end{center}
\caption{Network stretching. Consider a diamond quantum network $\mathcal{N}%
^{\diamond}=(\{\mathbf{p}_{0},\mathbf{p}_{1},\mathbf{p}_{2},\mathbf{p}%
_{3}\},E)$ with resource representation $\sigma(\mathcal{N}^{\diamond
})=\{\sigma_{01},\sigma_{02},\sigma_{12},\sigma_{13},\sigma_{23}\}$. Before
stretching, an arbitrary edge $(\mathbf{x},\mathbf{y})$ with channel
$\mathcal{E}_{\mathbf{xy}}$ is used $n_{\mathbf{xy}}$ times. After stretching,
the same edge $(\mathbf{x},\mathbf{y})$ is associated with $n_{\mathbf{xy}}$
copies of the resource state $\sigma_{\mathbf{xy}}$. The latter is the Choi
matrix $\sigma_{\mathcal{E}_{\mathbf{xy}}}$ if $\mathcal{E}_{\mathbf{xy}}$ is
teleportation-covariant. The global state of the network is expressed as in
Eq.~(\ref{LemmaNETstretching}), which may take an asymptotic form for a
network of bosonic channels.}%
\label{NET4}%
\end{figure}

Note that the decompositions of Eqs.~(\ref{NetREDUCTION}) and (\ref{NET2RED})
can be written for any adaptive network protocol (sequential or flooding). For
a sequential protocol $n_{\mathbf{xy}}=np_{\mathbf{xy}}\leq n$, where
$p_{\mathbf{xy}}$ is the probability of using edge $(\mathbf{x},\mathbf{y})$.
For a flooding protocol, we instead have $n_{\mathbf{xy}}=n$, because each
edge is used exactly once in each end-to-end transmission. In particular, in a
flooding protocol, we have the parallel use of several channels $\mathcal{E}%
_{\mathbf{x}_{1}\mathbf{y}_{1}}$, $\mathcal{E}_{\mathbf{x}_{2}\mathbf{y}_{2}}%
$, \ldots\ in each multicast, which means that trivial LOCCs (identities) are
applied between every two transmissions in the same multicast. We have
therefore proven the following result (see also Fig.~\ref{NET4} for a simple example).

\begin{lemma}
[Network stretching]\label{LemmaNET}Consider a quantum network $\mathcal{N}%
=(P,E)$ which is simulable with some resource representation $\sigma
(\mathcal{N})=\{\sigma_{\mathbf{xy}}\}_{(\mathbf{x},\mathbf{y})\in E}$. Then,
consider $n$ uses of an adaptive protocol so that edge $(\mathbf{x}%
,\mathbf{y})\in E$ is used $n_{\mathbf{xy}}$ times. We may write the global
output state of the network as%
\begin{equation}
\rho_{\mathbf{a\ldots b}}^{n}=\bar{\Lambda}\left[  \underset{(\mathbf{x}%
,\mathbf{y})\in E}{%
{\textstyle\bigotimes}
}~\sigma_{\mathbf{xy}}^{\otimes n_{\mathbf{xy}}}\right]  ,
\label{LemmaNETstretching}%
\end{equation}
for a trace-preserving LOCC $\bar{\Lambda}$. Similarly, Alice and Bob's output
state $\rho_{\mathbf{ab}}^{n}$ is given by Eq.~(\ref{LemmaNETstretching}) up
to a different trace-preserving LOCC $\bar{\Lambda}_{\mathbf{ab}}$. In
particular, we have $n_{\mathbf{xy}}\leq n$ ($n_{\mathbf{xy}}=n$) for a
sequential (flooding) protocol. Formulations may be asymptotic for bosonic channels.
\end{lemma}

As we state in the lemma, the stretching procedure also applies to networks of
bosonic channels with asymptotic simulations. This can be understood by
extending the argument already given for linear chains. For the sake of
clarity, we make this argument explicit here. Consider again the $j$th
transmission in the network occurring via channel $\mathcal{E}_{\mathbf{xy}}$
as in Fig.~\ref{NET3}. For the global state of the network, we may write
\begin{equation}
\rho_{\mathbf{a\ldots b}}^{j}=\Lambda_{j}\circ\mathcal{E}_{\mathbf{xy}}%
\circ\Lambda_{j-1}(\rho_{\mathbf{a\ldots b}}^{j-1})~.
\end{equation}
Suppose that we replace each channel $\mathcal{E}_{\mathbf{xy}}$ in the
network with an approximation $\mathcal{E}_{\mathbf{xy}}^{\mu}$, with
point-wise limit $\mathcal{E}_{\mathbf{xy}}=\lim_{\mu}\mathcal{E}%
_{\mathbf{xy}}^{\mu}$, meaning that $\left\Vert \mathcal{E}_{\mathbf{xy}}%
(\rho)-\mathcal{E}_{\mathbf{xy}}^{\mu}(\rho)\right\Vert \overset{\mu
}{\rightarrow}0$ for any state $\rho$. We may build the approximate network
state
\begin{equation}
\rho_{\mathbf{a\ldots b}}^{j,\mu}=\Lambda_{j}\circ\mathcal{E}_{\mathbf{xy}%
}^{\mu}\circ\Lambda_{j-1}(\rho_{\mathbf{a\ldots b}}^{j-1,\mu})~.
\end{equation}

Now assume that all the registers in the network are bounded by a large but
finite mean number of photons $\bar{E}$, so that we may write $\left\Vert
\mathcal{E}_{\mathbf{xy}}-\mathcal{E}_{\mathbf{xy}}^{\mu}\right\Vert
_{\diamond\bar{E}}\overset{\mu}{\rightarrow}0$ in the energy-bounded diamond
norm defined in Eq.~(\ref{bDIAMONDnorm}). By using the monotonicity under CPTP
maps and the triangular inequality, we then compute%
\begin{align}
&  \left\Vert \rho_{\mathbf{a\ldots b}}^{j}-\rho_{\mathbf{a\ldots b}}^{j,\mu
}\right\Vert \nonumber\\
&  \leq\left\Vert \mathcal{E}_{\mathbf{xy}}\circ\Lambda_{j-1}(\rho
_{\mathbf{a\ldots b}}^{j-1})-\mathcal{E}_{\mathbf{xy}}^{\mu}\circ\Lambda
_{j-1}(\rho_{\mathbf{a\ldots b}}^{j-1,\mu})\right\Vert \nonumber\\
&  \leq\left\Vert \mathcal{E}_{\mathbf{xy}}\circ\Lambda_{j-1}(\rho
_{\mathbf{a\ldots b}}^{j-1})-\mathcal{E}_{\mathbf{xy}}^{\mu}\circ\Lambda
_{j-1}(\rho_{\mathbf{a\ldots b}}^{j-1})\right\Vert \nonumber\\
&  +\left\Vert \mathcal{E}_{\mathbf{xy}}^{\mu}\circ\Lambda_{j-1}%
(\rho_{\mathbf{a\ldots b}}^{j-1})-\mathcal{E}_{\mathbf{xy}}^{\mu}\circ
\Lambda_{j-1}(\rho_{\mathbf{a\ldots b}}^{j-1,\mu})\right\Vert \nonumber\\
&  \leq\left\Vert \mathcal{E}_{\mathbf{xy}}-\mathcal{E}_{\mathbf{xy}}^{\mu
}\right\Vert _{\diamond\bar{E}}+\left\Vert \rho_{\mathbf{a\ldots b}}%
^{j-1}-\rho_{\mathbf{a\ldots b}}^{j-1,\mu}\right\Vert .
\end{align}
By iterating the previous formula for all the transmissions in the network, we
derive%
\begin{equation}
\left\Vert \rho_{\mathbf{a\ldots b}}^{n}-\rho_{\mathbf{a\ldots b}}^{n,\mu
}\right\Vert \leq%
{\textstyle\sum\limits_{(\mathbf{x},\mathbf{y})\in E}}
n_{\mathbf{xy}}\left\Vert \mathcal{E}_{\mathbf{xy}}-\mathcal{E}_{\mathbf{xy}%
}^{\mu}\right\Vert _{\diamond\bar{E}}~. \label{bbb1}%
\end{equation}
This distance goes to zero in $\mu$ for any number of uses $n$, any finite
number of edges $|E|$, and any energy $\bar{E}$.

Now suppose that the approximate channel $\mathcal{E}_{\mathbf{xy}}^{\mu}$ has
an LOCC simulation with some resource state $\sigma_{\mathbf{xy}}^{\mu}$.
Then, we may write the approximate network stretching
\begin{equation}
\rho_{\mathbf{a\ldots b}}^{n,\mu}=\bar{\Lambda}^{\mu}\left[  \underset
{(\mathbf{x},\mathbf{y})\in E}{%
{\textstyle\bigotimes}
}~\sigma_{\mathbf{xy}}^{\mu\otimes n_{\mathbf{xy}}}\right]  , \label{bbb2}%
\end{equation}
for a trace-preserving LOCC\ $\bar{\Lambda}^{\mu}$. Combining Eqs.~(\ref{bbb1}%
) and~(\ref{bbb2}), we may therefore write the asymptotic version of network
stretching%
\begin{equation}
\rho_{\mathbf{a\ldots b}}^{n}=\lim_{\mu}\bar{\Lambda}^{\mu}\left[
\underset{(\mathbf{x},\mathbf{y})\in E}{%
{\textstyle\bigotimes}
}~\sigma_{\mathbf{xy}}^{\mu\otimes n_{\mathbf{xy}}}\right]  ,
\label{totalASYMP}%
\end{equation}
where the limit in $\mu$\ is intended in trace norm and holds for any finite
$n$, $|E|$ and $\bar{E}$.

\subsection{Network stretching with entanglement cuts}

We may achieve a non-trivial simplification of previous Lemma~\ref{LemmaNET}
in such a way that we greatly reduce the number of resource states in the
decomposition of Alice and Bob's output state $\rho_{\mathbf{ab}}^{n}$. This
is possible using Alice-Bob entanglement cuts of the quantum network. These
types of cuts will enable us to include many resource states in Alice's and
Bob's LOs, while preserving the locality between the two end-points.

By definition, an Alice-Bob entanglement cut $C$ of the quantum network is a
bipartition $(\mathbf{A},\mathbf{B})$ of all the points $P$\ of the network
such that $\mathbf{a}\in\mathbf{A}$ and $\mathbf{b}\in\mathbf{B}$. Then, the
cut-set $\tilde{C}$ of $C$ is the set of edges with one end-point in each
subset of the bipartition, so that the removal of these edges disconnects the
network. Explicitly,
\begin{equation}
\tilde{C}=\{(\mathbf{x},\mathbf{y})\in E:\mathbf{x}\in\mathbf{A},\mathbf{y}%
\in\mathbf{B}\}.
\end{equation}
Note that the cut-set $\tilde{C}$\ identifies an ensemble of channels
$\{\mathcal{E}_{\mathbf{xy}}\}_{(\mathbf{x},\mathbf{y})\in\tilde{C}}$.
Similarly, we may define the following complementary sets
\begin{align}
\tilde{A}  &  =\{(\mathbf{x},\mathbf{y})\in E:\mathbf{x,y}\in\mathbf{A}\},\\
\tilde{B}  &  =\{(\mathbf{x},\mathbf{y})\in E:\mathbf{x,y}\in\mathbf{B}\},
\end{align}
so that $\tilde{A}\cup\tilde{B}\cup\tilde{C}=E$.

To simplify the stretching of the network, we then adopt the following
procedure. Given an arbitrary cut $C=(\mathbf{A},\mathbf{B})$, we extend Alice
and Bob to their corresponding partitions. This means that we consider
super-Alice with global register $\mathbf{A}$, and super-Bob with global
register $\mathbf{B}$. Then, all the resource states $\{\sigma_{\mathbf{xy}%
}\}_{(\mathbf{x},\mathbf{y})\in\tilde{A}}$ are included in the LOs of
super-Alice, and all those $\{\sigma_{\mathbf{xy}}\}_{(\mathbf{x}%
,\mathbf{y})\in\tilde{B}}$ are included in the LOs of super-Bob. Note that the
only resource states not absorbed in LOs are those in the cut-set
$\{\sigma_{\mathbf{xy}}\}_{(\mathbf{x},\mathbf{y})\in\tilde{C}}$. These states
are the only ones responsible for distributing entanglement between the
super-parties. The inclusion of all the other resource states into the global
LOCC $\bar{\Lambda}$ leads to another trace-preserving quantum operation
$\bar{\Lambda}_{\mathbf{AB}}$ which remains local with respect to $\mathbf{A}$
and $\mathbf{B}$. Thus, for any cut $C$, we may write the following output
state for super-Alice $\mathbf{A}$ and Bob $\mathbf{B}$ after $n$ uses of an
adaptive protocol%
\begin{equation}
\rho_{\mathbf{AB}}^{n}(C)=\bar{\Lambda}_{\mathbf{AB}}\left[  \underset
{(\mathbf{x},\mathbf{y})\in\tilde{C}}{%
{\textstyle\bigotimes}
}~\sigma_{\mathbf{xy}}^{\otimes n_{\mathbf{xy}}}\right]  .
\end{equation}

\begin{figure}[tbh]
\begin{center}
\vspace{-1.0cm} \includegraphics[width=0.48\textwidth]{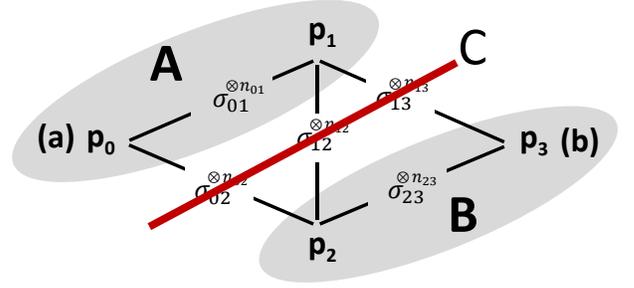}
\vspace{-1.7cm}
\end{center}
\caption{Network stretching with entanglement cuts. We show one of the
possible entanglement cuts $C$ of the diamond quantum network $\mathcal{N}%
^{\diamond}$. This cut creates super-Alice $\mathbf{A}=\{\mathbf{a}%
,\mathbf{p}_{1}\}$ and super-Bob $\mathbf{B}=\{\mathbf{b},\mathbf{p}_{2}\}$.
The resource states $\sigma_{01}^{\otimes n_{01}}$ are absorbed in the LOs of
$\mathbf{A}$, while the resource states $\sigma_{23}^{\otimes n_{23}}$ are
absorbed in the LOs of $\mathbf{B}$. The cut-set is composed by the set of
edges $\tilde{C}=\{(\mathbf{p}_{0},\mathbf{p}_{2}),(\mathbf{p}_{1}%
,\mathbf{p}_{2}),(\mathbf{p}_{1},\mathbf{p}_{3})\}$ with corresponding
resource states $\sigma_{02}^{\otimes n_{02}}$, $\sigma_{12}^{\otimes n_{12}}$
and $\sigma_{13}^{\otimes n_{13}}$. This subset of states can be used to
decompose the output state of Alice and Bob $\rho_{\mathbf{ab}}^{n}(C)$
according to Eq.~(\ref{cutEQ}).}%
\label{cut}%
\end{figure}

The next step is tracing out all registers but the original Alice's
$\mathbf{a}$ and Bob's $\mathbf{b}$. This operation preserves the locality
between $\mathbf{a}$ and $\mathbf{b}$. In other words, we may write the
following reduced output state for the two end-points%
\begin{align}
\rho_{\mathbf{ab}}^{n}(C)  &  =\mathrm{Tr}_{P\setminus\{\mathbf{a,b}\}}\left[
\rho_{\mathbf{AB}}^{n}(C)\right] \nonumber\\
&  =\bar{\Lambda}_{\mathbf{ab}}\left[  \underset{(\mathbf{x},\mathbf{y}%
)\in\tilde{C}}{%
{\textstyle\bigotimes}
}~\sigma_{\mathbf{xy}}^{\otimes n_{\mathbf{xy}}}\right]  ,
\end{align}
where $\bar{\Lambda}_{\mathbf{ab}}$ is a trace-preserving LOCC. All these
reasonings automatically transform Lemma~\ref{LemmaNET} into the following
improved Lemma. See also Fig.~\ref{cut} for an example.

\begin{lemma}
[Network stretching with cuts]\label{reduceCHOI}Consider a quantum network
$\mathcal{N}=(P,E)$ simulable with a resource representation $\sigma
(\mathcal{N})=\{\sigma_{\mathbf{xy}}\}_{(\mathbf{x},\mathbf{y})\in E}$. For a
teleportation-covariant network, $\sigma(\mathcal{N})$ is a
Choi-representation, i.e., $\sigma_{\mathbf{xy}}=\sigma_{\mathcal{E}%
_{\mathbf{xy}}}$. Then, consider $n$ uses of an adaptive protocol so that edge
$(\mathbf{x},\mathbf{y})\in E$ is used $n_{\mathbf{xy}}$ times. For any
entanglement cut $C$ and corresponding cut-set $\tilde{C}$, we may write Alice
and Bob's output state as%
\begin{equation}
\rho_{\mathbf{ab}}^{n}(C)=\bar{\Lambda}_{\mathbf{ab}}\left[  \underset
{(\mathbf{x},\mathbf{y})\in\tilde{C}}{%
{\textstyle\bigotimes}
}~\sigma_{\mathbf{xy}}^{\otimes n_{\mathbf{xy}}}\right]  , \label{cutEQ}%
\end{equation}
for a trace-preserving LOCC $\bar{\Lambda}_{\mathbf{ab}}$. In particular, we
have $n_{\mathbf{xy}}\leq n$ ($n_{\mathbf{xy}}=n$) for a sequential (flooding)
protocol. Formulations may be asymptotic for bosonic channels.
\end{lemma}

As stated in this improved lemma,\ the decomposition in Eq.~(\ref{cutEQ}) can
be extended to networks of bosonic channels with asymptotic simulations. We
can adapt the previous reasoning to find the cut-version of
Eq.~(\ref{totalASYMP}), i.e., the trace-norm limit%
\begin{equation}
\left\Vert \rho_{\mathbf{ab}}^{n}(C)-\bar{\Lambda}_{\mathbf{ab}}^{\mu}\left[
\underset{(\mathbf{x},\mathbf{y})\in\tilde{C}}{%
{\textstyle\bigotimes}
}~\sigma_{\mathbf{xy}}^{\mu\otimes n_{\mathbf{xy}}}\right]  \right\Vert
\overset{\mu}{\rightarrow}0, \label{cutASYpproof}%
\end{equation}
for suitable sequences of trace-preserving LOCC $\bar{\Lambda}_{\mathbf{ab}%
}^{\mu}$ and resource states $\sigma_{\mathbf{xy}}^{\mu}$ (with the result
holding for any $n$, number of edges $|E|$ and mean number of photons $\bar
{E}$).

With Lemma~\ref{reduceCHOI} in our hands, we have the necessary tool to derive
our single-letter upper bounds for the single- and multi-path capacities of an
arbitrary quantum network. This tool needs to be combined with a general weak
converse upper bound based on the REE. In the following section, we derive our
results for the case of single-path routing over the network. The results for
multipath routing will be given in Sec.~\ref{secMULTI}. In these sections, the
upper bounds will be combined with suitable lower bounds that are derived by
mixing point-to-point quantum protocols with classical routing strategies
(widest path and maximum flow of a network). 

\section{Results for single-path routing\label{secSINGLE}}

\subsection{Converse part (upper bound)}

In order to write a single-letter upper bound for the single-path capacity of
the quantum network, we need to introduce the notion of REE flowing through a
cut under some simulation. Consider an arbitrary quantum network
$\mathcal{N}=(P,E)$ with a resource representation $\sigma(\mathcal{N}%
)=\{\sigma_{\mathbf{xy}}\}_{(\mathbf{x},\mathbf{y})\in E}$. Then, consider an
arbitrary entanglement cut $C$ with corresponding cut-set $\tilde{C}$. Under
the simulation considered, we define the single-edge flow of REE through the
cut as the following quantity%
\begin{equation}
E_{\mathrm{R}}(C):=\max_{(\mathbf{x},\mathbf{y})\in\tilde{C}}E_{\mathrm{R}%
}(\sigma_{\mathbf{xy}})~. \label{ERC}%
\end{equation}
By minimizing $E_{\mathrm{R}}(C)$ over all possible entanglement cuts of the
network, we build our upper bound for the single-path capacity. In fact, we
may prove the following.

\begin{theorem}
[Converse for single-path capacity]\label{theoUBnet}Consider an arbitrary
quantum network $\mathcal{N}=(P,E)$ with some resource representation
$\sigma(\mathcal{N})=\{\sigma_{\mathbf{xy}}\}_{(\mathbf{x},\mathbf{y})\in E}$.
In particular, $\sigma(\mathcal{N})$ may be a Choi-representation for a
teleportation-covariant network. Then, the single-path capacity of
$\mathcal{N}$ must satisfy the single-letter bound%
\begin{equation}
\mathcal{C}(\mathcal{N})\leq\min_{C}E_{\mathrm{R}}(C)~, \label{UBnetmain}%
\end{equation}
where the single-edge flow of REE in Eq.~(\ref{ERC}) is minimized across all
cuts of the network. Formulations may be asymptotic for networks of bosonic channels.
\end{theorem}

\textbf{Proof.}~~According to Eq.~(\ref{necCAPdef}) the single-path capacity
is defined by the following optimization of the asymptotic rate over the
sequential protocols
\begin{equation}
\mathcal{C}(\mathcal{N}):=\sup_{\mathcal{P}_{\text{\textrm{seq}}}}\lim
_{n}R_{n}. \label{ccc1}%
\end{equation}
We can directly extend the general weak-converse in Theorem~\ref{generalWEAK}
from channels to networks. This means to replace the supremum over
point-to-point protocols $\mathcal{P}$ with a supremum over network sequential
protocols $\mathcal{P}_{\text{\textrm{seq}}}$, i.e.,%
\begin{equation}
\mathcal{C}(\mathcal{N})\leq E_{\mathrm{R}}^{\bigstar}(\mathcal{N}%
):=\sup_{\mathcal{P}_{\text{\textrm{seq}}}}\underset{n}{\lim}\frac
{E_{\mathrm{R}}(\rho_{\mathbf{ab}}^{n})}{n}. \label{ccc2}%
\end{equation}
The other elements of this bound are unchanged because they are exclusively
based on the fact that the output state $\rho_{\mathbf{ab}}^{n}$ is (by
definition) epsilon close to a target state.

According to previous Lemma~\ref{reduceCHOI}, for any sequential protocol
$\mathcal{P}_{\text{\textrm{seq}}}$ and entanglement cut $C$ of the network,
we may write Eq.~(\ref{cutEQ}). Computing the REE on this decomposition and
exploiting basic properties (monotonicity of REE under $\bar{\Lambda
}_{\mathbf{ab}}$ and subadditivity over tensor products), we derive the
following inequality%
\begin{equation}
E_{\mathrm{R}}\left[  \rho_{\mathbf{ab}}^{n}(C)\right]  \leq\sum
_{(\mathbf{x},\mathbf{y})\in\tilde{C}}n_{\mathbf{xy}}E_{\mathrm{R}}%
(\sigma_{\mathbf{xy}}), \label{ccc3}%
\end{equation}
where $n_{\mathbf{xy}}=np_{\mathbf{xy}}$\ and $p_{\mathbf{xy}}$\ being the
probability of using edge $(\mathbf{x},\mathbf{y})$\ according to protocol
$\mathcal{P}_{\text{\textrm{seq}}}$.\ By maximizing over the convex
combination, we get rid of $p_{\mathbf{xy}}$ and write%
\begin{equation}
E_{\mathrm{R}}\left[  \rho_{\mathbf{ab}}^{n}(C)\right]  \leq n\max
_{(\mathbf{x},\mathbf{y})\in\tilde{C}}E_{\mathrm{R}}(\sigma_{\mathbf{xy}%
})=nE_{\mathrm{R}}(C)~. \label{mmllll}%
\end{equation}
By using Eq.~(\ref{mmllll}) in Eq.~(\ref{ccc2}), we see that both the
optimization over $\mathcal{P}_{\text{\textrm{seq}}}$ and the limit over $n$
disappear, and we are left with the bound%
\begin{equation}
\mathcal{C}(\mathcal{N})\leq E_{\mathrm{R}}(C),~~\text{for any~}C\text{.}
\label{preRESS}%
\end{equation}
By minimizing over all cuts, we therefore prove Eq.~(\ref{UBnetmain}).

Note that, from Eq.~(\ref{ccc3}) we may also derive
\begin{equation}
\mathcal{C}(\mathcal{N})\leq\bar{E}_{\mathrm{R}}(C):=\sum_{(\mathbf{x}%
,\mathbf{y})\in\tilde{C}}\bar{p}_{\mathbf{xy}}E_{\mathrm{R}}(\sigma
_{\mathbf{xy}}), \label{boundAVE}%
\end{equation}
where $\bar{p}_{\mathbf{xy}}$ is the optimal use of edge $(\mathbf{x}%
,\mathbf{y})$ over all possible $\mathcal{P}_{\text{\textrm{seq}}}$. Here
$\bar{E}_{\mathrm{R}}(C)$ represents the \textit{average flow} of REE through
$C$ under the chosen simulation and optimized over $\mathcal{P}%
_{\text{\textrm{seq}}}$. By minimizing over all cuts, we get%
\begin{equation}
\mathcal{C}(\mathcal{N})\leq\min_{C}\bar{E}_{\mathrm{R}}(C).
\end{equation}
This may be tighter than Eq.~(\ref{UBnetmain}) but difficult to compute due to
residual optimization over the protocols.

Finally, note that Eq.~(\ref{UBnetmain}) can be extended to considering
asymptotic simulations, following the same ideas in the proof of
Theorem~\ref{singleLETTtheorem}. Let us compute the REE on the asymptotic
state $\rho_{\mathbf{ab}}^{n}(C)$ of Eq.~(\ref{cutASYpproof}). We may write
\begin{align}
&  E_{\mathrm{R}}[\rho_{\mathbf{ab}}^{n}(C)]\nonumber\\
&  =\inf_{\gamma\in\mathrm{SEP}}S[\rho_{\mathbf{ab}}^{n}(C)||\gamma
]\nonumber\\
&  \overset{(1)}{\leq}\inf_{\gamma^{\mu}}S\left.  \left\{  \lim_{\mu}%
\bar{\Lambda}_{\mathbf{ab}}^{\mu}\left[  \underset{(\mathbf{x},\mathbf{y}%
)\in\tilde{C}}{%
{\textstyle\bigotimes}
}~\sigma_{\mathbf{xy}}^{\mu\otimes n_{\mathbf{xy}}}\right]  \right\Vert
~\lim_{\mu}\gamma^{\mu}\right\} \nonumber\\
&  \overset{(2)}{\leq}\inf_{\gamma^{\mu}}\underset{\mu\rightarrow+\infty}%
{\lim\inf}S\left.  \left\{  \bar{\Lambda}_{\mathbf{ab}}^{\mu}\left[
\underset{(\mathbf{x},\mathbf{y})\in\tilde{C}}{%
{\textstyle\bigotimes}
}~\sigma_{\mathbf{xy}}^{\mu\otimes n_{\mathbf{xy}}}\right]  \right\Vert
~\gamma^{\mu}\right\} \nonumber\\
&  \overset{(3)}{\leq}\inf_{\gamma^{\mu}}\underset{\mu\rightarrow+\infty}%
{\lim\inf}S\left.  \left\{  \bar{\Lambda}_{\mathbf{ab}}^{\mu}\left[
\underset{(\mathbf{x},\mathbf{y})\in\tilde{C}}{%
{\textstyle\bigotimes}
}~\sigma_{\mathbf{xy}}^{\mu\otimes n_{\mathbf{xy}}}\right]  \right\Vert
~\bar{\Lambda}_{\mathbf{ab}}^{\mu}(\gamma^{\mu})\right\} \nonumber\\
&  \overset{(4)}{\leq}\inf_{\gamma^{\mu}}\underset{\mu\rightarrow+\infty}%
{\lim\inf}S\left.  \left[  \underset{(\mathbf{x},\mathbf{y})\in\tilde{C}}{%
{\textstyle\bigotimes}
}~\sigma_{\mathbf{xy}}^{\mu\otimes n_{\mathbf{xy}}}\right\Vert ~\gamma^{\mu
}\right] \nonumber\\
&  \overset{(5)}{=}E_{\mathrm{R}}\left[  \underset{(\mathbf{x},\mathbf{y}%
)\in\tilde{C}}{%
{\textstyle\bigotimes}
}~\sigma_{\mathbf{xy}}^{\otimes n_{\mathbf{xy}}}\right] \nonumber\\
&  \overset{(6)}{\leq}\sum_{(\mathbf{x},\mathbf{y})\in\tilde{C}}%
n_{\mathbf{xy}}E_{\mathrm{R}}(\sigma_{\mathbf{xy}}),
\end{align}
where: (1)$~\gamma^{\mu}$ is a generic sequence of separable states converging
in trace norm, i.e., such that there is a separable state $\gamma:=\lim_{\mu
}\gamma^{\mu}$ so that $\Vert\gamma-\gamma^{\mu}\Vert\overset{\mu}%
{\rightarrow}0$; (2)~we use the lower semi-continuity of the relative
entropy~\cite{HolevoBOOK}; (3)~we use that $\bar{\Lambda}_{\mathbf{ab}}^{\mu
}(\gamma^{\mu})$ are specific types of converging separable sequences within
the set of all such sequences; (4)~we use the monotonicity of the relative
entropy under trace-preserving LOCCs; (5)~we use the definition of REE for
asymptotic states $\sigma_{\mathbf{xy}}:=\lim_{\mu}\sigma_{\mathbf{xy}}^{\mu}%
$; (6)~we use the subadditivity over tensor products.

Therefore, we have again Eq.~(\ref{ccc3}) but where the REE is written as in
the weaker formulation for asymptotic states given in Eq.~(\ref{REE_weaker}).
The next steps of the proof are exactly as before, and they lead to
Eq.~(\ref{preRESS}).~$\blacksquare$

\subsection{Direct part (achievable rate)}

In this section, we derive an achievable asymptotic rate for the end-to-end
quantum/private communication via single-path routing. This rate will provide
a lower bound to the single-path capacity of an arbitrary quantum network,
i.e., with arbitrary topology and arbitrary quantum channels. The non-trivial
result is that the achievable rate can be written in terms of a capacity
minimized over the entanglement cuts in the network. This step will allow us
to exactly establish the single-path capacity of distillable networks in the
next subsection.

Consider an arbitrary quantum network $\mathcal{N}=(P,E)$ where edge
$(\mathbf{x},\mathbf{y})\in E$ is connected by channel $\mathcal{E}%
_{\mathbf{xy}}$ with associated two-way capacity $\mathcal{C}_{\mathbf{xy}%
}=\mathcal{C}(\mathcal{E}_{\mathbf{xy}})$. Given an arbitrary entanglement cut
$C$ of the network, we define its single-edge capacity as the maximum number
of target bits distributed by a single edge across the cut, i.e.,%
\begin{equation}
\mathcal{C}(C):=\max_{(\mathbf{x},\mathbf{y})\in\tilde{C}}\mathcal{C}%
_{\mathbf{xy}}~. \label{cutEQUIV}%
\end{equation}
A minimum cut $C_{\text{min}}$ is such that
\begin{equation}
\mathcal{C}(C_{\text{min}})=\min_{C}\mathcal{C}(C). \label{minCUUTT}%
\end{equation}
Then, given a route $\omega\in\Omega$ with an associated chain of channels
$\{\mathcal{E}_{i}^{\omega}\}$, we define its capacity as the minimum capacity
among its channels, i.e.,%
\begin{equation}
\mathcal{C}(\omega):=\min_{i}\mathcal{C}(\mathcal{E}_{i}^{\omega})~.
\label{genRR}%
\end{equation}
An optimal route $\omega_{\ast}$\ is such that%
\begin{equation}
\mathcal{C}(\omega_{\ast})=\max_{\omega\in\Omega}\mathcal{C}(\omega)~.
\end{equation}

It is clear that $\mathcal{C}(\omega_{\ast})$ is an achievable end-to-end
rate. In fact, consider independent point-to-point protocols between pairs of
consecutive points along route $\omega_{\ast}$. An optimal adaptive protocol
between points $\mathbf{r}_{i}^{\omega_{\ast}}$ and $\mathbf{r}_{i+1}%
^{\omega_{\ast}}$ (connected by $\mathcal{E}_{i}^{\omega_{\ast}}$) achieves
the capacity value $\mathcal{C}(\mathcal{E}_{i}^{\omega_{\ast}})$. Then, by
composing all outputs via a network LOCCs (e.g., swapping the distilled states
or relaying the secret keys via one-time pad sessions), Alice and Bob obtain
an achievable rate of $\min_{i}\mathcal{C}(\mathcal{E}_{i}^{\omega_{\ast}%
})=\mathcal{C}(\omega_{\ast})$.

Thus, we may write the lower bound $\mathcal{C}(\mathcal{N})\geq
\mathcal{C}(\omega_{\ast})=\max_{\omega}\mathcal{C}(\omega)$. The crucial
observation is that this bound is also equal to the minimization in
Eq.~(\ref{minCUUTT}) over all entanglement cuts. In fact, we may prove the following.

\begin{theorem}
[Lower bound]\label{LBtheor}Consider an arbitrary quantum network
$\mathcal{N}=(P,E)$ where two end-points are connected by an ensemble of
routes $\Omega=\{\omega\}$ and may be disconnected by an entanglement cut $C$.
The single-path capacity of the network satisfies%
\begin{equation}
\mathcal{C}(\mathcal{N})\geq\max_{\omega\in\Omega}\mathcal{C}(\omega)=\min
_{C}\mathcal{C}(C).
\end{equation}
Thus, the capacity $\mathcal{C}(\omega_{\ast})$ of an optimal route
$\omega_{\ast}$ not only is an achievable rate but it is also equal to the
single-edge capacity $\mathcal{C}(C_{\text{min}})$ of a minimum cut
$C_{\text{min}}$. Furthermore, the optimal route $\omega_{\ast}$ is a simple
path within a maximum spanning tree of the network.
\end{theorem}

\textbf{Proof.}~It is easy to show the inequality $\mathcal{C}(\omega_{\ast
})\geq\mathcal{C}(C_{\text{min}})$. In fact, an edge $(\mathbf{\tilde{x}%
},\mathbf{\tilde{y}})$ of the optimal route $\omega_{\ast}$ must belong to the
cut-set $\tilde{C}_{\text{min}}$. Thus, the capacity of that edge must
simultaneously satisfy $\mathcal{C}_{\mathbf{\tilde{x}\tilde{y}}}%
\geq\mathcal{C}(\omega_{\ast})$ and $\mathcal{C}_{\mathbf{\tilde{x}\tilde{y}}%
}\leq\mathcal{C}(C_{\text{min}})$. In order to show the opposite inequality
$\mathcal{C}(\omega_{\ast})\leq\mathcal{C}(C_{\text{min}})$, we need to
exploit some basic results from graph theory. Consider the maximum spanning
tree of the connected undirected graph $(P,E)$. This is a subgraph
$\mathcal{T}=(P,E_{\text{tree}})$ which connects all the points in such a way
that the sum of the capacities associated with each edge $(\mathbf{x}%
,\mathbf{y})\in E_{\text{tree}}$ is the maximum. In other words, it maximizes
the following quantity%
\begin{equation}
\mathcal{C}(\mathcal{T}):=%
{\textstyle\sum\nolimits_{(\mathbf{x},\mathbf{y})\in E_{\text{tree}}}}
\mathcal{C}_{\mathbf{xy}}~.
\end{equation}

Note that the optimal route $\omega_{\ast}$ between Alice and Bob is the
unique path between Alice and Bob within this tree~\cite{tree}. Let us call
$e(\omega_{\ast})$ the critical edge in $\omega_{\ast}$, i.e., that specific
edge which realizes the minimization
\begin{equation}
\mathcal{C}_{e(\omega_{\ast})}=\mathcal{C}(\omega_{\ast})=\min_{i}%
\mathcal{C}(\mathcal{E}_{i}^{\omega_{\ast}})~.
\end{equation}
Since this edge is part of a spanning tree, there is always an Alice-Bob cut
$C_{\ast}$ of the network which crosses $e(\omega_{\ast})$ and no other edges
of the spanning tree. In fact, this condition would fail only if there was a
cycle in the tree, which is not possible by definition. \begin{figure}[ptbh]
\vspace{-2.7cm}
\par
\begin{center}
\includegraphics[width=0.6\textwidth] {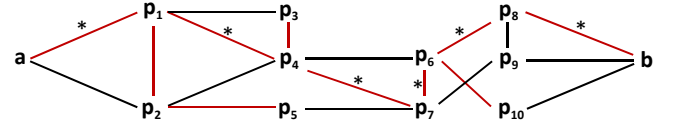}
\end{center}
\par
\vspace{-3.1cm}\caption{Example of a network and its maximum spanning tree
(red edges). The optimal route $\omega_{\ast}$ between Alice and Bob is a
unique path within this tree (highlighted by the asterisks). The critical edge
$e(\omega_{\ast})$ is the one maximizing the capacity, i.e., realizing the
condition $\mathcal{C}_{e(\omega_{\ast})}=\mathcal{C}(\omega_{\ast})$.
Wherever the critical edge might be along the optimal route, we can always
make an Alice-Bob entanglement cut $C_{\ast}$ which crosses that specific edge
and no other edge of the spanning tree. This property leads to $\mathcal{C}%
(C_{\ast})=\mathcal{C}(\omega_{\ast})$.}%
\label{cutP}%
\end{figure}

Then, we must also have that $e(\omega_{\ast})$ is the optimal edge in the
cut-set $\tilde{C}_{\ast}$, i.e., $\mathcal{C}_{e(\omega_{\ast})}%
=\mathcal{C}(C_{\ast})$. By absurd, assume this is not the case. This implies
that there is another edge $e^{\prime}\in\tilde{C}_{\ast}$, not belonging to
$\mathcal{T}$, such that $\mathcal{C}_{e^{\prime}}=\mathcal{C}(C_{\ast})$. For
the cut property of the maximum spanning trees~\cite{Dijkstra}, we have that
an edge in $C_{\ast}$ with maximum capacity must belong to all the maximum
spanning trees of the network. Therefore $e^{\prime}$ must belong to
$\mathcal{T}$ which leads to a contradiction. In conclusion, we have found an
Alice-Bob cut $C_{\ast}$ which realizes the condition $\mathcal{C}(C_{\ast
})=\mathcal{C}(\omega_{\ast})$. For an example see Fig.~\ref{cutP}%
.~$\blacksquare$

Note that the previous result applies not only to quantum networks but to any
graphical weighted network. It is sufficient to replace the capacity of the
edge with a generic weight. In fact, Theorem~\ref{LBtheor} can be restated as
follows, which represents a \textquotedblleft single-flow\textquotedblright%
\ formulation of the max-flow min-cut
theorem~\cite{Harris,Ford,ShannonFLOW,netflow}.

\begin{proposition}
[Cut property of the widest path]\label{widePROPO}Consider a network described
by an undirected graph $\mathcal{N}=(P,E)$, whose edge $e\in E$\ has weight
$W(e)$. Denote by $\Omega=\{\omega\}$\ the ensemble of undirected paths
between the end-points, Alice and Bob. Define the weight of a path
$\omega=\{e_{i}\}$ as $W(\omega)=\min_{i}W(e_{i})$, and the weight of an
Alice-Bob cut $C$ as $W(C)=\max_{e\in\tilde{C}}W(e)$. The weight of the widest
path is equal to that of the minimum cut%
\begin{align}
W(\omega_{\text{wide}})  &  :=\max_{\omega}W(\omega)\nonumber\\
&  =\min_{C}W(C):=W(C_{\text{min}}).
\end{align}
\
\end{proposition}

Finding the optimal route $\omega_{\ast}$ in a quantum network
(Theorem~\ref{LBtheor}) is equivalent to finding the widest path
$\omega_{\text{wide}}$\ in a weighted network (Proposition~\ref{widePROPO}),
i.e.,\ solving the well-known widest path problem. Using a modified Dijkstra's
shortest path algorithm~\cite{MITp}, the solution is found in time
$O(\left\vert E\right\vert \log_{2}\left\vert P\right\vert )$. In practical
cases, this algorithm can be optimized and its asymptotic performance becomes
$O(\left\vert E\right\vert +\left\vert P\right\vert \log_{2}\left\vert
P\right\vert )$~\cite{Fred}. Another possibility is using an algorithm for
finding a maximum spanning tree of the network, such as the Kruskal's
algorithm~\cite{Kruskal,MITp}. The latter has the asymptotic complexity
$O(\left\vert E\right\vert \log_{2}\left\vert P\right\vert )$ for building the
tree. This step is then followed by the search of the route within the tree
which takes linear time $O(\left\vert P\right\vert )$~\cite{tree}.

\subsection{Formulas for teleportation-covariant and distillable
networks\label{SecDISTILLABLENET}}

The results of Theorems~\ref{theoUBnet} and~\ref{LBtheor} can be specified for
quantum networks which are connected by teleportation-covariant channels.
Given a teleportation-covariant network $\mathcal{N}=(P,E)$ whose
teleportation simulation has an associated Choi-representation $\sigma
(\mathcal{N})=\{\sigma_{\mathcal{E}_{\mathbf{xy}}}\}_{(\mathbf{x}%
,\mathbf{y})\in E}$, we may write the following for the single-path capacity%
\begin{equation}
\min_{C}\mathcal{C}(C)\leq\mathcal{C}(\mathcal{N})\leq\min_{C}E_{\mathrm{R}%
}(C)~, \label{sandmmm}%
\end{equation}
with $\mathcal{C}(C)$ being defined in Eq.~(\ref{cutEQUIV}), and%
\begin{equation}
E_{\mathrm{R}}(C)=\max_{(\mathbf{x},\mathbf{y})\in\tilde{C}}E_{\mathrm{R}%
}(\sigma_{\mathcal{E}_{\mathbf{xy}}})~.
\end{equation}
The latter may have an asymptotic formulation for networks of bosonic
channels, with the REE\ taking the form as in Eq.~(\ref{REE_weaker}) over
$\sigma_{\mathcal{E}_{\mathbf{xy}}}:=\lim_{\mu}\sigma_{\mathcal{E}%
_{\mathbf{xy}}}^{\mu}$, where $\sigma_{\mathcal{E}_{\mathbf{xy}}}^{\mu}$ is a
sequence of Choi approximating states with finite energy.

In particular, consider a network connected by distillable channels. This
means that for any edge $(\mathbf{x},\mathbf{y})\in E$, we may write (exactly
or asymptotically)
\begin{equation}
\mathcal{C}_{\mathbf{xy}}:=\mathcal{C}(\mathcal{E}_{\mathbf{xy}}%
)=E_{\mathrm{R}}(\sigma_{\mathcal{E}_{\mathbf{xy}}})=D_{1}(\sigma
_{\mathcal{E}_{\mathbf{xy}}}). \label{dueEELL}%
\end{equation}
By imposing this condition in Eq.~(\ref{sandmmm}), we find that upper and
lower bounds coincide. We have therefore the following result which
establishes the single-path capacity $\mathcal{C}(\mathcal{N})$\ of a
distillable network and fully extends the widest path problem~\cite{Pollack}
to quantum communications.

\begin{corollary}
[Single-path capacities]\label{coroNETseq}Consider a distillable network
$\mathcal{N}=(P,E)$, where two end-points are connected by an ensemble\ of
routes $\Omega=\{\omega\}$ and may be disconnected by an entanglement cut $C$.
An arbitrary edge $(\mathbf{x},\mathbf{y})\in E$\ is connected by a
distillable channel $\mathcal{E}_{\mathbf{xy}}$\ with two-way capacity
$\mathcal{C}_{\mathbf{xy}}$\ and Choi matrix $\sigma_{\mathcal{E}%
_{\mathbf{xy}}}$. Then, the single-path capacity of the network is equal to
\begin{equation}
\mathcal{C}(\mathcal{N})=\min_{C}E_{\mathrm{R}}(C)=\min_{C}\max_{(\mathbf{x}%
,\mathbf{y})\in\tilde{C}}E_{\mathrm{R}}(\sigma_{\mathcal{E}_{\mathbf{xy}}}),
\label{StretcCOR}%
\end{equation}
with an implicit asymptotic formulation for bosonic channels. Equivalently,
$\mathcal{C}(\mathcal{N})$ is also equal to the minimum (single-edge) capacity
of the entanglement cuts and the maximum capacity of the routes, i.e.,%
\begin{equation}
\mathcal{C}(\mathcal{N})=\min_{C}\mathcal{C}(C)=\max_{\omega}\mathcal{C}%
(\omega)~.
\end{equation}
The optimal end-to-end route $\omega_{\ast}$ achieving the capacity can be
found in time $O(\left\vert E\right\vert \log_{2}\left\vert P\right\vert )$,
where $\left\vert E\right\vert $ is the number of edges and $\left\vert
P\right\vert $ is the number of points. Over this route, a capacity-achieving
protocol is based on one-way entanglement distillation sessions between
consecutive points, followed by entanglement swapping.
\end{corollary}

The proof of this corollary is a direct application of the previous
reasonings. We see that it first reduces the routing problem to a classical
optimization problem, i.e., finding the widest path. Then, over this optimal
route, the single-path capacity is achieved by a non-adaptive protocol based
on one-way CCs. In fact, we have that any two consecutive points
$\mathbf{r}_{i}$ and $\mathbf{r}_{i+1}$ along $\omega_{\ast}$ may distill
ebits at the rate of $D_{1}(\sigma_{\mathcal{E}_{i}^{\omega_{\ast}}})$, where
$\mathcal{E}_{i}^{\omega_{\ast}}$ is the connecting channel. Then, sessions of
entanglement swapping (also based on one-way CCs), distribute ebits at the
end-points with a rate of at least $\min_{i}D_{1}(\sigma_{\mathcal{E}%
_{i}^{\omega_{\ast}}})$. Due to Eq.~(\ref{dueEELL}), this rate is equal to
$\min_{i}\mathcal{C}(\mathcal{E}_{i}^{\omega_{\ast}})=\mathcal{C}(\omega
_{\ast})$, which corresponds to the capacity $\mathcal{C}(\mathcal{N})$.

\subsection{Single-path capacities of fundamental networks}

Let us specify the result of Corollary~\ref{coroNETseq}\ to fundamental
scenarios such as bosonic networks subject to pure-loss or quantum-limited
amplification, or spin networks affected by dephasing or erasure. These are in
fact all distillable networks. We find extremely simple formulas for their
single-path capacities, setting their ultimate limit for quantum
communication, entanglement distribution, key generation and private
communication under single-path routing.

Start with a network connected by lossy channels $\mathcal{N}_{\text{loss}}$,
which well describes both free-space or fiber-based optical communications.
According to Corollary~\ref{coroNETseq}, we may compute its capacity
$\mathcal{C}(\mathcal{N}_{\text{loss}})$ by minimizing over the cuts or
maximizing over the routes. Generic edge $(\mathbf{x},\mathbf{y})\in E$ has an
associated lossy channel with transmissivity $\eta_{\mathbf{xy}}$ and capacity
$\mathcal{C}_{\mathbf{xy}}=-\log_{2}(1-\eta_{\mathbf{xy}})$. Therefore, an
entanglement cut has single-edge capacity
\begin{align}
\mathcal{C}(C)  &  =\max_{(\mathbf{x},\mathbf{y})\in\tilde{C}}\left[
-\log_{2}(1-\eta_{\mathbf{xy}})\right]  =-\log_{2}(1-\eta_{C}),\nonumber\\
\eta_{C}  &  :=\max_{(\mathbf{x},\mathbf{y})\in\tilde{C}}\eta_{\mathbf{xy}},
\end{align}
where $\eta_{C}$ may be identified as the (single-edge) transmissivity of the
cut. By minimizing over the cuts, we may write the single-path capacity of the
lossy network as
\begin{equation}
\mathcal{C}(\mathcal{N}_{\text{loss}})=-\log_{2}(1-\tilde{\eta}_{C}%
),~\tilde{\eta}_{C}:=\min_{C}\eta_{C},
\end{equation}
where $\tilde{\eta}_{C}$ is the minimum transmissivity of the cuts.

Consider now a generic end-to-end route $\omega$ along the lossy network. This
route is associated with a sequence of lossy channels with transmissivities
$\{\eta_{i}^{\omega}\}$. We then compute the route capacity as%
\begin{align}
\mathcal{C}_{\omega}  &  =\min_{i}\left[  -\log_{2}(1-\eta_{i}^{\omega
})\right]  =-\log_{2}(1-\eta_{\omega}),\nonumber\\
\eta_{\omega}  &  :=\min_{i}\eta_{i}^{\omega},
\end{align}
where $\eta_{\omega}$ is the route transmissivity. By maximizing over the
routes, we may equivalently write the single-path capacity of the lossy
network as%
\begin{equation}
\mathcal{C}(\mathcal{N}_{\text{loss}})=-\log_{2}(1-\tilde{\eta}),~\tilde{\eta
}:=\max_{\omega}\eta_{\omega},
\end{equation}
where $\tilde{\eta}$ is the maximum transmissivity of the routes.

Similar conclusions can be derived for bosonic networks which are composed of
other distillable Gaussian channels, such as multiband lossy channels,
quantum-limited amplifiers or even hybrid combinations. In particular,
consider a network of quantum-limited amplifiers $\mathcal{N}_{\text{amp}}$,
where the generic edge $(\mathbf{x},\mathbf{y})\in E$ has gain $g_{\mathbf{xy}%
}$ with capacity $\mathcal{C}_{\mathbf{xy}}=-\log_{2}(1-g_{\mathbf{xy}}^{-1}%
)$, and the generic end-to-end route $\omega$ is associated with a sequence of
gains $\{g_{i}^{\omega}\}$. We can repeat the previous steps of the lossy
network but setting $g^{-1}=\eta$, so that $\max\eta=\min g$. Thus, for an
entanglement cut $C$, we may write
\begin{align}
\mathcal{C}(C)  &  =\max_{(\mathbf{x},\mathbf{y})\in\tilde{C}}\left[
-\log_{2}(1-g_{\mathbf{xy}}^{-1})\right]  =-\log_{2}(1-g_{C}^{-1}),\nonumber\\
g_{C}  &  :=\min_{(\mathbf{x},\mathbf{y})\in\tilde{C}}g_{\mathbf{xy}}~.
\end{align}
For a route $\omega$, we have the capacity%
\begin{align}
\mathcal{C}_{\omega}  &  =\min_{i}\{-\log_{2}[1-(g_{i}^{\omega})^{-1}%
]\}=-\log_{2}(1-g_{\omega}^{-1}),\nonumber\\
g_{\omega}  &  :=\max_{i}g_{i}^{\omega}~.
\end{align}
By minimizing over the cuts or maximizing over the routes, we derive the two
equivalent formulas%
\begin{equation}
\mathcal{C}(\mathcal{N}_{\text{amp}})=-\log_{2}(1-\tilde{g}_{C}^{-1}%
)=-\log_{2}(1-\tilde{g}^{-1}),
\end{equation}
where $\tilde{g}_{C}:=\max_{C}g_{C}$ and $\tilde{g}:=\min_{\omega}g_{\omega}$.

We can also compute the single-path capacities of DV networks where links
between qudits are affected by dephasing or erasure or a mix of the two
errors. For simplicity, consider the case of qubits, such as spin $1/2$ or
polarized photons. In a qubit network with dephasing channels $\mathcal{N}%
_{\text{deph}}$, the generic edge $(\mathbf{x},\mathbf{y})\in E$ has a
dephasing probability $p_{\mathbf{xy}}\leq1/2$ and capacity $\mathcal{C}%
_{\mathbf{xy}}=1-H_{2}(p_{\mathbf{xy}})$. The generic end-to-end route
$\omega$ is associated with a sequence of such dephasing probabilities
$\{p_{i}^{\omega}\}$. For an entanglement cut $C$, we have
\begin{align}
\mathcal{C}(C)  &  =\max_{(\mathbf{x},\mathbf{y})\in\tilde{C}}\left[
1-H_{2}(p_{\mathbf{xy}})\right]  =1-H_{2}(p_{C}),\nonumber\\
p_{C}  &  :=\min_{(\mathbf{x},\mathbf{y})\in\tilde{C}}p_{\mathbf{xy}}.
\label{pcDEP}%
\end{align}
For a generic route $\omega$, we may write%
\begin{align}
\mathcal{C}_{\omega}  &  =\min_{i}\left[  1-H_{2}(p_{i}^{\omega})\right]
=1-H_{2}(p_{\omega}),\nonumber\\
p_{\omega}  &  :=\max_{i}p_{i}^{\omega}~. \label{pomegaDEP}%
\end{align}
By minimizing over the cuts or maximizing over the routes, we then derive the
single-path capacity
\begin{equation}
\mathcal{C}(\mathcal{N}_{\text{deph}})=1-H_{2}(\tilde{p}_{C})=1-H_{2}%
(\tilde{p}),
\end{equation}
where we have set
\begin{equation}
\tilde{p}_{C}:=\max_{C}p_{C},~~\tilde{p}:=\min_{\omega}p_{\omega}.
\label{pcset}%
\end{equation}

Finally, for a qubit network affected by erasures $\mathcal{N}_{\text{erase}}
$ we have that edge $(\mathbf{x},\mathbf{y})\in E$ is associated with an
erasure channel with probability $p_{\mathbf{xy}}$ and corresponding capacity
$\mathcal{C}_{\mathbf{xy}}=1-p_{\mathbf{xy}}$. As a result, we may repeat all
the previous derivation for the dephasing network $\mathcal{N}_{\text{deph}}$
up to replacing $H_{2}(p)$ with $p$. For a cut and a route, we have%
\begin{equation}
\mathcal{C}(C)=1-p_{C},~~\mathcal{C}_{\omega}=1-p_{\omega},
\end{equation}
where $p_{C}$ and $p_{\omega}$ are defined as in Eqs.~(\ref{pcDEP})
and~(\ref{pomegaDEP}). Thus, the single-path capacity of the erasure network
simply reads%
\begin{equation}
\mathcal{C}(\mathcal{N}_{\text{erase}})=1-\tilde{p}_{C}=1-\tilde{p},~.
\end{equation}
where $\tilde{p}_{C}$ and $\tilde{p}$ are defined as in Eq.~(\ref{pcset}).

\section{Results for multipath routing\label{secMULTI}}

\subsection{Converse part (upper bound)}

In order to write a single-letter upper bound for the multipath capacity of a
quantum network, we need to introduce the concept of \textit{multi-edge} flow
of REE through a cut, under some simulation of the network. Consider an
arbitrary quantum network $\mathcal{N}=(P,E)$ whose simulation has an
associate resource representation $\sigma(\mathcal{N})=\{\sigma_{\mathbf{xy}%
}\}_{(\mathbf{x},\mathbf{y})\in E}$. Then, consider an arbitrary entanglement
cut $C$ with corresponding cut-set $\tilde{C}$. Under the simulation
considered, we define the multi-edge flow of REE through the cut as the
following quantity%
\begin{equation}
E_{\mathrm{R}}^{\text{m}}(C):=\sum_{(\mathbf{x},\mathbf{y})\in\tilde{C}%
}E_{\mathrm{R}}(\sigma_{\mathbf{xy}})~. \label{multiooo}%
\end{equation}
By minimizing $E_{\mathrm{R}}^{\text{m}}(C)$ over all possible entanglement
cuts of the network, we build our upper bound for the multipath capacity. In
fact, we may prove the following.

\begin{theorem}
[Converse for multi-path capacity]\label{TheoMP1}Consider an arbitrary quantum
network $\mathcal{N}=(P,E)$ with some resource representation $\sigma
(\mathcal{N})=\{\sigma_{\mathbf{xy}}\}_{(\mathbf{x},\mathbf{y})\in E}$. In
particular, $\sigma(\mathcal{N})$ may be a Choi-representation for a
teleportation-covariant network. Then, the multipath capacity of $\mathcal{N}$
must satisfy the single-letter bound%
\begin{equation}
\mathcal{C}^{\text{m}}(\mathcal{N})\leq\min_{C}E_{\mathrm{R}}^{\text{m}}(C)~,
\label{theommll}%
\end{equation}
where the multi-edge flow of REE in Eq.~(\ref{multiooo}) is minimized across
all cuts of the network. Formulations may be asymptotic for networks of
bosonic channels.
\end{theorem}

\textbf{Proof.}~~According to Eq.~(\ref{bbnetCAPdef}) the multipath capacity
is defined by the following optimization of the asymptotic rate over the
flooding protocols
\begin{equation}
\mathcal{C}^{\text{m}}(\mathcal{N}):=\sup_{\mathcal{P}_{\text{\textrm{flood}}%
}}\lim_{n}R_{n}. \label{ccc11}%
\end{equation}
Let us apply the weak-converse in Theorem~\ref{generalWEAK} by replacing the
supremum over point-to-point protocols $\mathcal{P}$ with a supremum over
flooding protocols $\mathcal{P}_{\text{\textrm{flood}}}$, i.e.,%
\begin{equation}
\mathcal{C}^{\text{m}}(\mathcal{N})\leq E_{\mathrm{R}}^{\bigstar}%
(\mathcal{N}):=\sup_{\mathcal{P}_{\text{\textrm{flood}}}}\underset{n}{\lim
}\frac{E_{\mathrm{R}}(\rho_{\mathbf{ab}}^{n})}{n}. \label{m2m2}%
\end{equation}

According to previous Lemma~\ref{reduceCHOI}, for any flooding protocol
$\mathcal{P}_{\text{\textrm{flood}}}$ and entanglement cut $C$, we may write
Eq.~(\ref{cutEQ}) with $n_{\mathbf{xy}}=n$. Computing the REE on this
decomposition and exploiting basic properties of the REE, we derive%
\begin{equation}
E_{\mathrm{R}}\left[  \rho_{\mathbf{ab}}^{n}(C)\right]  \leq n\sum
_{(\mathbf{x},\mathbf{y})\in\tilde{C}}E_{\mathrm{R}}(\sigma_{\mathbf{xy}%
})=nE_{\mathrm{R}}^{\text{m}}(C). \label{ddd3}%
\end{equation}
By using Eq.~(\ref{ddd3}) in Eq.~(\ref{m2m2}), both the supremum and the limit
disappear, and we are left with the bound%
\begin{equation}
\mathcal{C}^{\text{m}}(\mathcal{N})\leq E_{\mathrm{R}}^{\text{m}%
}(C),~~\text{for any~}C\text{.} \label{preRESS22}%
\end{equation}
By minimizing over all cuts, we therefore prove Eq.~(\ref{theommll}). The
extension to asymptotic simulations follows the same derivation in the proof
of Theorem~\ref{theoUBnet} but setting $n_{\mathbf{xy}}=n$. We find again
Eq.~(\ref{theommll}) but where the REE takes the weaker formulation for
asymptotic states of Eq.~(\ref{REE_weaker}).~$\blacksquare$

\subsection{Direct part (achievable rate)}

We now provide a general lower bound to the multipath capacity. Consider an
arbitrary quantum network $\mathcal{N}=(P,E)$ where edge $(\mathbf{x}%
,\mathbf{y})\in E$ is connected by channel $\mathcal{E}_{\mathbf{xy}}$ with
two-way capacity $\mathcal{C}_{\mathbf{xy}}=\mathcal{C}(\mathcal{E}%
_{\mathbf{xy}})$. Given an arbitrary entanglement cut $C$ of the network, we
define its multi-edge capacity as the total number of target bits distributed
by all the edges across the cut, i.e.,%
\begin{equation}
\mathcal{C}^{\text{m}}(C):=\sum_{(\mathbf{x},\mathbf{y})\in\tilde{C}%
}\mathcal{C}_{\mathbf{xy}}~. \label{Cmpform}%
\end{equation}
In this setting, a minimum cut $C_{\text{min}}$ is such that
\begin{equation}
\mathcal{C}^{\text{m}}(C_{\text{min}})=\min_{C}\mathcal{C}^{\text{m}}(C).
\label{achievTARGET}%
\end{equation}
We now prove that the later is an achievable rate for multipath
quantum/private communication.

\begin{theorem}
[Lower bound]\label{TheoMP2}Consider an arbitrary quantum network
$\mathcal{N}=(P,E)$ where two end-points may be disconnected by an
entanglement cut $C$. The multipath capacity of the network satisfies%
\begin{equation}
\mathcal{C}^{\text{m}}(\mathcal{N})\geq\min_{C}\mathcal{C}^{\text{m}}(C).
\label{cut2}%
\end{equation}
In other words, the minimum multi-edge capacity of the entanglement cuts is an
achievable rate. This rate is achieved by a flooding protocol whose\ multipath
routing can be found in $O(|P|\times|E|)$ time by solving the classical
maximum flow problem.
\end{theorem}

\textbf{Proof.}~~To show the achievability of the rate in
Eq.~(\ref{achievTARGET}), we resort to the classical max-flow min-cut
theorem~\cite{Ford}. In the literature, this theorem has been widely adopted
for the study of directed graphs. In general, it can also be applied to
directed multi-graphs as well as undirected graphs/multi-graphs (e.g.,
see~\cite[Sec.~6]{netflow}). The latter cases can be treated by splitting the
undirected edges into directed ones (e.g., see~\cite[Sec.~2.4]{netflow}).

\begin{figure}[ptbh]
\begin{center}
\vspace{-2.2cm} \includegraphics[width=0.48\textwidth] {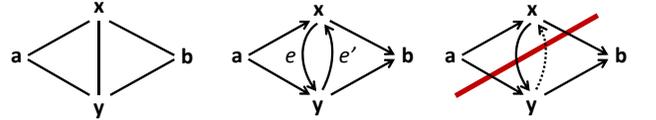}
\vspace{-2.8cm}
\end{center}
\caption{Manipulations of the undirected diamond network. (Left)~Original
undirected quantum network $\mathcal{N}^{\diamond}$. (Middle)~Flow network
$\mathcal{N}_{\text{flow}}^{\diamond}$ with Alice $\mathbf{a}$ as source and
Bob $\mathbf{b}$ as sink, where the middle undirected edge $(\mathbf{x}%
,\mathbf{y})$ has been split in two directed edges $e$ and $e^{\prime}$ with
the same capacity. (Right)~Assuming the displayed Alice-Bob cut, the dotted
edge does not belong to the directed cut-set $\tilde{C}_{D}$.}%
\label{diamondsPLOT}%
\end{figure}

Our first step is therefore the transformation of the undirected graph of the
quantum network $\mathcal{N}=(P,E)$ into a suitable directed graph (in
general, these may be multi-graphs, in which case the following derivation
still holds but with more technical notation). Starting from $(P,E)$, we
consider the directed graph where Alice's edges are all out-going (so that she
is a source), while Bob's edges are all in-going (so that he is a sink). Then,
for any pair $\mathbf{x}$ and $\mathbf{y}$ of intermediate points
$P\backslash\{\mathbf{a},\mathbf{b}\}$, we split the undirected edge
$(\mathbf{x},\mathbf{y})\in E$ into two directed edges $e:=(\mathbf{x}%
,\mathbf{y})\in E_{D}$ and $e^{\prime}:=(\mathbf{y},\mathbf{x})\in E_{D}$,
having capacities equal to the capacity $\mathcal{C}_{\mathbf{xy}}$ of the
original undirected edge~\cite{OneDIR}. These manipulations generate our flow
network $\mathcal{N}_{\text{flow}}=(P,E_{D})$. See Fig.~\ref{diamondsPLOT} for
an example. 

We then adopt the standard definition of cut-set for flow networks, here
called \textquotedblleft directed cut-set\textquotedblright. Given an
Alice-Bob cut $C$ of the flow network, with bipartition $(\mathbf{A}%
,\mathbf{B})$ of the points $P$, its directed cut-set is defined as $\tilde
{C}_{D}=\{(\mathbf{x},\mathbf{y})\in E_{D}:\mathbf{x}\in\mathbf{A}%
,\mathbf{y}\in\mathbf{B}\}$. This means that directed edges of the type
$(\mathbf{y}\in\mathbf{B},\mathbf{x}\in\mathbf{A})$ do not belong to this set
(see Fig.~\ref{diamondsPLOT}). Using this definition, the cut-properties of
the flow network $\mathcal{N}_{\text{flow}}$ are exactly the same as those of
the original undirected graph $\mathcal{N}$, for which we used the
\textquotedblleft undirected\textquotedblright\ definition of cut-set. For
this reason, we have
\begin{equation}
\left[  \min_{C}\sum_{(\mathbf{x},\mathbf{y})\in\tilde{C}}\mathcal{C}%
_{\mathbf{xy}}\right]  _{\mathcal{N}}=\left[  \min_{C}\sum_{(\mathbf{x}%
,\mathbf{y})\in\tilde{C}_{D}}\mathcal{C}_{\mathbf{xy}}\right]  _{\mathcal{N}%
_{\text{flow}}}, \label{bbACH0}%
\end{equation}
where the first quantity is computed on $\mathcal{N}$, while the second one is
computed on the flow network $\mathcal{N}_{\text{flow}}$. We aim to show that
the latter is an achievable rate.

Let us now define the \textquotedblleft flow\textquotedblright\ in the network
$\mathcal{N}_{\text{flow}}$ as the number of qubits per use which are reliably
transmitted from $\mathbf{x}$ to $\mathbf{y}$ along the directed edge
$e=(\mathbf{x},\mathbf{y})\in E_{D}$, denoted by $R_{\mathbf{xy}}^{e}\geq0$.
This quantum transmission is performed by means of a point-to-point protocol
where $\mathbf{x}$ and $\mathbf{y}$ exploit adaptive LOCCs, i.e., unlimited
two-way CCs and adaptive LOs, without the help of the other points of the
network. It is therefore bounded by the two-way quantum capacity of the
associated channel $\mathcal{E}_{\mathbf{xy}}$, i.e., $R_{\mathbf{xy}}^{e}\leq
Q_{2}(\mathcal{E}_{\mathbf{xy}})$. The actual physical direction of the
quantum channel does not matter since it is used with two-way CCs, so that the
two points $\mathbf{x}$ and $\mathbf{y}$ first distill entanglement and then
they teleport qubits in the \textquotedblleft logical
direction\textquotedblright\ specified by the directed edge.

Since every directed edge $e=(\mathbf{x},\mathbf{y})$ between two intermediate
points $\mathbf{x},\mathbf{y}\in P\backslash\{\mathbf{a},\mathbf{b}\}$ has an
opposite counterpart\ $e^{\prime}:=(\mathbf{y},\mathbf{x})$, we may
simultaneously consider an opposite flow of qubits from $\mathbf{y}$ to
$\mathbf{x}$ with rate $0\leq R_{\mathbf{yx}}^{e^{\prime}}\leq Q_{2}%
(\mathcal{E}_{\mathbf{xy}})$. As a result, there will be an \textquotedblleft
effective\textquotedblright\ point-to-point\ rate between $\mathbf{x}$ and
$\mathbf{y}$ which is defined by the difference of the two \textquotedblleft
directed\textquotedblright\ rates
\begin{equation}
R_{\mathbf{xy}}:=R_{\mathbf{xy}}^{e}-R_{\mathbf{yx}}^{e^{\prime}}.
\end{equation}
Its absolute value $|R_{\mathbf{xy}}|$ provides the effective number of qubits
transmitted between $\mathbf{x}$ to $\mathbf{y}$ per use of the undirected
edge. For $R_{\mathbf{xy}}\geq0$, effective qubits flow from $\mathbf{x}$ to
$\mathbf{y}$, while $R_{\mathbf{xy}}\leq0$ means that effective qubits flow
from $\mathbf{y}$ to\textbf{\ }$\mathbf{x}$. The effective rate is correctly
bounded $|R_{\mathbf{xy}}|\leq Q_{2}(\mathcal{E}_{\mathbf{xy}})$ and we set
$R_{\mathbf{xy}}=0$ if two points are not connected. The ensemble of positive
directed rates $\{R_{\mathbf{xy}}^{e}\}_{e\in E_{D}}$ represents a flow vector
in $\mathcal{N}_{\text{flow}}$. For any choice of this vector, there is a
corresponding ensemble of effective rates $\{R_{\mathbf{xy}}\}_{(\mathbf{x}%
,\mathbf{y})\in E}$ for the original network $\mathcal{N}$. The signs
$\{\mathrm{sgn}(R_{\mathbf{xy}})\}_{(\mathbf{x},\mathbf{y})\in E}$ specify an
orientation $\mathcal{N}_{D}=(P,E_{D}^{\prime})$ for $\mathcal{N}$, and the
absolute values $\{|R_{\mathbf{xy}}|\}_{(\mathbf{x},\mathbf{y})\in E}$ provide
point-to-point quantum communication rates for the associated protocol.

It is important to note that $\{R_{\mathbf{xy}}^{e}\}_{e\in E_{D}}$ represents
a \textquotedblleft legal\textquotedblright\ flow vector in $\mathcal{N}%
_{\text{flow}}$\ only if we impose the property of flow
conservation~\cite{netflow}. This property can be stated for $\{R_{\mathbf{xy}%
}^{e}\}_{e\in E_{D}}$\ or, equivalently, for the effective vector
$\{R_{\mathbf{xy}}\}_{(\mathbf{x},\mathbf{y})\in E}$. At any intermediate
point, the number of qubits simultaneously received must be equal to the
number of qubits simultaneously transmitted through all the point-to-point
communications with neighbor points. In other words, for any $\mathbf{x}\in
P\backslash\{\mathbf{a},\mathbf{b}\}$, we must impose
\begin{equation}
\sum_{\mathbf{y}\in P}R_{\mathbf{xy}}=0.
\end{equation}

This property does not hold for Alice $\mathbf{a}$ (source) and Bob
$\mathbf{b}$ (sink), for which we impose%
\[
\sum_{\mathbf{y}\in P}R_{\mathbf{ay}}=-\sum_{\mathbf{y}\in P}R_{\mathbf{by}%
}:=|R|,
\]
where $|R|$ is known as the value of the flow. This is an achievable
end-to-end rate since it represents the total number of qubits per network use
which are transmitted by Alice and correspondingly received by Bob via all the
end-to-end routes, where the intermediate points quantum-communicate at the
rates $\{R_{\mathbf{xy}}\}_{(\mathbf{x},\mathbf{y})\in E}$.

Now, from the classical max-flow min-cut theorem, we know that the maximum
value of the flow in the network $|R|_{\max}$ is equal to the capacity of the
minimum cut~\cite{Ford,netflow}, i.e., we may write
\begin{equation}
|R|_{\max}=\min_{C}\sum_{(\mathbf{x},\mathbf{y})\in\tilde{C}_{D}}%
Q_{2}(\mathcal{E}_{\mathbf{xy}})~. \label{maxPROO}%
\end{equation}
Thus, by construction, we have that $|R|_{\max}$ is an achievable rate for
quantum communication. The previous reasoning can be repeated for private bits
by defining a corresponding flow of private information through the network.
Thus, in general, we may write that
\begin{equation}
|R|_{\max}=\min_{C}\sum_{(\mathbf{x},\mathbf{y})\in\tilde{C}_{D}}%
\mathcal{C}_{\mathbf{xy}}%
\end{equation}
is an achievable rate for any of the quantum tasks. This proves that
Eq.~(\ref{bbACH0}) is an achievable rate.

In order to better understand the flooding protocol that achieves $|R|_{\max}%
$, call $\{\tilde{R}_{\mathbf{xy}}^{e}\}_{e\in E_{D}}$ the optimal flow vector
in $\mathcal{N}_{\text{flow}}$. There is a corresponding vector $\{\tilde
{R}_{\mathbf{xy}}\}_{(\mathbf{x},\mathbf{y})\in E}$ which determines an
optimal orientation $\mathcal{N}_{D}=(P,E_{D}^{\prime})$ for the quantum
network $\mathcal{N}=(P,E)$, besides providing the optimal rates $\{|\tilde
{R}_{\mathbf{xy}}|\}_{(\mathbf{x},\mathbf{y})\in E}$ to be reached by the
point-to-point connections. In other words, starting from the capacities
$\mathcal{C}_{\mathbf{xy}}$, the points solve the maximum flow problem and
establish an optimal multipath routing $\mathcal{R}_{\text{opt}}^{\text{m}}$.
After this, each point $\mathbf{x}\in P$ multicasts to its out-neighborhood
$N^{\text{out}}(\mathbf{x})$, according to the optimal rates and the optimal orientation.

In conclusion, let us discuss the complexity of finding the optimal multipath
routing $\mathcal{R}_{\text{opt}}^{\text{m}}$. By construction, the flow
network $\mathcal{N}_{\text{flow}}=\{P,E_{D}\}$ has only a small overhead with
respect to the original network $\mathcal{N}=\{P,E\}$. In fact, we just have
$|E_{D}|\leq2|E|$. Within $\mathcal{N}_{\text{flow}}$, the maximum flow can be
found with classical algorithms. If the capacities are rational, we can apply
the Ford-Fulkerson algorithm~\cite{Ford} or the Edmonds--Karp
algorithm~\cite{Karp}, the latter running in $O(|P|\times|E_{D}|^{2})$ time.
An alternative is Dinic's algorithm~\cite{Dinic}, which runs in $O(|P|^{2}%
\times|E_{D}|)$ time. More powerful algorithms are
available~\cite{Alon,Ahuja,Cheriyan} and the best running performance is
currently $O(|P|\times|E_{D}|)$ time~\cite{King,Orlin}. Thus, adopting Orlin's
algorithm~\cite{Orlin}, we find the solution in $O(|P|\times|E_{D}%
|)=O(|P|\times|E|)$ time.~$\blacksquare$

\subsection{Formulas for teleportation-covariant and distillable networks}

Consider a teleportation-covariant quantum network $\mathcal{N}=(P,E)$ whose
teleportation simulation has an associated Choi-representation $\sigma
(\mathcal{N})=\{\sigma_{\mathcal{E}_{\mathbf{xy}}}\}_{(\mathbf{x}%
,\mathbf{y})\in E}$. Then, from Theorems~\ref{TheoMP1} and~\ref{TheoMP2}, we
may write the following sandwich for the multipath capacity%
\begin{equation}
\min_{C}\mathcal{C}^{\text{m}}(C)\leq\mathcal{C}^{\text{m}}(\mathcal{N}%
)\leq\min_{C}E_{\mathrm{R}}^{\text{m}}(C)~, \label{sandmp}%
\end{equation}
with $\mathcal{C}^{\text{m}}(C)$ being defined in Eq.~(\ref{Cmpform}), and%
\begin{equation}
E_{\mathrm{R}}^{\text{m}}(C)=\sum_{(\mathbf{x},\mathbf{y})\in\tilde{C}%
}E_{\mathrm{R}}(\sigma_{\mathcal{E}_{\mathbf{xy}}})~.
\end{equation}
As usual, the latter may have an asymptotic formulation for networks of
bosonic channels, with the REE\ taking the form as in Eq.~(\ref{REE_weaker})
over $\sigma_{\mathcal{E}_{\mathbf{xy}}}:=\lim_{\mu}\sigma_{\mathcal{E}%
_{\mathbf{xy}}}^{\mu}$, where $\sigma_{\mathcal{E}_{\mathbf{xy}}}^{\mu}$ is a
sequence of states with finite energy.

In particular, consider now a distillable network. This means that, for any
edge $(\mathbf{x},\mathbf{y})\in E$, we may write Eq.~(\ref{dueEELL}), exactly
or asymptotically. By imposing this condition in Eq.~(\ref{sandmp}), we find
that upper and lower bounds coincide. We have therefore the following result
which establishes the multipath capacity $\mathcal{C}^{\text{m}}(\mathcal{N}%
)$\ of a distillable network and fully extends the max-flow min-cut
theorem~\cite{Harris,Ford,ShannonFLOW} to quantum communications.

\begin{corollary}
[Multi-path capacities]\label{coroNETmp}Consider a distillable network
$\mathcal{N}=(P,E)$, whose arbitrary edge $(\mathbf{x},\mathbf{y})\in E$\ is
connected by a distillable channel $\mathcal{E}_{\mathbf{xy}}$\ with two-way
capacity $\mathcal{C}_{\mathbf{xy}}$\ and Choi matrix $\sigma_{\mathcal{E}%
_{\mathbf{xy}}}$. Then, the multipath capacity of the network is equal to
\begin{equation}
\mathcal{C}^{\text{m}}(\mathcal{N})=\min_{C}E_{\mathrm{R}}^{\text{m}}%
(C)=\min_{C}\sum_{(\mathbf{x},\mathbf{y})\in\tilde{C}}E_{\mathrm{R}}%
(\sigma_{\mathcal{E}_{\mathbf{xy}}}),
\end{equation}
with an implicit asymptotic formulation for bosonic channels. Equivalently,
$\mathcal{C}^{\text{m}}(\mathcal{N})$ is also equal to the minimum
(multi-edge) capacity of the entanglement cuts%
\begin{equation}
\mathcal{C}^{\text{m}}(\mathcal{N})=\min_{C}\mathcal{C}^{\text{m}}(C).
\label{coroMAINkk}%
\end{equation}
The optimal multipath routing can be found in $O(|P|\times|E|)$ time by
solving the classical maximum flow problem. A capacity-achieving flooding
protocol corresponds to performing one-way entanglement distillation between
neighbor points, followed by multiple sessions of teleportation in the
direction of the optimal network orientation.
\end{corollary}

The proof is a direct application of the previous reasonings. In particular,
from Theorem~\ref{TheoMP2}, we have that the routing problem is reduced to the
solution of a classical optimization problem, i.e., finding the maximum flow
in a flow network. This solution provides an optimal orientation
$\mathcal{N}_{D}$ of the quantum network and also the point-to-point rates
$|\tilde{R}_{\mathbf{xy}}|$ to be used in the various multicasts. Under this
optimal routing, the multipath capacity is achieved by a non-adaptive flooding
protocol based on one-way CCs. In fact, because the channels are distillable,
each pair of points $\mathbf{x}$ and $\mathbf{y}$ may distill $n|\tilde
{R}_{\mathbf{xy}}|$ ebits. By using the distilled ebits, Alice's qubits are
teleported to Bob along the multipath routes associated with the maximum flow.
Since Alice's qubits can be part of ebits and, therefore, private bits, this
protocol can also distill entanglement and keys at the same end-to-end rate.

Thus, Corollary~\ref{coroNETmp} reduces the computation of the multipath
capacity of a distillable quantum network to the determination of the maximum
flow in a classical network. In this sense the max-flow min-cut theorem is
extended from classical to quantum communications. In particular, the
distillable network can always be transformed in a teleportation network,
where quantum information is teleported as a flow from Alice to Bob.

\subsection{Multipath capacities of fundamental networks}

Consider the practical scenario of quantum optical communications affected by
loss, e.g., free-space or fiber-based. A specific distillable network is a
bosonic network connected by lossy channels $\mathcal{N}_{\text{loss}}$, so
that each undirected edge $(\mathbf{x},\mathbf{y})$ has an associated lossy
channel $\mathcal{E}_{\mathbf{xy}}$ with transmissivity $\eta_{\mathbf{xy}}$
or equivalent \textquotedblleft loss parameter\textquotedblright%
\ $1-\eta_{\mathbf{xy}}$. We may then apply Corollary~\ref{coroNETmp} and
express the multipath capacity $\mathcal{C}^{\text{m}}(\mathcal{N}%
_{\text{loss}})$\ in terms of the loss parameters of the network.

Let us define the loss of an Alice-Bob entanglement cut $C$ as the product of
the loss parameters of the channels in the cut-set, i.e., we set%
\begin{equation}
l(C):=%
{\textstyle\prod\limits_{(\mathbf{x},\mathbf{y})\in\tilde{C}}}
(1-\eta_{\mathbf{xy}}).
\end{equation}
This quantity determines the multi-edge capacity of the cut, since we have
$\mathcal{C}^{\text{m}}(C)=-\log_{2}l(C)$. By applying Eq.~(\ref{coroMAINkk}),
we find that the multipath capacity of the lossy network is given by%
\begin{equation}
\mathcal{C}^{\text{m}}(\mathcal{N}_{\text{loss}})=\min_{C}\left[  -\log
_{2}l(C)\right]  =-\log_{2}\left[  \max_{C}l(C)\right]  .
\end{equation}
Thus, we may define the total loss of the network as the maximization of
$l(C)$ over all cuts, i.e.,%
\begin{equation}
l(\mathcal{N}_{\text{loss}}):=\max_{C}l(C),
\end{equation}
and write the simple formula%
\begin{equation}
\mathcal{C}^{\text{m}}(\mathcal{N}_{\text{loss}})=-\log_{2}l(\mathcal{N}%
_{\text{loss}}). \label{toGENmulti}%
\end{equation}

In general, we may consider a multiband lossy network $\mathcal{N}%
_{\text{loss}}^{\text{band}}$, where each edge $(\mathbf{x},\mathbf{y})$
represents a multiband lossy channel $\mathcal{E}_{\mathbf{xy}}^{\text{band}}$
with bandwidth $M_{\mathbf{xy}}$\ and constant transmissivity $\eta
_{\mathbf{xy}}$. In other words, each single edge $(\mathbf{x},\mathbf{y})$
corresponds to $M_{\mathbf{xy}}$ independent lossy channels with the same
transmissivity $\eta_{\mathbf{xy}}$. In this case, we have $\mathcal{C}%
(\mathcal{E}_{\mathbf{xy}}^{\text{band}})=-M_{\mathbf{xy}}\log_{2}%
(1-\eta_{\mathbf{xy}})$ and we write
\begin{equation}
\mathcal{C}^{\text{m}}(\mathcal{N}_{\text{loss}}^{\text{band}})=-\log
_{2}\left[  \max_{C}%
{\textstyle\prod\limits_{(\mathbf{x},\mathbf{y})\in\tilde{C}}}
(1-\eta_{\mathbf{xy}})^{M_{\mathbf{xy}}}\right]  ,
\end{equation}
which directly generalizes Eq.~(\ref{toGENmulti}).

In particular, suppose that we have the same loss in each edge of the
multiband network, i.e., $\eta_{\mathbf{xy}}:=\eta$ for any $(\mathbf{x}%
,\mathbf{y})\in E$, which may occur when points $\mathbf{x}$ and $\mathbf{y}$
are equidistant. Then, we may simply write%
\begin{align}
\mathcal{C}^{\text{m}}(\mathcal{N}_{\text{loss}}^{\text{band}})  &  =-M_{\min
}\log_{2}(1-\eta),\\
M_{\min}  &  :=\min_{C}\sum_{(\mathbf{x},\mathbf{y})\in\tilde{C}%
}M_{\mathbf{xy}},
\end{align}
where $M_{\min}$ is the effective bandwidth of the network.

Consider now other types of distillable networks. Start with a bosonic network
of quantum-limited amplifiers $\mathcal{N}_{\text{amp}}$, where the generic
edge $(\mathbf{x},\mathbf{y})$\ has an associated gain $g_{\mathbf{xy}}$. Its
multipath capacity is given by
\begin{equation}
\mathcal{C}^{\text{m}}(\mathcal{N}_{\text{amp}})=-\log_{2}\left[  \max_{C}%
{\textstyle\prod\limits_{(\mathbf{x},\mathbf{y})\in\tilde{C}}}
(1-g_{\mathbf{xy}}^{-1})\right]  .
\end{equation}
For a qubit network of dephasing channels $\mathcal{N}_{\text{deph}}$, where
the generic edge $(\mathbf{x},\mathbf{y})$ has dephasing probability
$p_{\mathbf{xy}}$, we may write the multipath capacity%
\begin{equation}
\mathcal{C}^{\text{m}}(\mathcal{N}_{\text{deph}})=\min_{C}%
{\textstyle\sum\limits_{(\mathbf{x},\mathbf{y})\in\tilde{C}}}
\left[  1-H_{2}(p_{\mathbf{xy}})\right]  .
\end{equation}
Finally, for a qubit network of erasure channels $\mathcal{N}_{\text{erase}}$
with erasure probabilities $p_{\mathbf{xy}}$, we simply have%
\begin{equation}
\mathcal{C}^{\text{m}}(\mathcal{N}_{\text{erase}})=\min_{C}%
{\textstyle\sum\limits_{(\mathbf{x},\mathbf{y})\in\tilde{C}}}
(1-p_{\mathbf{xy}}).
\end{equation}
Similar expressions may be derived for qudit networks of dephasing and erasure
channels in arbitrary dimension.

\section{Generalization to multiple senders and
receivers\label{SECmultipleNETs}}

Previous results have been derived in the unicast setting, with a single
sender $\mathbf{a}$ and a single receiver $\mathbf{b}$. In general, we may
consider the presence of multiple senders $\{\mathbf{a}_{i}\}$ and receivers
$\{\mathbf{b}_{j}\}$, which may simultaneously communicate according to
various configurations. For simplicity, these sets are intended to be disjoint
$\{\mathbf{a}_{i}\}\cap\{\mathbf{b}_{j}\}=\emptyset$, so that an end-point
cannot be sender and receiver at the same time. It is clear that all previous
results derived for the two basic routing strategies provide general upper
bounds which are still valid for the individual end-to-end capacities
associated with each sender-receiver pair $(\mathbf{a}_{i},\mathbf{b}_{i})$ in
the various settings with multiple end-points.

In the following sections, we start with the multiple-unicast quantum network.
This consists of $M$ Alices $\{\mathbf{a}_{1},\ldots,\mathbf{a}_{M}\}$ and $M$
Bobs $\{\mathbf{b}_{1},\ldots,\mathbf{b}_{M}\}$, with the generic $i$th Alice
$\mathbf{a}_{i}$ communicating with a corresponding $i$th Bob $\mathbf{b}_{i}%
$. This case can be studied by assuming single-path routing
(\ref{SecMULTIunicast}) or multipath routing (\ref{SecMULTIunicast2}). Besides
the general bounds inherited from the unicast scenario, we also derive a
specific set of upper bounds for the rates that are simultaneously achievable
by all parties.

Another important case is the multicast quantum network, where a single sender
simultaneously communicates with $M\geq1$ receivers, e.g., for distributing
$M$ different keys. By its nature, this is studied under multipath routing
(Sec.~\ref{SECmulticastSINGLE}). In this setting, an interesting variant is
the distribution of the same key to all receivers, which may be assisted by
network coding~\cite{Gamal} (Sec.~\ref{SecNETCODING}).

\begin{figure}[ptbh]
\vspace{-1.0cm}
\par
\begin{center}
\includegraphics[width=0.37\textwidth] {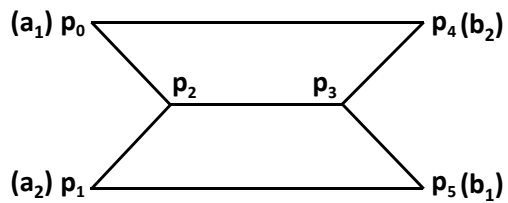} \vspace{-1.3cm}
\end{center}
\caption{Butterfly quantum network. (i)~An example of multiple-unicast is
considering two sender-receiver pairs, e.g., Alice $\mathbf{a}_{1}$
communicating with Bob $\mathbf{b}_{1}$, and Alice $\mathbf{a}_{2}$ with Bob
$\mathbf{b}_{2}$. Single-path routing corresponds to the simultaneous use of
two end-to-end routes, e.g., $(\mathbf{a}_{1})\mathbf{p}_{0}-\mathbf{p}%
_{2}-\mathbf{p}_{3}-\mathbf{p}_{5}(\mathbf{b}_{1})$ and $(\mathbf{a}%
_{2})\mathbf{p}_{1}-\mathbf{p}_{2}-\mathbf{p}_{3}-\mathbf{p}_{4}%
(\mathbf{b}_{2})$. Multipath routing corresponds to choosing a network
orientation, where the end-points may also act as relays. Each point of the
network multicasts to its out-neighborhood. For instance, we may have the
following point-to-point multicasts: $\mathbf{p}_{0}\rightarrow\{\mathbf{p}%
_{2},\mathbf{p}_{4}\}$, $\mathbf{p}_{1}\rightarrow\{\mathbf{p}_{2}%
,\mathbf{p}_{5}\}$, $\mathbf{p}_{2}\rightarrow\mathbf{p}_{3}$, and
$\mathbf{p}_{3}\rightarrow\{\mathbf{p}_{4},\mathbf{p}_{5}\}$. (ii)~An example
of end-to-end multicast is Alice $\mathbf{a}_{1}$ communicating wih both Bobs
$\{\mathbf{b}_{1},\mathbf{b}_{2}\}$ via multipath routing. (iii)~Finally, in a
multiple-multicast, Alice $\mathbf{a}_{1}$ communicates with $\{\mathbf{b}%
_{1},\mathbf{b}_{2}\}$, and Alice $\mathbf{a}_{2}$ communicates with the same
destination set $\{\mathbf{b}_{1},\mathbf{b}_{2}\}$.}%
\label{butterfly}%
\end{figure}

More generally, we may consider a multiple-multicast quantum network. Here we
have $M_{A}\geq1$ senders and $M_{B}\geq1$ receivers, and each sender
communicates simultaneously with the entire set of receivers
(Sec.~\ref{SECmulticastMANY}). In a private communication scenario, this
corresponds to the distribution of $M_{A}M_{B}$ different keys. For a
description of these configurations, see the simple example of the butterfly
quantum network in Fig.~\ref{butterfly}.

\subsection{Multiple-unicast quantum networks with single-path
routing\label{SecMULTIunicast}}

The generalization to a multiple-unicast setting is relatively easy. Let us
start by considering two Alice-Bob pairs $(\mathbf{a}_{1},\mathbf{b}_{1})$ and
$(\mathbf{a}_{2},\mathbf{b}_{2})$, since the extension to arbitrary number of
pairs is immediate. We may easily formulate network protocols which are based
on single-path routing. In this case, each sequential use of the network
involves the transmission of quantum systems along two
(potentially-overlapping) routes%
\begin{equation}
\omega_{1}:\mathbf{a}_{1}-\cdots-\mathbf{b}_{1},~~~\omega_{2}:\mathbf{a}%
_{2}-\cdots-\mathbf{b}_{2},
\end{equation}
where each transmission through an edge is assisted by network LOCCs. The
routes are updated use after use.

After $n$ uses, the output of the double-unicast network protocol
$\mathcal{P}_{\text{2-unicast}}$ is a state $\rho_{\mathbf{a}_{1}%
\mathbf{a}_{2}\mathbf{b}_{1}\mathbf{b}_{2}}^{n}$ which is $\varepsilon$-close
in trace norm to a target state
\begin{equation}
\phi:=\phi_{\mathbf{a}_{1}\mathbf{b}_{1}}^{\otimes nR_{1}^{n}}\otimes
\phi_{\mathbf{a}_{2}\mathbf{b}_{2}}^{\otimes nR_{2}^{n}}, \label{targettt}%
\end{equation}
where $\phi_{\mathbf{a}_{i}\mathbf{b}_{i}}$ is a one-bit state (private bit or
ebit) for the pair $(\mathbf{a}_{i},\mathbf{b}_{i})$ and $nR_{i}^{n}$ the
number of its copies. Taking the limit of large $n$ and optimizing over all
protocols $\mathcal{P}_{\text{2-unicast}}$, we define the capacity region as
the closure of the set of the achievable asymptotic rates $(R_{1},R_{2})$. In
general, for $M$ sender-receiver pairs, we have an $M$-tuple of achievable
rates $(R_{1},\ldots,R_{M})$. Depending on the task of the protocol (i.e., the
target state), these rates refer to end-to-end entanglement distillation
(equivalently, error-free quantum communication) or secret-key generation.

Before proceeding, let us first introduce more general types of entanglement
cuts of the quantum network. Given two sets of senders $\{\mathbf{a}_{i}\}$
and receivers $\{\mathbf{b}_{i}\}$, we adopt the notation $C:\{\mathbf{a}%
_{i}\}|\{\mathbf{b}_{i}\}$ for a cut $C=(\mathbf{A},\mathbf{B})$ such that
$\{\mathbf{a}_{i}\}\subset\mathbf{A}$ and $\{\mathbf{b}_{i}\}\subset
\mathbf{B}$. Similarly, we write $C:\mathbf{a}_{i}|\mathbf{b}_{i}$ for a cut
with $\mathbf{a}_{i}\in\mathbf{A}$ and $\mathbf{b}_{i}\in\mathbf{B}$, and
$C:\mathbf{a}_{i}\mathbf{a}_{j}|\mathbf{b}_{i}\mathbf{b}_{j}$ for a cut with
$\{\mathbf{a}_{i},\mathbf{a}_{j}\}\subset\mathbf{A}$ and $\{\mathbf{b}%
_{i},\mathbf{b}_{j}\}\subset\mathbf{B}$. As usual, we consider the single-edge
flow of REE trough a cut as%
\begin{equation}
E_{\mathrm{R}}(C):=\max_{(\mathbf{x},\mathbf{y})\in\tilde{C}}E_{R}%
(\sigma_{\mathbf{xy}}),
\end{equation}
where $\sigma_{\mathbf{xy}}$ is a resource state associated with an edge
$(\mathbf{x},\mathbf{y})$ in the cut-set $\tilde{C}$, under some simulation of
the network. We can now state the following result.

\begin{theorem}
[Multi-unicast with single paths]\label{TheomultipleUNICAST}Let us consider a
multiple-unicast quantum network $\mathcal{N}=(P,E)$ with $M$ sender-receiver
pairs $(\mathbf{a}_{i},\mathbf{b}_{i})$ communicating by means of single-path
routing. Adopt a simulation of the network with a resource representation
$\sigma(\mathcal{N})=\{\sigma_{\mathbf{xy}}\}_{(\mathbf{x},\mathbf{y})\in E}$.
In particular, $\sigma(\mathcal{N})$\ can be a Choi-representation for a
teleportation-covariant $\mathcal{N}$. We have the following outer bounds for
the capacity region%
\begin{align}
R_{i}  &  \leq\min_{C:\mathbf{a}_{i}|\mathbf{b}_{i}}E_{\mathrm{R}%
}(C)~~\text{for any }i,\label{unicvvv}\\
R_{i}+R_{j}  &  \leq\min_{C:\mathbf{a}_{i}\mathbf{a}_{j}|\mathbf{b}%
_{i}\mathbf{b}_{j}}E_{\mathrm{R}}(C)~~\text{for any }i\neq j
\label{doublehhhh}\\
&  \vdots\nonumber\\
\sum\limits_{i=1}^{M}R_{i}  &  \leq\min_{C:\{\mathbf{a}_{i}\}|\{\mathbf{b}%
_{i}\}}E_{\mathrm{R}}(C),
\end{align}
where $E_{\mathrm{R}}(C)$\ is the single-edge flow of REE through cut $C$. It
is understood that formulations may be asymptotic for quantum networks with
bosonic channels.
\end{theorem}

\textbf{Proof.}~~For simplicity consider first the case $M=2$, since the
generalization to arbitrary $M$ is straightforward. Let us also consider key
generation, since it automatically provides an upper bound for all the other
tasks. Considering the bipartition $\mathbf{a}_{1}\mathbf{a}_{2}%
|\mathbf{b}_{1}\mathbf{b}_{2}$, the distillable key of the target state $\phi$
in Eq.~(\ref{targettt}) is equal to%
\begin{equation}
K_{\mathrm{D}}(\mathbf{a}_{1}\mathbf{a}_{2}|\mathbf{b}_{1}\mathbf{b}%
_{2})_{\phi}=n(R_{1}^{n}+R_{2}^{n}).
\end{equation}
Using the REE with respect to the same bipartition, we may write the upper
bound%
\begin{align}
n(R_{1}^{n}+R_{2}^{n})  &  \leq E_{\mathrm{R}}(\mathbf{a}_{1}\mathbf{a}%
_{2}|\mathbf{b}_{1}\mathbf{b}_{2})_{\phi}\nonumber\\
&  \leq E_{\mathrm{R}}(\mathbf{a}_{1}\mathbf{a}_{2}|\mathbf{b}_{1}%
\mathbf{b}_{2})_{\rho^{n}}+\delta(\varepsilon,d),
\end{align}
where the latter inequality comes from the fact that $\rho^{n}:=\rho
_{\mathbf{a}_{1}\mathbf{a}_{2}\mathbf{b}_{1}\mathbf{b}_{2}}^{n}$ is
$\varepsilon$-close to $\phi$. The extra term $\delta(\varepsilon,d)$ depends
the $\varepsilon$-closeness, and the dimension $d$ of the total Hilbert space.
In the limit of large $n$ and small $\varepsilon$ (weak converse), we can
neglect $\delta(\varepsilon,d)/n$. This is a straightforward application of
the exponential scaling of the dimension $d$ shown in
Refs.~\cite{Matthias1a,Matthias2a} for DV systems and extended to CV systems
via standard truncation arguments, e.g., see Ref.~\cite[arXiv version 2 (Dec
2015)]{QKDpaper} for a simple proof and the discussion in Ref.~\cite{Notaa1}.
As a result we may write
\begin{equation}
\lim_{n}(R_{1}^{n}+R_{2}^{n})\leq\underset{n\rightarrow+\infty}{\lim}%
n^{-1}E_{\mathrm{R}}(\mathbf{a}_{1}\mathbf{a}_{2}|\mathbf{b}_{1}\mathbf{b}%
_{2})_{\rho^{n}}~. \label{todecppp}%
\end{equation}

By simulating and stretching the network, we may write the following
decomposition of the output state%
\begin{equation}
\rho_{\mathbf{a}_{1}\mathbf{a}_{2}\mathbf{b}_{1}\mathbf{b}_{2}}^{n}%
=\bar{\Lambda}_{\mathbf{a}_{1}\mathbf{a}_{2}\mathbf{b}_{1}\mathbf{b}_{2}%
}\left[  \underset{(\mathbf{x},\mathbf{y})\in E}{%
{\textstyle\bigotimes}
}~\sigma_{\mathbf{xy}}^{\otimes n_{\mathbf{xy}}}\right]  , \label{cmtoo}%
\end{equation}
where $n_{\mathbf{xy}}=np_{\mathbf{xy}}$ is the number of uses of edge
$(\mathbf{x},\mathbf{y})$ and $\bar{\Lambda}_{\mathbf{a}_{1}\mathbf{a}%
_{2}\mathbf{b}_{1}\mathbf{b}_{2}}$ is a trace-preserving LOCC, which is local
with respect to the bipartition $\mathbf{a}_{1}\mathbf{a}_{2}|\mathbf{b}%
_{1}\mathbf{b}_{2}$. By inserting entanglement cuts which disconnect the
senders and receivers, we reduce the number of resource states appearing in
Eq.~(\ref{cmtoo}) while preserving the locality of the LOCC with respect to
the bipartition of the end-points. In other words, for any cut $C:\mathbf{a}%
_{1}\mathbf{a}_{2}|\mathbf{b}_{1}\mathbf{b}_{2}$ we may write%
\begin{equation}
\rho_{\mathbf{a}_{1}\mathbf{a}_{2}\mathbf{b}_{1}\mathbf{b}_{2}}^{n}%
(C)=\bar{\Lambda}_{\mathbf{a}_{1}\mathbf{a}_{2}\mathbf{b}_{1}\mathbf{b}_{2}%
}^{C}\left[  \underset{(\mathbf{x},\mathbf{y})\in\tilde{C}}{%
{\textstyle\bigotimes}
}~\sigma_{\mathbf{xy}}^{\otimes n_{\mathbf{xy}}}\right]  .
\end{equation}

Using the latter decomposition in Eq.~(\ref{todecppp}), we obtain%
\begin{align}
\lim_{n}(R_{1}^{n}+R_{2}^{n})  &  \leq\underset{n\rightarrow+\infty}{\lim
}~n^{-1}E_{\mathrm{R}}(\mathbf{a}_{1}\mathbf{a}_{2}|\mathbf{b}_{1}%
\mathbf{b}_{2})_{\rho^{n}(C)}\nonumber\\
&  \leq\underset{n\rightarrow+\infty}{\lim}~n^{-1}\sum\limits_{(\mathbf{x}%
,\mathbf{y})\in\tilde{C}}n_{\mathbf{xy}}E_{\mathrm{R}}(\sigma_{\mathbf{xy}%
})\nonumber\\
&  =\sum\limits_{(\mathbf{x},\mathbf{y})\in\tilde{C}}p_{\mathbf{xy}%
}E_{\mathrm{R}}(\sigma_{\mathbf{xy}})\nonumber\\
&  \leq\max_{(\mathbf{x},\mathbf{y})\in\tilde{C}}E_{\mathrm{R}}(\sigma
_{\mathbf{xy}}):=E_{\mathrm{R}}(C)
\end{align}
By minimizing over the cuts, we derive%
\begin{equation}
\lim_{n}(R_{1}^{n}+R_{2}^{n})\leq\min_{C:\mathbf{a}_{1}\mathbf{a}%
_{2}|\mathbf{b}_{1}\mathbf{b}_{2}}E_{\mathrm{R}}(C). \label{outerCCC}%
\end{equation}
It is important to note that this bound holds for any protocol $\mathcal{P}%
_{\text{2-unicast}}$, whose details are all collapsed in the LOCC
$\bar{\Lambda}_{\mathbf{a}_{1}\mathbf{a}_{2}\mathbf{b}_{1}\mathbf{b}_{2}}$ and
therefore discarded. Thus, the same bound applies if we optimize over all
protocols, which means that Eq.~(\ref{outerCCC}) provides the following outer
bound for the capacity region%
\begin{equation}
R_{1}+R_{2}\leq\min_{C:\mathbf{a}_{1}\mathbf{a}_{2}|\mathbf{b}_{1}%
\mathbf{b}_{2}}E_{\mathrm{R}}(C).
\end{equation}

Note that, besides this bound, we also have the following unicast bounds for
the individual rates
\begin{equation}
R_{1}\leq\min_{C:\mathbf{a}_{1}|\mathbf{b}_{1}}E_{\mathrm{R}}(C),~~R_{2}%
\leq\min_{C:\mathbf{a}_{2}|\mathbf{b}_{2}}E_{\mathrm{R}}(C).
\end{equation}
These follows directly from Theorem~\ref{theoUBnet} on the converse for
unicast quantum networks. Equivalently, we may re-derive these bounds here, by
setting $R_{2}=0$ or $R_{1}=0$ in the target state of Eq.~(\ref{targettt}) and
repeating the previous derivation. For instance, for $R_{2}=0$, we have
$\phi:=\phi_{\mathbf{a}_{1}\mathbf{b}_{1}}^{\otimes nR_{1}^{n}}\otimes
\sigma_{\mathbf{a}_{2}\mathbf{b}_{2}}$, where $\sigma_{\mathbf{a}%
_{2}\mathbf{b}_{2}}$ does not contain target bits and may be taken to be
separable. Therefore, we start from $K_{\mathrm{D}}(\mathbf{a}_{1}%
|\mathbf{b}_{1})_{\phi}=nR_{1}^{n}$ and we repeat all the derivation with
respect to the bipartition $\mathbf{a}_{1}|\mathbf{b}_{1}$.

It is clear that the generalization from $M=2$ to arbitrary $M$ is immediate.
For any integer $M$, we have the target state%
\begin{equation}
\phi:=%
{\textstyle\bigotimes_{i=1}^{M}}
\phi_{\mathbf{a}_{i}\mathbf{b}_{i}}^{\otimes nR_{i}^{n}}.
\end{equation}
Considering the bipartition $\{\mathbf{a}_{i}\}|\{\mathbf{b}_{i}\}$ and the
corresponding cuts of the network leads to
\begin{equation}
\sum\limits_{i=1}^{M}R_{i}\leq\min_{C:\{\mathbf{a}_{i}\}|\{\mathbf{b}_{i}%
\}}E_{\mathrm{R}}(C),
\end{equation}
where we note that increasing the number of rates reduces the number of
possible cuts in the minimization. In order to get all the remaining
inequalities of the theorem, we just need to set some of the rates to zero.
For instance, for $R_{i}\neq0$ and $R_{j\neq i}=0$, we get the unicast bounds
of Eq.~(\ref{unicvvv}). For $R_{i}\neq0$, $R_{j\neq i}\neq0$ and $R_{k\neq
i,j}=0$ we get the double-unicast bounds of Eq.~(\ref{doublehhhh}), and so on.
The extension to asymptotic simulations of bosonic channels is achieved by
adopting the weaker definition of the REE as in Eq.~(\ref{REE_weaker}%
).~$\blacksquare$

Once we have proven the previous theorem, it is immediate to specify the
results for the case of multiple-unicast distillable networks, for which we
may write $E_{\mathrm{R}}(\sigma_{\mathbf{xy}})=E_{\mathrm{R}}(\sigma
_{\mathcal{E}_{\mathbf{xy}}})=\mathcal{C}_{\mathbf{xy}}$ for each edge
$(\mathbf{x,y})\in E$, where $\mathcal{C}_{\mathbf{xy}}$ is the two-way
capacity of the associated quantum channel $\mathcal{E}_{\mathbf{xy}}$. In
this case, we may directly write
\begin{equation}
E_{\mathrm{R}}(C)=\mathcal{C}(C):=\max_{(\mathbf{x},\mathbf{y})\in\tilde{C}%
}\mathcal{C}_{\mathbf{xy}},
\end{equation}
where $\mathcal{C}(C)$ is the single-edge capacity of cut $C$, already
introduced in Eq.~(\ref{cutEQUIV}) for the unicast quantum network. Thus, we
may express the bounds of Theorem~\ref{TheomultipleUNICAST}\ directly in terms
of the capacities of the cuts, i.e., we automatically prove the following.

\begin{corollary}
Consider a multiple-unicast quantum network $\mathcal{N}$\ with $M$
sender-receiver pairs $(\mathbf{a}_{i},\mathbf{b}_{i})$ communicating by means
of single-path routing. If the network is distillable, then we may write the
following outer bounds for the capacity region%
\begin{align}
R_{i}  &  \leq\min_{C:\mathbf{a}_{i}|\mathbf{b}_{i}}\mathcal{C}(C)~~\text{for
any }i,\label{distooo}\\
R_{i}+R_{j}  &  \leq\min_{C:\mathbf{a}_{i}\mathbf{a}_{j}|\mathbf{b}%
_{i}\mathbf{b}_{j}}\mathcal{C}(C)~~\text{for any }i\neq j\\
&  \vdots\nonumber\\
\sum\limits_{i=1}^{M}R_{i}  &  \leq\min_{C:\{\mathbf{a}_{i}\}|\{\mathbf{b}%
_{i}\}}\mathcal{C}(C), \label{distoooo}%
\end{align}
where $\mathcal{C}(C)$ is the single-edge capacity of cut $C$.
\end{corollary}

Note that we cannot establish the achievability of the outer bounds in
Eqs.~(\ref{distooo})-(\ref{distoooo}), apart from the case $M=1$. This case in
fact corresponds to a unicast distillable network for which the bound is
achievable by solving the widest path problem (see Corollary~\ref{coroNETseq}%
). In general, for $M>1$, achievable lower bounds can be established by
combining the point-to-point composition strategies with classical routing
algorithms that solve the multiple-version of the widest path problem.

\subsection{Multiple-unicast quantum networks with mutipath
routing\label{SecMULTIunicast2}}

Here we consider a quantum network where $M$\ senders $\{\mathbf{a}_{i}\}$ and
$M$ receivers $\{\mathbf{b}_{i}\}$ communicate in a pairwise fashion
$(\mathbf{a}_{i},\mathbf{b}_{i})$ by means of multipath routing. As usual in a
multipath protocol, the points first agree an orientation for the quantum
network. For multiple-unicasts note that both the senders and receivers may
assists one with each other as relays of the network. This means that
$\{\mathbf{a}_{i}\}$ are not necessarily sources and $\{\mathbf{b}_{i}\}$ are
not necessarily sinks, i.e., these sets may have both incoming and outgoing
edges. Given an orientation, each point multicasts to its out-neighborhood
with the assistance of network LOCCs. This flooding process ends when each
edge of the network has been exploited. For the next use, the points may agree
a different orientation, and so on.

The sequence of the orientations together with the sequence of all network
LOCCs (exploited in each orientation) define a multiple-unicast flooding
protocol $\mathcal{P}_{\text{M-unicast}}^{\text{flood}}$. Its output will be a
shared state $\rho_{\{\mathbf{a}_{i}\}\{\mathbf{b}_{i}\}}^{n}$ which is
$\varepsilon$-close to a target state
\begin{equation}
\phi:=%
{\textstyle\bigotimes_{i=1}^{M}}
\phi_{\mathbf{a}_{i}\mathbf{b}_{i}}^{\otimes nR_{i}^{n}}.
\end{equation}
where $\phi_{\mathbf{a}_{i}\mathbf{b}_{i}}$ is a one-bit state (private bit or
ebit) for the pair $(\mathbf{a}_{i},\mathbf{b}_{i})$ and $nR_{i}^{n}$ the
number of its copies. By taking the limit of large $n$ and optimizing over
$\mathcal{P}_{\text{M-unicast}}^{\text{flood}}$, we define the capacity region
associated with the achievable rates $(R_{1}^{\text{m}},\ldots,R_{M}%
^{\text{m}})$ for the various quantum tasks. We can state the following result.

\begin{theorem}
[Multi-unicast with multipaths]\label{TheomultipleUNICAST2}Let us consider a
multiple-unicast quantum network $\mathcal{N}=(P,E)$\ with $M$ sender-receiver
pairs $(\mathbf{a}_{i},\mathbf{b}_{i})$ communicating via multipath routing.
Adopt a simulation of the network with a resource representation
$\sigma(\mathcal{N})=\{\sigma_{\mathbf{xy}}\}_{(\mathbf{x},\mathbf{y})\in E}$.
In particular, $\sigma(\mathcal{N})$\ can be a Choi-representation for a
teleportation-covariant $\mathcal{N}$. We have the following outer bounds for
the capacity region%
\begin{align}
R_{i}^{\text{m}}  &  \leq\min_{C:\mathbf{a}_{i}|\mathbf{b}_{i}}E_{\mathrm{R}%
}^{\text{m}}(C)~~\text{for any }i,\label{rtg}\\
R_{i}^{\text{m}}+R_{j}^{\text{m}}  &  \leq\min_{C:\mathbf{a}_{i}\mathbf{a}%
_{j}|\mathbf{b}_{i}\mathbf{b}_{j}}E_{\mathrm{R}}^{\text{m}}(C)~~\text{for any
}i\neq j\label{rtg2}\\
&  \vdots\nonumber\\
\sum\limits_{i=1}^{M}R_{i}^{\text{m}}  &  \leq\min_{C:\{\mathbf{a}%
_{i}\}|\{\mathbf{b}_{i}\}}E_{\mathrm{R}}^{\text{m}}(C), \label{rtg3}%
\end{align}
where $E_{\mathrm{R}}^{\text{m}}(C):=\sum_{(\mathbf{x},\mathbf{y})\in\tilde
{C}}E_{\mathrm{R}}(\sigma_{\mathbf{xy}})$ is the multi-edge flow of REE across
cut $C$ as in Eq.~(\ref{multiooo}). It is understood that formulations may be
asymptotic for quantum networks with bosonic channels.
\end{theorem}

\textbf{Proof.}~~The proof follows the main steps of the one of
Theorem~\ref{TheomultipleUNICAST}. As before, consider key generation. For the
bipartition $\{\mathbf{a}_{i}\}|\{\mathbf{b}_{i}\}$, the distillable key of
the target state $\phi$ is equal to%
\begin{align}
K_{\mathrm{D}}(\{\mathbf{a}_{i}\}|\{\mathbf{b}_{i}\})_{\phi}  &
=n\sum\limits_{i=1}^{M}R_{i}^{n}\\
&  \leq E_{R}(\{\mathbf{a}_{i}\}|\{\mathbf{b}_{i}\})_{\phi}\\
&  \leq E_{R}(\{\mathbf{a}_{i}\}|\{\mathbf{b}_{i}\})_{\rho^{n}}+\delta
(\varepsilon,d),
\end{align}
which leads to the inequality%
\begin{equation}
\lim_{n}\sum\limits_{i=1}^{M}R_{i}^{n}\leq\underset{n\rightarrow+\infty}%
{\lim~}n^{-1}E_{R}(\{\mathbf{a}_{i}\}|\{\mathbf{b}_{i}\})_{\rho^{n}}~.
\label{repkkkj}%
\end{equation}

For any cut $C:\{\mathbf{a}_{i}\}|\{\mathbf{b}_{i}\}$ of the (simulated)
network, we may write the following decomposition of the output state%
\begin{equation}
\rho_{\{\mathbf{a}_{i}\}\{\mathbf{b}_{i}\}}^{n}(C)=\bar{\Lambda}%
_{\{\mathbf{a}_{i}\}\{\mathbf{b}_{i}\}}^{C}\left[  \underset{(\mathbf{x}%
,\mathbf{y})\in\tilde{C}}{%
{\textstyle\bigotimes}
}~\sigma_{\mathbf{xy}}^{\otimes n}\right]  ,
\end{equation}
for some trace-preserving LOCC $\bar{\Lambda}_{\{\mathbf{a}_{i}\}\{\mathbf{b}%
_{i}\}}^{C}$. Note that here we have $n_{\mathbf{xy}}=n$.\ By replacing
$\rho^{n}=\rho_{\{\mathbf{a}_{i}\}\{\mathbf{b}_{i}\}}^{n}(C)$ in
Eq.~(\ref{repkkkj}), we therefore get%
\begin{equation}
\lim_{n}\sum\limits_{i=1}^{M}R_{i}^{n}\leq\sum\limits_{(\mathbf{x}%
,\mathbf{y})\in\tilde{C}}E_{\mathrm{R}}(\sigma_{\mathbf{xy}}):=E_{\mathrm{R}%
}^{\text{m}}(C).
\end{equation}
The next step is to minimize over the cuts, leading to%
\begin{equation}
\lim_{n}\sum\limits_{i=1}^{M}R_{i}^{n}\leq\min_{C:\{\mathbf{a}_{i}%
\}|\{\mathbf{b}_{i}\}}E_{\mathrm{R}}^{\text{m}}(C).
\end{equation}

Since the latter inequality holds for any protocol $\mathcal{P}%
_{\text{M-unicast}}^{\text{flood}}$, it can be extended to the achievable
rates, i.e., we write
\begin{equation}
\sum\limits_{i=1}^{M}R_{i}^{\text{m}}\leq\min_{C:\{\mathbf{a}_{i}%
\}|\{\mathbf{b}_{i}\}}E_{\mathrm{R}}^{\text{m}}(C).
\end{equation}
Finally, by setting some of the rates equal to zero in the target state, we
may repeat the procedure with respect to different bipartitions and derive all
the remaining conditions in Eqs.~(\ref{rtg})-(\ref{rtg3}). The extension to
asymptotic simulations of bosonic channels is achieved by adopting the weaker
definition of the REE as in Eq.~(\ref{REE_weaker}).~$\blacksquare$

It is immediate to specify the result for distillable networks for which we
may directly write
\begin{equation}
E_{\mathrm{R}}^{\text{m}}(C)=\mathcal{C}^{\text{m}}(C):=\sum
\limits_{(\mathbf{x},\mathbf{y})\in\tilde{C}}\mathcal{C}_{\mathbf{xy}},
\end{equation}
where $\mathcal{C}^{\text{m}}(C)$ is the multi-edge capacity of cut $C$,
already introduced in Eq.~(\ref{Cmpform}). We may write the following.

\begin{corollary}
Consider a multiple-unicast quantum network $\mathcal{N}$\ with $M$
sender-receiver pairs $(\mathbf{a}_{i},\mathbf{b}_{i})$ communicating via
multipath routing. If the network is distillable, then we may write the
following outer bounds for the capacity region%
\begin{align}
R_{i}^{\text{m}}  &  \leq\min_{C:\mathbf{a}_{i}|\mathbf{b}_{i}}\mathcal{C}%
^{\text{m}}(C)~~\text{for any }i,\\
R_{i}^{\text{m}}+R_{j}^{\text{m}}  &  \leq\min_{C:\mathbf{a}_{i}\mathbf{a}%
_{j}|\mathbf{b}_{i}\mathbf{b}_{j}}\mathcal{C}^{\text{m}}(C)~~\text{for any
}i\neq j\\
&  \vdots\nonumber\\
\sum\limits_{i=1}^{M}R_{i}^{\text{m}}  &  \leq\min_{C:\{\mathbf{a}%
_{i}\}|\{\mathbf{b}_{i}\}}\mathcal{C}^{\text{m}}(C),
\end{align}
where $\mathcal{C}^{\text{m}}(C)$ is the multi-edge capacity of cut $C$.
\end{corollary}

Achievable lower bounds may be determined by combining the point-to-point
composition strategy with classical routing algorithms based on the
maximization of multiple flows. For the specific case $M=1$, the outer bound
is achievable and we retrieve the max-flow min-cut theorem for quantum
communications (see Corollary~\ref{coroNETmp}). For $M>2$, achievable lower
bounds may be found by exploiting classical literature on multicommodity flow
algorithms, e.g., Ref.~\cite{TCHu} who showed a version of max-flow min-cut
theorem for undirected networks with two commodities, and
Ref.~\cite{Schrijver} which discusses extensions to more than two commodities.

\subsection{Multicast quantum networks\label{SECmulticastSINGLE}}

Let us now consider a multicast scenario, where Alice $\mathbf{a}$ aims at
simultaneously communicate with a set of $M$\ receivers, i.e., a set of Bobs
$\mathbf{\{b}_{i}\}$. Because of the implicit parallel nature of this
communication process, it is directly formulated under the assumption of
multipath routing. We can easily generalize the description of the one-sender
one-receiver flooding protocol to the present case of multiple receivers.

In a $1$-to-$M$ multicast network protocol, the quantum network $\mathcal{N}$
is subject to an orientation where Alice is treated as a source, while the
various Bobs are destination points, each one being a receiver but also a
potential relay for another receiver (so that they are not necessarily sinks
in the general case). Each end-to-end simultaneous communication between Alice
and the Bobs consists of a sequence of multicasts from each point of the
network to its out-neighborhood, assisted by network LOCCs. This is done in a
flooding fashion so that each edge of the network is exploited. The
orientation of the network may be updated and optimized at each round of the protocol.

The sequence of orientations and the network LOCCs define the multicast
flooding protocol $\mathcal{P}_{\text{multicast}}^{\text{flood}}$. After $n$
uses of the network, Alice and the $M$ Bobs will share an output state
$\rho_{\mathbf{a\{b}_{i}\}}^{n}$ which is $\varepsilon$-close to a target
state
\begin{equation}
\phi:=%
{\textstyle\bigotimes_{i=1}^{M}}
\phi_{\mathbf{ab}_{i}}^{\otimes nR_{i}^{n}}.
\end{equation}
where $\phi_{\mathbf{ab}_{i}}$ is a one-bit state (private bit or ebit) for
the pair of points $(\mathbf{a},\mathbf{b}_{i})$ and $nR_{i}^{n}$ the number
of its copies. Note that this is a compact notation which involves countable
sets of systems $\mathbf{a}=(a,a^{\prime},a^{\prime\prime},\ldots)$ and
$\mathbf{b}_{i}=(b_{i},b_{i}^{\prime},b_{i}^{\prime\prime},\ldots)$.
Therefore, the tensor product $\phi_{\mathbf{ab}_{1}}^{\otimes nR_{1}^{n}%
}\otimes\phi_{\mathbf{ab}_{2}}^{\otimes nR_{2}^{n}}$ explicitly means
$\phi_{ab_{1}}^{\otimes nR_{1}^{n}}\otimes\phi_{a^{\prime}b_{2}^{\prime}%
}^{\otimes nR_{2}^{n}}$, so that there are different systems involved in
Alice's side.

By taking the limit of large $n$ and optimizing over $\mathcal{P}%
_{\text{multicast}}^{\text{flood}}$, we define the capacity region associated
with the achievable rates $(R_{1},\ldots,R_{M})$. In particular, we may define
a unique capacity which is associated with the symmetric condition
$R_{1}=\ldots=R_{M}$. In fact, we may consider a symmetric type of protocol
$\mathcal{\tilde{P}}_{\text{multicast}}^{\text{flood}}$ whose target state
$\phi$ must have $nR_{i}^{n}\geq nR_{n}$ bits for any $i$. Then, by taking the
asymptotic limit of large $n$ and maximizing over all such protocols, we may
define the multicast network capacity%
\begin{equation}
\mathcal{C}^{M}(\mathcal{N})=\sup_{\mathcal{\tilde{P}}_{\text{multicast}%
}^{\text{flood}}}\lim_{n}R_{n}~. \label{multiCC}%
\end{equation}
This rate quantifies the maximum number of target bits per network use
(multipath transmission) that Alice may simultaneously share with each Bob in
the destination set $\mathbf{\{b}_{i}\}$. We have the usual hierarchy
$Q_{2}^{M}(\mathcal{N})=D_{2}^{M}(\mathcal{N})\leq K^{M}(\mathcal{N})$ when we
specify the target state. We can now state the following general bound.

\begin{theorem}
[Quantum multicast]\label{TheomultiCAST}Let us consider a multicast quantum
network $\mathcal{N}$ with one sender and $M$ receivers $\mathbf{\{b}_{i}%
\}$.\ Adopt a simulation of the network with a resource representation
$\sigma(\mathcal{N})=\{\sigma_{\mathbf{xy}}\}_{(\mathbf{x},\mathbf{y})\in E}$.
In particular, $\sigma(\mathcal{N})$\ can be a Choi-representation for a
teleportation-covariant $\mathcal{N}$. Then we have the following outer bounds
for the capacity region
\begin{align}
R_{i}  &  \leq E_{\mathrm{R}}^{\text{m}}(i):=\min_{C:\mathbf{a}|\mathbf{b}%
_{i}}E_{\mathrm{R}}^{\text{m}}(C)~~\text{for any }i,\label{cvcc}\\
R_{i}+R_{j}  &  \leq\min_{C:\mathbf{a}|\mathbf{b}_{i}\mathbf{b}_{j}%
}E_{\mathrm{R}}^{\text{m}}(C)~~\text{for any }i\neq j\\
&  \vdots\nonumber\\
\sum\limits_{i=1}^{M}R_{i}  &  \leq\min_{C:\mathbf{a}|\{\mathbf{b}_{i}%
\}}E_{\mathrm{R}}^{\text{m}}(C), \label{cvcc3}%
\end{align}
where $E_{\mathrm{R}}^{\text{m}}(C)$ is the multi-edge flow of REE\ through
cut $C$. In particular, the multicast network capacity satisfies%
\begin{equation}
\mathcal{C}^{M}(\mathcal{N})\leq\min_{i\in\{1,M\}}E_{\mathrm{R}}^{\text{m}%
}(i). \label{CMNmultics}%
\end{equation}
It is understood that formulations may be asymptotic for quantum networks with
bosonic channels.
\end{theorem}

\textbf{Proof.}~~Consider the upper bound given by secret-key generation. With
respect to the bipartition $\mathbf{a}|\{\mathbf{b}_{i}\}$, we may write the
usual steps starting form the distillable key of the target state
\begin{align}
K_{\mathrm{D}}(\mathbf{a}|\{\mathbf{b}_{i}\})_{\phi}  &  =n\sum\limits_{i=1}%
^{M}R_{i}^{n}\\
&  \leq E_{\mathrm{R}}(\mathbf{a}|\{\mathbf{b}_{i}\})_{\phi}\\
&  \leq E_{\mathrm{R}}(\mathbf{a}|\{\mathbf{b}_{i}\})_{\rho^{n}}%
+\delta(\varepsilon,d),
\end{align}
leading to the asymptotic limit%
\begin{equation}
\lim_{n}\sum\limits_{i=1}^{M}R_{i}^{n}\leq\underset{n\rightarrow+\infty}{\lim
}~n^{-1}E_{\mathrm{R}}(\mathbf{a}|\{\mathbf{b}_{i}\})_{\rho^{n}}.
\label{repkkkj2}%
\end{equation}

For any cut $C:\mathbf{a}|\{\mathbf{b}_{i}\}$ of the (simulated) network, we
may write the decomposition%
\begin{equation}
\rho_{\mathbf{a}\{\mathbf{b}_{i}\}}^{n}(C)=\bar{\Lambda}_{\mathbf{a}%
\{\mathbf{b}_{i}\}}^{C}\left[  \underset{(\mathbf{x},\mathbf{y})\in\tilde{C}}{%
{\textstyle\bigotimes}
}~\sigma_{\mathbf{xy}}^{\otimes n}\right]  ,
\end{equation}
for some trace-preserving LOCC $\bar{\Lambda}_{\mathbf{a}\{\mathbf{b}_{i}%
\}}^{C}$. By replacing $\rho^{n}=\rho_{\mathbf{a}\{\mathbf{b}_{i}\}}^{n}(C)$
in Eq.~(\ref{repkkkj2}), we therefore get%
\begin{equation}
\lim_{n}\sum\limits_{i=1}^{M}R_{i}^{n}\leq\sum\limits_{(\mathbf{x}%
,\mathbf{y})\in\tilde{C}}E_{\mathrm{R}}(\sigma_{\mathbf{xy}}):=E_{\mathrm{R}%
}^{\text{m}}(C).
\end{equation}
By minimizing over the cuts and maximizing over the protocols, we may write%
\begin{equation}
\sum\limits_{i=1}^{M}R_{i}\leq\min_{C:\mathbf{a}|\{\mathbf{b}_{i}%
\}}E_{\mathrm{R}}^{\text{m}}(C).
\end{equation}

The other conditions in Eqs.~(\ref{cvcc})-(\ref{cvcc3}) are obtained by
setting part of the rates $R_{i}^{n}$ to zero in the target state (as in the
previous proofs). In particular, set $R_{i}^{n}\neq0$ for some $i$, while
$R_{j}^{n}=0$ for any $j\neq i$. The target state becomes $\phi:=\phi
_{\mathbf{ab}_{i}}^{\otimes nR_{i}^{n}}\otimes\sigma_{\text{sep}}$ and we
repeat the derivation with respect to the bipartition $\mathbf{a|b}_{i}$. This
leads to
\begin{equation}
\lim_{n}R_{i}^{n}\leq\underset{n\rightarrow+\infty}{\lim}~n^{-1}E_{\mathrm{R}%
}(\mathbf{a}|\mathbf{b}_{i})_{\rho^{n}}, \label{multi11}%
\end{equation}
where we may directly consider the reduced state
\begin{equation}
\rho^{n}=\rho_{\mathbf{ab}_{i}}^{n}=\mathrm{Tr}_{\{\mathbf{b}_{j\neq i}%
\}}\left[  \rho_{\mathbf{a}\{\mathbf{b}_{1},\ldots,\mathbf{b}_{M}\}}%
^{n}\right]  .
\end{equation}

For any cut $C:\mathbf{a}|\mathbf{b}_{i}$, we therefore have%
\begin{equation}
\rho_{\mathbf{ab}_{i}}^{n}(C)=\bar{\Lambda}_{\mathbf{ab}_{i}}^{C}\left[
\underset{(\mathbf{x},\mathbf{y})\in\tilde{C}}{%
{\textstyle\bigotimes}
}~\sigma_{\mathbf{xy}}^{\otimes n}\right]  , \label{multi22}%
\end{equation}
which leads to $\lim_{n}R_{i}^{n}\leq E_{\mathrm{R}}^{\text{m}}(C)$. By
minimizing over the cuts, one gets
\begin{equation}
\lim_{n}R_{i}^{n}\leq E_{\mathrm{R}}^{\text{m}}(i):=\min_{C:\mathbf{a}%
|\mathbf{b}_{i}}E_{\mathrm{R}}^{\text{m}}(C). \label{bbbmmm}%
\end{equation}
Since this is true for any protocol $\mathcal{P}^{M}$, it can be extended to
the achievable rates, i.e., we get Eq.~(\ref{cvcc}).

For the multicast network capacity, just note that
\begin{equation}
\lim_{n}R^{n}\leq\min_{i}\{\lim_{n}R_{i}^{n}\}.
\end{equation}
Therefore, from Eq.~(\ref{bbbmmm}), we may write%
\begin{equation}
\lim_{n}R^{n}\leq\min_{i}E_{\mathrm{R}}^{\text{m}}(i).
\end{equation}
This is true for any symmetric protocol $\mathcal{P}_{\text{sym}}^{M}$ which
leads to the result of Eq.~(\ref{CMNmultics}). Results are extended to
asymptotic simulations of bosonic channels in the usual way.~$\blacksquare$

As usual, in the case of distillable networks, we may prove stronger results.
As a direct consequence of Theorem~\ref{TheomultiCAST}, we may write the
following cutset bound.

\begin{corollary}
Consider a multicast quantum network $\mathcal{N}$ with one sender and $M$
receivers $\mathbf{\{b}_{i}\}$.\ If the network is distillable, then we have
the following outer bounds for the capacity region
\begin{align}
R_{i}  &  \leq\mathcal{C}^{\text{m}}(i)=\min_{C:\mathbf{a}|\mathbf{b}_{i}%
}\mathcal{C}^{\text{m}}(C)~~\text{for any }i,\\
R_{i}+R_{j}  &  \leq\min_{C:\mathbf{a}|\mathbf{b}_{i}\mathbf{b}_{j}%
}\mathcal{C}^{\text{m}}(C)~~\text{for any }i\neq j\\
&  \vdots\nonumber\\
\sum\limits_{i=1}^{M}R_{i}  &  \leq\min_{C:\mathbf{a}|\{\mathbf{b}_{i}%
\}}\mathcal{C}^{\text{m}}(C),
\end{align}
where $\mathcal{C}^{\text{m}}(C)$ is the multi-edge capacity of cut $C$ and
$\mathcal{C}^{\text{m}}(i)$ is the multipath capacity between the sender and
the $i$th receiver (in a unicast setting). In particular, the multicast
network capacity must satisfy the bound%
\begin{equation}
\mathcal{C}^{M}(\mathcal{N})\leq\min_{i\in\{1,M\}}\mathcal{C}^{\text{m}}(i)~.
\label{cutsetDIST}%
\end{equation}

\end{corollary}

\subsection{Network coding for quantum key distribution \label{SecNETCODING}}

Our previous results refer to the general case of multiple independent
messages. In a multicast quantum network, this means that Alice distributes
$M$ different sequences of\ target bits to the $M$ Bobs $\{\mathbf{b}_{i}\}$.
For instance, these may represent $M$ different secret keys, one for each Bob
in the destination set. For this specific task (key distribution), the
multicast capacity of the network $\mathcal{C}^{M}(\mathcal{N})$ becomes a
multicast secret-key capacity $\mathcal{K}^{M}(\mathcal{N})$.

In quantum key distribution, it is also interesting to consider the variant
scenario where Alice distributes exactly the same secret key to all Bobs
$\{\mathbf{b}_{i}\}$, for instance, to enable a quantum-secured conference
among these parties. For this particular task, we may define a single-key
version for the multicast secret-key capacity, that we denote as
$\mathcal{K}_{\text{1-key}}^{M}(\mathcal{N})$. This represents the maximum
rate at which Alice$\mathbf{\ }$may distribute the same secret key to all Bobs
in each parallel use of the network.

Here some considerations are in order. First of all, it is clear that
$\mathcal{K}_{\text{1-key}}^{M}(\mathcal{N})\geq\mathcal{K}^{M}(\mathcal{N})$
just because the various Bobs may compose their keys to distil a common key.
This also means that the cutset bound in Eq.~(\ref{cutsetDIST}) does not
automatically apply to $\mathcal{K}_{\text{1-key}}^{M}(\mathcal{N})$. Another
reason for this is because, more generally, we may include the possibility of
network coding~\cite{Gamal} in the definition of $\mathcal{K}_{\text{1-key}%
}^{M}(\mathcal{N})$. This means that the intermediate nodes of the network not
only help with the multipath routing of the quantum systems from Alice to the
Bobs, but they may also apply operations on the incoming systems before their
subsequent transmission. For this reason, $\mathcal{K}_{\text{1-key}}%
^{M}(\mathcal{N})$ may be greater than a routing capacity.

In a multicast distillable network, we may therefore combine point-to-point
key generation sessions with linear network coding~\cite{Gamal}. In this way
we may find a lower bound to the single-key multicast capacity $\mathcal{K}%
_{\text{1-key}}^{M}(\mathcal{N})$ by exploiting the network coding
theorem~\cite{netcod1,netcod2,netcod3}.

\begin{theorem}
[Network coding for QKD]Consider a distillable network $\mathcal{N}$. Then,
the single-key multicast capacity between one sender and $M$ receivers
satisfies%
\begin{equation}
\mathcal{K}_{\text{1-key}}^{M}(\mathcal{N})\geq\min_{i\in\{1,M\}}%
\mathcal{K}_{i}^{\text{m}}~, \label{cutsetnnn}%
\end{equation}
where $\mathcal{K}_{i}^{\text{m}}$ is the multipath secret-key capacity
between the sender and the $i$th receiver (in a unicast setting).
\end{theorem}

\textbf{Proof.}~~The proof repeats some of the steps of the previous proofs
for multipath routing. First of all we transform the quantum network into a
directed network where each undirected edge is split in two directed edges.
The Alice-Bob cut properties of the original quantum network and the new
directed graphical network are exactly the same if we consider a corresponding
\textquotedblleft directed\textquotedblright\ definition for the cut-sets. In
particular, the cutset bound in Eq.~(\ref{cutsetnnn}) remains the same for the
directed network under the re-definition of the cut-sets.

An optimal key distribution protocol goes as follows. The points distill
$n\mathcal{C}_{\mathbf{xy}}$ ebits along each (undirected) edge. These ebits
are then used to teleport orthogonal states along the directed edges of the
oriented graphical network. Let us call $k$ Alice's secret variable, uniformly
chosen and encoding $R$ bits. After $n$ extractions of $k$, we have a sequence
$k^{n}$ of $nR$ bits. Let us split this sequence into $m$ blocks
$k^{n}:=(k_{1}^{n},\ldots,k_{m}^{n})$, where each block $k_{i}^{n}$ contains
$nm^{-1}R$ bits. For large $n$, we may always assume that $q:=nm^{-1}R$ is an
integer, so that each block corresponds to an element of the finite field
$\mathrm{GF}(q)$.

The blocks are then subject to a linear coding transformation, i.e., Alice
computes the output
\begin{equation}
k_{\mathbf{a}\rightarrow}^{n}:=\left(
\begin{array}
[c]{c}%
\tilde{k}_{1}^{n}\\
\vdots\\
\tilde{k}_{m}^{n}%
\end{array}
\right)  =\left(
\begin{array}
[c]{ccc}%
\alpha_{11} & \cdots & \alpha_{1m}\\
\vdots & \ddots & \vdots\\
\alpha_{m1} & \cdots & \alpha_{mm}%
\end{array}
\right)  \left(
\begin{array}
[c]{c}%
k_{1}^{n}\\
\vdots\\
k_{m}^{n}%
\end{array}
\right)  ,
\end{equation}
with some coefficients $\alpha_{ij}\in\mathrm{GF}(q)$. The generic block
$\tilde{k}_{i}^{n}$ is encoded into an orthogonal set of pure states
$|\tilde{k}_{i}^{n}\rangle$ and teleported to a neighbor point $\mathbf{y}\in
N^{\text{out}}(\mathbf{a})$ by means of the $n\mathcal{C}_{\mathbf{ay}}$
shared ebits. Alice communicates both the dimension of the basis $\{|\tilde
{k}_{i}^{n}\rangle\}$ and the outcome of the Bell detection to point
$\mathbf{y}$. The latter will apply the correction unitary and then detect the
state with the POVM $\{|\tilde{k}_{i}^{n}\rangle\langle\tilde{k}_{i}^{n}|\}$,
so to extract $\tilde{k}_{i}^{n}$ without errors. In this way, the blocks of
the sequence $k_{\mathbf{a}\rightarrow}^{n}$ are all teleported from Alice to
her neighborhood $N^{\text{out}}(\mathbf{a})$.

In turn, each point $\mathbf{x}$ of the network will receive a number of
teleported states which will be measured and decoded into the blocks of an
input sequence $k_{\rightarrow\mathbf{x}}^{n}$. The latter will be subject to
linear coding with coefficients $\alpha_{ij}^{\mathbf{x}}$ and transformed
into an output sequence $k_{\mathbf{x}\rightarrow}^{n}$ whose blocks are
encoded into orthogonal states and then teleported to neighbor points, and so
on. In this way, we have transformed the original network into a teleportation
network where orthogonal states are used to securely transfer blocks of the
secret key through the points of the network, with the only limitation being
provided by the point-to-point capacities $\mathcal{C}_{\mathbf{xy}}$.

Security is provided by the pre-distillation of the ebits, while the effective
secret-key transfer has become equivalent to solving the transfer of classical
bits in a directed network, thanks to teleportation. For this reason we can
apply the classical network coding theorem~\cite{netcod1,netcod2,netcod3},
which states that the optimal achievable rate $R$ is equal to the cutset bound
(e.g., see Theorem~15.3 of Ref.~\cite{Gamal}). Here, this means that the
single-key multicast capacity $\mathcal{K}_{\text{1-key}}^{M}(\mathcal{N})$
satisfies the lower bound in Eq.~(\ref{cutsetnnn}).~$\blacksquare$

\subsection{Multiple-multicast quantum networks\label{SECmulticastMANY}}

In the multiple-multicast quantum network, we have $M_{A}$ Alices
$\{\mathbf{a}_{1},\ldots,\mathbf{a}_{i},\ldots,\mathbf{a}_{M_{A}}\}$, each of
them communicating with the same destination set of $M_{B}$ Bobs
$\{\mathbf{b}_{1},\ldots,\mathbf{b}_{j},\ldots,\mathbf{b}_{M_{B}}\}$ by means
of multipath routing. Each end-to-end multicast $\mathbf{a}_{i}\rightarrow
\{\mathbf{b}_{j}\}$ is associated with the distribution of $M_{B}$ independent
sequences of target bits (e.g., secret keys) between the $i$th Alice
$\mathbf{a}_{i}$ and each Bob $\mathbf{b}_{j}$ in the destination set. The
description of a multiple-multicast protocol for a quantum network follows the
same main features discussed for the case of a single-multicast network
($M_{A}=1$). Because we have multiple senders and receivers, here we need to
consider all possible orientations of the network. Each use of the quantum
network is performed under some orientation which is adopted by the points for
their point-to-point out-neighborhood multicasts, suitably assisted by network
LOCCs. Use after use, these steps define a multiple-multicast flooding
protocol $\mathcal{P}_{\text{M-multicast}}^{\text{flood}}$.

After $n$ uses, the ensembles of Alices and Bobs share an output state
$\rho_{\{\mathbf{a}_{i}\}\mathbf{\{b}_{j}\}}^{n}$ which is $\varepsilon$-close
to a target state
\begin{equation}
\phi:=%
{\textstyle\bigotimes_{i=1}^{M_{A}}}
{\textstyle\bigotimes_{j=1}^{M_{B}}}
\phi_{\mathbf{a}_{i}\mathbf{b}_{j}}^{\otimes nR_{ij}^{n}}.
\end{equation}
where $\phi_{\mathbf{a}_{i}\mathbf{b}_{j}}$ is a one-bit state (private bit or
ebit) for the pair $(\mathbf{a}_{i},\mathbf{b}_{j})$ and $nR_{ij}^{n}$ the
number of its copies. By taking the limit of large $n$ and optimizing over
$\mathcal{P}_{\text{M-multicast}}^{\text{flood}}$, we define the capacity
region for the achievable rates $\{R_{ij}\}$. Assume the symmetric case where
the $i$th Alice $\mathbf{a}_{i}$ achieves the same rate $R_{i1}=\ldots
=R_{iM_{B}}$ with all Bobs $\{\mathbf{b}_{j}\}$. This means to consider
symmetric protocols whose target state $\phi$ must have $\min_{j}R_{ij}%
^{n}\geq R_{i}^{n}$ bits for any $i$. By taking the asymptotic limit of
$R_{i}^{n}$ for large $n$ and maximizing over all these symmetric protocols,
we may define the capacity region for the achievable multicast rates
$(R_{1},\ldots,R_{M_{A}})$. In the latter set, rate $R_{i}$ provides the
minimum number of target bits per use that the $i$th Alice may share with each
Bob in the destination set $\mathbf{\{b}_{j}\}$ (in the multi-message setting,
i.e., assuming independent sequences shared with the various Bobs). We have
the following outer bounds to the capacity region.

\begin{theorem}
[Quantum multiple-multicast]\label{Theomultimulti}Let us consider a
multiple-multicast quantum network $\mathcal{N}=(P,E)$ where each of the
$M_{A}$ senders $\mathbf{\{a}_{i}\}$ communicates with $M_{B}$ receivers
$\mathbf{\{b}_{j}\}$ at the multicast rate $R_{i}$. Adopt a simulation of
$\mathcal{N}$ with some resource representation $\sigma(\mathcal{N}%
)=\{\sigma_{\mathbf{xy}}\}_{(\mathbf{x},\mathbf{y})\in E}$, which may be a
Choi-representation for a teleportation-covariant $\mathcal{N}$. Then, we have
the following outer bounds for the capacity region%
\begin{align}
R_{i}  &  \leq\min_{\substack{C|\mathbf{a}_{i}\in\mathbf{A}\\\{\mathbf{b}%
_{j}\}\cap\mathbf{B}\neq\emptyset}}E_{\mathrm{R}}^{\text{m}}%
(C),\label{setCVB0}\\
R_{i}+R_{j}  &  \leq\min_{\substack{C|\mathbf{a}_{i},\mathbf{a}_{j}%
\in\mathbf{A}\\\{\mathbf{b}_{j}\}\cap\mathbf{B}\neq\emptyset}}E_{\mathrm{R}%
}^{\text{m}}(C),\\
&  \vdots\nonumber\\
\sum\limits_{i=1}^{M_{A}}R_{i}  &  \leq\min_{\substack{C|\{\mathbf{a}%
_{i}\}\subseteq\mathbf{A}\\\{\mathbf{b}_{j}\}\cap\mathbf{B}\neq\emptyset
}}E_{\mathrm{R}}^{\text{m}}(C), \label{setCVB}%
\end{align}
where $E_{\mathrm{R}}^{\text{m}}(C)$ is the multi-edge flow of REE\ through
cut $C$. For a distillable network, we may write the bounds in
Eqs.~(\ref{setCVB0})-(\ref{setCVB}) with $E_{\mathrm{R}}^{\text{m}%
}(C)=\mathcal{C}^{\text{m}}(C)$, i.e., in terms of the multi-edge capacity of
the cuts.
\end{theorem}

\textbf{Proof.}~~The proof is again similar to previous ones. Consider the
upper bound given by secret-key generation. With respect to the bipartition
$\{\mathbf{a}_{i}\}|\{\mathbf{b}_{j}\}$, we can manipulate the distillable key
$K_{\mathrm{D}}$ of the target state $\phi$ as follows
\begin{align}
K_{\mathrm{D}}(\{\mathbf{a}_{i}\}|\{\mathbf{b}_{j}\})_{\phi}  &
=n\sum\limits_{i=1}^{M_{A}}\sum\limits_{j=1}^{M_{B}}R_{ij}^{n}\\
&  \leq E_{\mathrm{R}}(\{\mathbf{a}_{i}\}|\{\mathbf{b}_{j}\})_{\phi}\\
&  \leq E_{\mathrm{R}}(\{\mathbf{a}_{i}\}|\{\mathbf{b}_{j}\})_{\rho^{n}%
}+\delta(\varepsilon,d),
\end{align}
leading to the asymptotic limit%
\begin{equation}
\lim_{n}\sum\limits_{i=1}^{M_{A}}\sum\limits_{j=1}^{M_{B}}R_{ij}^{n}%
\leq\underset{n\rightarrow+\infty}{\lim}~n^{-1}E_{\mathrm{R}}(\{\mathbf{a}%
_{i}\}|\{\mathbf{b}_{j}\})_{\rho^{n}}. \label{repkkkj3}%
\end{equation}

For any cut $C:\{\mathbf{a}_{i}\}|\{\mathbf{b}_{j}\}$ of the (simulated)
network, we may write the decomposition%
\begin{equation}
\rho_{\{\mathbf{a}_{i}\}\{\mathbf{b}_{j}\}}^{n}(C)=\bar{\Lambda}%
_{\{\mathbf{a}_{i}\}\{\mathbf{b}_{j}\}}^{C}\left[  \underset{(\mathbf{x}%
,\mathbf{y})\in\tilde{C}}{%
{\textstyle\bigotimes}
}\sigma_{\mathbf{xy}}^{\otimes n}\right]  ,
\end{equation}
and manipulate Eq.~(\ref{repkkkj3}) into the following%
\begin{equation}
\lim_{n}\sum\limits_{i=1}^{M_{A}}\sum\limits_{j=1}^{M_{B}}R_{ij}^{n}\leq
\sum\limits_{(\mathbf{x},\mathbf{y})\in\tilde{C}}E_{\mathrm{R}}(\sigma
_{\mathbf{xy}}):=E_{\mathrm{R}}^{\text{m}}(C).
\end{equation}
By minimizing over the cuts and maximizing over the protocols, we may write%
\begin{equation}
\sum\limits_{i=1}^{M_{A}}\sum\limits_{j=1}^{M_{B}}R_{ij}\leq\min
_{C:\{\mathbf{a}_{i}\}|\{\mathbf{b}_{j}\}}E_{\mathrm{R}}^{\text{m}}(C)~.
\end{equation}

By setting part of the rates $R_{ij}^{n}$ to zero in the target state, we
derive the full set of conditions
\begin{align}
\sum\limits_{i=1}^{M_{A}}\sum\limits_{j=1}^{M_{B}}R_{ij}  &  \leq
\min_{C:\{\mathbf{a}_{i}\}|\{\mathbf{b}_{j}\}}E_{\mathrm{R}}^{\text{m}%
}(C),\label{xzx1}\\
&  \vdots\nonumber\\
R_{ij}+R_{kl}  &  \leq\min_{C:\mathbf{a}_{i}\mathbf{a}_{k}|\mathbf{b}%
_{j}\mathbf{b}_{l}}E_{\mathrm{R}}^{\text{m}}(C),\label{xzx2}\\
R_{ij}  &  \leq\min_{C:\mathbf{a}_{i}|\mathbf{b}_{j}}E_{\mathrm{R}}^{\text{m}%
}(C). \label{xzx3}%
\end{align}

The latter conditions are valid for the end-to-end rates $R_{ij}$ achievable
between each pair $(\mathbf{a}_{i},\mathbf{b}_{j})$. We are interested in the
achievable multicast rates $\{R_{i}\}$ between each sender $\mathbf{a}_{i}$
and all receivers $\{\mathbf{b}_{j}\}$. Corresponding conditions can be
derived by considering a subset of protocols with target state of the type
\begin{equation}
\phi_{k}:=%
{\textstyle\bigotimes_{i=1}^{M_{A}}}
\phi_{\mathbf{a}_{i}\mathbf{b}_{k}}^{\otimes nR_{ik}^{n}}\otimes
\sigma_{\text{sep}},
\end{equation}
for some $k$, where all Alices $\{\mathbf{a}_{i}\}$\ aim to optimize their
rates $\{R_{ik}^{n}\}$ with some fixed Bob $\mathbf{b}_{k}$, so that
$R_{ij}^{n}=0$ for any $j\neq k$. By repeating the previous steps with respect
to the bipartition $\{\mathbf{a}_{i}\}|\mathbf{b}_{k}$, we obtain%
\begin{equation}
\lim_{n}\sum\limits_{i=1}^{M_{A}}R_{ik}^{n}\leq\min_{C:\{\mathbf{a}%
_{i}\}|\mathbf{b}_{k}}E_{\mathrm{R}}^{\text{m}}(C).
\end{equation}
Since we have $R_{i}^{n}\leq\min_{j}R_{ij}^{n}\leq R_{ik}^{n}$ for any $k$, we
can then write the same inequality for $\lim_{n}\sum\nolimits_{i=1}^{M_{A}%
}R_{i}^{n}$. Then, by optimizing over the protocols, we get%
\begin{equation}
\sum\limits_{i=1}^{M_{A}}R_{i}\leq\min_{C:\{\mathbf{a}_{i}\}|\mathbf{b}_{k}%
}E_{\mathrm{R}}^{\text{m}}(C).
\end{equation}
Because the latter expression is true for any $k$, we may equivalently write%
\begin{equation}
\sum\limits_{i=1}^{M_{A}}R_{i}\leq\min_{C}E_{\mathrm{R}}^{\text{m}}(C),
\label{yyu1}%
\end{equation}
with $C=(\mathbf{A},\mathbf{B})$ such that $\{\mathbf{a}_{i}\}\subseteq
\mathbf{A}$ and $\{\mathbf{b}_{j}\}\cap\mathbf{B}\neq\emptyset$.

Now, for any fixed $k$, impose that the rates $\{R_{ik}^{n}\}$ are zero for
some of the Alices $\{\mathbf{a}_{i}\}$. If we only have $R_{ik}^{n}\neq0$ for
a pair $(\mathbf{a}_{i},\mathbf{b}_{k})$, then the condition $R_{i}^{n}\leq
R_{ik}^{n}$ leads to%
\begin{equation}
R_{i}\leq\min_{C:\mathbf{a}_{i}|\mathbf{b}_{k}}E_{\mathrm{R}}^{\text{m}}(C).
\end{equation}
Because the latter is true for any $k$, we may then write%
\begin{equation}
R_{i}\leq\min_{C}E_{\mathrm{R}}^{\text{m}}(C), \label{yyu2}%
\end{equation}
with $C=(\mathbf{A},\mathbf{B})$ such that $\mathbf{a}_{i}\in\mathbf{A}$ and
$\{\mathbf{b}_{j}\}\cap\mathbf{B}\neq\emptyset$. Extending the previous
reasoning to two non-zero rates $R_{ik}^{n}\neq0$ and $R_{jk}^{n}\neq0$ leads
to%
\begin{equation}
R_{i}+R_{j}\leq\min_{C}E_{\mathrm{R}}^{\text{m}}(C), \label{yyu3}%
\end{equation}
with $C=(\mathbf{A},\mathbf{B})$ such that $\mathbf{a}_{i},\mathbf{a}_{j}%
\in\mathbf{A}$ and $\{\mathbf{b}_{j}\}\cap\mathbf{B}\neq\emptyset$. Other
similar conditions can be derived for the multicast rates, so that we get the
result of Eqs.~(\ref{setCVB0})-(\ref{setCVB}). Finally, for a distillable
network we have $E_{\mathrm{R}}^{\text{m}}(C)=\mathcal{C}^{\text{m}}(C)$ and,
therefore, it is immediate to express these results in terms of the multi-edge
capacities of the cuts.~$\blacksquare$

\section{Conclusions\label{SECconclusions}}

This extended information-theoretic work has investigated the ultimate
end-to-end rates for transmitting quantum information, distributing
entanglement and generating secret keys between two end-points of a repeater
chain and, more generally, of an arbitrary quantum network under single- or
multi-path routing strategies. We have established a general single-letter REE
upper bound for these end-to-end capacities which applies to chains and
networks of any topology and type, i.e., connected by completely arbitrary
channels of any dimension (finite or infinite). In fundamental cases, such
bound is so tight that it coincides with suitably-constructed lower bounds. In
this way, we have determined the end-to-end capacities of chains and networks
affected by the most relevant models of decoherence for CV and DV systems,
including loss, quantum-limited amplification, dephasing and erasure. Our
theory can also be extended to simultaneous quantum communication between
multiple senders and receivers in a network, even though we cannot prove the
achievability of the upper bound in this general case.

From a methodological point of view, we have shown how to simulate a quantum
network so as to replace all its quantum channels with an ensemble of resource
states; this is what we have called a resource representation of the network.
This is particularly simple to identify when the channels are
teleportation-covariant, so that the simulation is done via teleportation and
the resource states are just the Choi matrices of the channels (Choi-based
representation). Starting from a simulation of the network, we have then
applied teleportation stretching and reduced any adaptive network protocol
into a much simpler block version, where the output state is expressed in
terms of tensor products of resource states. More powerfully, we have combined
this technique with suitable entanglement cuts of the network so that the
decomposition of the output state undergoes a drastic reduction in the number
of resource states. Using this improved decomposition with a general weak
converse bound based on the relative entropy of entanglement, we have then
derived single-letter \textquotedblleft cutset\textquotedblright\ upper bounds
for the various end-to-end capacities. This result holds for chains and
networks connected by arbitrary channels at any dimension. It holds for
different types of routing and can also be extended to multiple senders and receivers.

In order to derive lower bounds, we have combined point-to-point quantum
protocols with classical routing strategies. For single-path routing between
two end-points, the optimal solution is reduced to finding the widest path in
a network. For multipath routing, we need to maximize the flow of qubits from
the \textquotedblleft source\textquotedblright\ (Alice) to the
\textquotedblleft sink\textquotedblright\ (Bob) and the optimal solution is
provided by the max-flow min-cut theorem. In this setting, let us remark that
the \textquotedblleft flooding\textquotedblright\ condition is crucial to
achieve a maximum flow of quantum information and, therefore, to extend the
max-flow min-cut theorem to quantum communications. Remarkably, these lower
bounds coincide with the upper bounds in the case of distillable networks,
i.e., networks connected by distillable channels such as bosonic lossy
channels, quantum-limited amplifiers, dephasing and erasure channels. Thus,
the end-to-end capacities of (unicast) distillable networks are completely
established with extremely simple formulas.

An important practical application is clearly for optical and telecom quantum
communications, where bosonic loss is the main cause of decoherence in fibers
and free-space links, especially at long distances, e.g., in connections with
satellites. In the specific optical/telecom setting, our results establish the
fundamental rate-loss scaling affecting repeater-assisted and network-based
quantum and private communications. This trade-off sets an exact limit to the
optimal performance of any end-to-end QKD protocol, which is performed in
repeater chains or quantum networks, therefore generalizing the fundamental
repeaterless limit discovered in Ref.~\cite{QKDpaper}. In particular, we now
have the full \textquotedblleft meter\textquotedblright\ for assessing the
performance of quantum repeaters: Not only we can establish if a repeater is
beating the point-to-point benchmark, i.e., the PLOB bound~\cite{QKDpaper},
but we may also analyze how far it is from the optimal rate allowed by quantum mechanics.

\textbf{Acknowledgments}.~This work has been supported by the EPSRC\ via the
`UK Quantum Communications HUB' (EP/M013472/1) and `qDATA' (EP/L011298/1).
S.P. would like to thank Richard Wilson, Edwin Hancock, Rod Van Meter, Marco
Lucamarini, Riccardo Laurenza, Carlo Ottaviani, Gaetana Spedalieri, Cosmo
Lupo, Samuel Braunstein, Seth Lloyd and Saikat Guha.


\begin{thebibliography}{999}
\bibitem {Acyclic}Without loss of generality, the graph may be considered to
be acyclic.
\bibitem {OneDIR}Note that one may always enforce a single direction between
$\mathbf{x}$ and $\mathbf{y}$ by introducing an artificial point $\mathbf{z}%
$\ in one of the two directed edges. For instance, we may keep $(\mathbf{x}%
,\mathbf{y})$ as is, while replacing $(\mathbf{y},\mathbf{x})$ with
$(\mathbf{y},\mathbf{z})$ and $(\mathbf{z},\mathbf{x})$, both having the same
capacity of $(\mathbf{y},\mathbf{x})$. This further modification does not
affect the maximum flow value and the minimum cut capacity, but increases the
complexity of the network.
\end{thebibliography}
\end{document}